\documentclass[english]{llncs}
\usepackage[latin9]{inputenc}
\usepackage{array}
\usepackage{url}
\usepackage{multirow}
\usepackage{amsmath}
\usepackage{amssymb}
\usepackage{graphicx}
\usepackage{rotating}

\makeatletter

\providecommand{\tabularnewline}{\\}

\usepackage{tikz}
\usepackage{comment}
\usepackage{color}

\usepackage{epsfig}

\usepackage[pagebackref=true,breaklinks=true,colorlinks,bookmarks=false]{hyperref}
\usepackage{colortbl}
\usepackage{hhline}
\usepackage{caption}
\usepackage[export]{adjustbox}

\definecolor{gray0}{gray}{0.95}
\definecolor{gray1}{gray}{0.85}
\definecolor{gray2}{gray}{0.75}

\newcolumntype{d}{>{\columncolor{gray0}}c}
\newcolumntype{a}{>{\columncolor{gray1}}c}
\newcolumntype{b}{>{\columncolor{gray2}}c}
\newcolumntype{?}{!{\color{blue}\vline width 2pt}}

\@ifundefined{showcaptionsetup}{}{%
 \PassOptionsToPackage{caption=false}{subfig}}
\usepackage{subfig}
\makeatother

\usepackage{babel}
\begin{document}
\renewcommand{\contentsname}{Table of Content for Appendix}
\pagestyle{plain}
\let\oldaddcontentsline\addcontentsline 
\def\addcontentsline#1#2#3{}

\title{Towards Effective and Robust Neural Trojan Defenses via Input Filtering}
\author{Kien Do\inst{1} \and Haripriya Harikumar\inst{1} \and Hung Le\inst{1} \and Dung Nguyen\inst{1} \and Truyen Tran\inst{1} \and Santu Rana\inst{1} \and Dang Nguyen\inst{1} \and Willy Susilo\inst{2} \and Svetha Venkatesh\inst{1}}
\institute{Applied Artificial Intelligence Institute (A2I2), Deakin University, Australia \and University of Wollongong, Australia\\
\email{\{k.do, h.harikumar, thai.le, dung.nguyen, truyen.tran, santu.rana, d.nguyen, svetha.venkatesh\}@deakin.edu.au}\\
\email{wsusilo@uow.edu.au}}
\maketitle

\global\long\def\Expect{\mathbb{E}}%
\global\long\def\Real{\mathbb{R}}%
\global\long\def\Data{\mathcal{D}}%
\global\long\def\Loss{\mathcal{L}}%
\global\long\def\Classifier{\mathtt{C}}%
\global\long\def\Generator{\mathtt{G}}%
\global\long\def\Filter{\mathtt{F}}%
\global\long\def\argmin#1{\underset{#1}{\text{argmin }}}%
\global\long\def\argmax#1{\underset{#1}{\text{argmax }}}%

\def\addcontentsline#1#2#3{\oldaddcontentsline{#1}{#2}{#3}}
\thispagestyle{plain}
\begin{abstract}
Trojan attacks on deep neural networks are both dangerous and surreptitious.
Over the past few years, Trojan attacks have advanced from using only
a single input-agnostic trigger and targeting only one class to using
multiple, input-specific triggers and targeting multiple classes.
However, Trojan defenses have not caught up with this development.
Most defense methods still make inadequate assumptions about Trojan
triggers and target classes, thus, can be easily circumvented by modern
Trojan attacks. To deal with this problem, we propose two novel \emph{``filtering''}
defenses called \emph{Variational Input Filtering (VIF)} and \emph{Adversarial
Input Filtering (AIF)} which leverage lossy data compression and adversarial
learning respectively to effectively purify potential Trojan triggers
in the input at run time without making assumptions about the number
of triggers/target classes or the input dependence property of triggers.
In addition, we introduce a new defense mechanism called \emph{``Filtering-then-Contrasting''}
(FtC) which helps avoid the drop in classification accuracy on clean
data caused by ``filtering'', and combine it with VIF/AIF to derive
new defenses of this kind. Extensive experimental results and ablation
studies show that our proposed defenses significantly outperform well-known
baseline defenses in mitigating five advanced Trojan attacks including
two recent state-of-the-art while being quite robust to small amounts
of training data and large-norm triggers.

\end{abstract}
\addtocontents{toc}{\protect\setcounter{tocdepth}{-1}}

\section{Introduction}

Deep neural networks (DNNs) have achieved superhuman performance in
recent years and have been increasingly employed to make decisions
on our behalf in various critical applications in computer vision
including object detection \cite{redmon2016you}, face recognition
\cite{parkhi2015deep,schroff2015facenet}, medical imaging \cite{moeskops2017adversarial,yang2017automatic},
surveillance \cite{sultani2018real} and so on. However, many recent
works have shown that besides the powerful modeling capability, DNNs
are highly vulnerable to adversarial attacks \cite{fawzi2018adversarial,goodfellow2014explaining,gu2017badnets,liu2017neural,thys2019fooling}.
Currently, there are two major types of attacks on DNNs. The first
is \emph{evasion/adversarial attacks} which cause a \emph{successfully
trained} model to misclassify by perturbing the model's input with
imperceptible adversarial noise \cite{goodfellow2014explaining,madry2017towards}.
The second is \emph{Trojan/backdoor attacks} in which attackers \emph{interfere
with the training process} of a model in order to insert hidden malicious
features (referred to as \emph{Trojans/backdoors}) into the model
\cite{chen2017targeted,gu2017badnets,liu2017neural,shafahi2018poison}.
These Trojans do not cause any harm to the model under normal conditions.
However, once they are triggered, they will force the model to output
the target classes specified by the attackers. Unfortunately, only
the attackers know exactly the Trojan triggers and the target classes.
Such stealthiness makes Trojan attacks difficult to defend against.

In this work, we focus on defending against Trojan attacks. Most existing
Trojan defenses assume that attacks use only \emph{one} \emph{input-agnostic}
Trojan trigger and/or target only \emph{one} class \cite{chen2019deepinspect,doan2020februus,Gao_etal_19Strip,guo2019tabor,harikumar2020scalable,Wang_etal_19Neural}.
By constraining the space of possible triggers, these defenses are
able to find the true trigger of some simple Trojan attacks satisfying
their assumptions and mitigate the attacks \cite{chen2017targeted,gu2017badnets}.
However, these defenses often do not perform well against other advanced
attacks that use \emph{multiple} \emph{input-specific} Trojan triggers
and/or target \emph{multiple} classes \cite{dong2021black,nguyen2020input,nguyen2021wanet}.
To address this problem, we propose two novel \emph{filtering''}
defenses named \emph{Variational Input Filtering (VIF)} and \emph{Adversarial
Input Filtering (AIF)}. Both defenses aim at learning a filter network
$\Filter$ that can purify potential Trojan triggers in the model's
input at run time without making any of the above assumptions about
attacks. VIF treats $\Filter$ as a variational autoencoder (VAE)
\cite{kingma2013auto} and utilizes the lossy data compression property
of VAE to discard noisy information in the input including triggers.
AIF, on the other hand, uses an auxiliary generator $\Generator$
to reveal hidden triggers in the input and leverages adversarial learning
\cite{goodfellow2014generative} between $\Generator$ and $\Filter$
to encourage $\Filter$ to remove potential triggers found by $\Generator$.
In addition, to overcome the issue that input filtering may hurt the
model's prediction on clean data, we introduce a new defense mechanism
called \emph{``Filtering-then-Contrasting'' (FtC)}. The key idea
behind FtC is comparing the two outputs of the model with and without
input filtering to determine whether the input is clean or not. If
the two outputs are different, the input will be marked as containing
triggers, otherwise clean. We equip VIF and AIF with FtC to arrive
at the two defenses dubbed VIFtC and AIFtC respectively. Through extensive
experiments and ablation studies, we demonstrate that our proposed
defenses are more effective than many well-known defenses \cite{doan2020februus,Gao_etal_19Strip,liu2018fine,Wang_etal_19Neural}
in mitigating various advanced Trojan attacks including two recent
state-of-the-art (SOTA) \cite{nguyen2020input,nguyen2021wanet} while
being quite robust to small amounts of training data and large trigger's
norms.

\section{Standard Trojan Attack\label{sec:Standard-Trojan-Attacks}}

We consider image classification as the task of interest. We denote
by $\mathbb{I}$ the real interval {[}0, 1{]}. In standard Trojan
attack scenarios \cite{chen2017targeted,gu2017badnets}, an attacker
(usually a service provider) \emph{fully controls} the training process
of an image classifier $\Classifier:\mathcal{X}\rightarrow\mathcal{Y}$
where $\mathcal{X}\subset\mathbb{I}^{c\times h\times w}$ is the input
image domain, and $\mathcal{Y}=\{0,...,K-1\}$ is the set of $K$
classes. The attacker's goal is to insert a \emph{Trojan }into the
classifier $\Classifier$ so that given an input image $x\in\mathcal{X}$,
$\Classifier$ will misclassify $x$ as belonging to a target class
$t\in\mathcal{Y}$ specified by the attacker if $x$ contains the
\emph{Trojan trigger} $\psi$, and will predict the true label $y\in\mathcal{Y}$
of $x$ otherwise. A common attack strategy to achieve this goal is
poisoning a small portion of the training data with the Trojan trigger
$\psi$. At each training step, the attacker randomly replaces each
clean training pair $(x,y)$ in the current mini-batch by a poisoned
one $(\tilde{x},t)$ with a probability $\rho$ ($0<\rho<1$) and
trains $\Classifier$ as normal using the modified mini-batch. $\tilde{x}$
is an image embedded with Trojan triggers (or \emph{Trojan image}
for short) corresponding to $x$. $\tilde{x}$ is constructed by combining
$x$ with $\psi$ via a Trojan injection function $T(x,\psi)$. A
common choice of $T$ is the image blending function \cite{chen2017targeted,gu2017badnets}
given below:
\begin{equation}
\tilde{x}=T(x,\psi)=(1-m)\odot x+m\odot p,\label{eq:Trojan_img_atk}
\end{equation}
where $\psi\triangleq(m,p)$, $m\in\mathbb{I}^{c\times h\times w}$
is the trigger mask, $p\in\mathbb{I}^{c\times h\times w}$ is the
trigger pattern, and $\odot$ is the element-wise product. To ensure
$\tilde{x}$ cannot be detected by human inspection at test time,
$\|m\|$ must be small. Some recent works use more advanced variants
of $T$ such as reflection \cite{liu2020reflection} and warping \cite{nguyen2021wanet}
to craft better natural-looking Trojan images.

\begin{figure}[t]
\begin{centering}
\subfloat[\label{fig:gen_rev_eng}]{\begin{centering}
\includegraphics[scale=0.062]{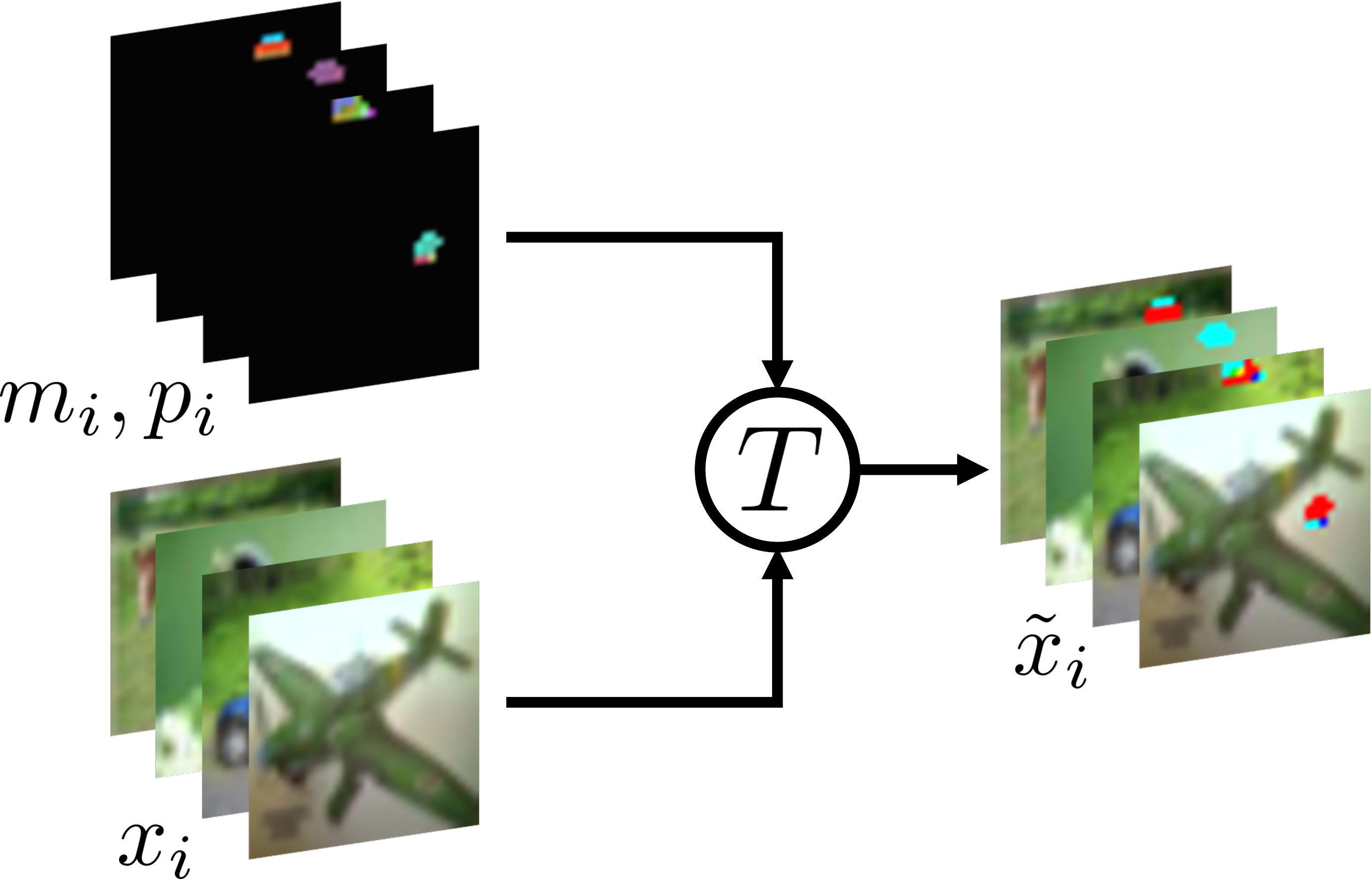}
\par\end{centering}
}\hspace{0.02\textwidth}\subfloat[\label{fig:gen_trigger_gen}]{\begin{centering}
\includegraphics[scale=0.062]{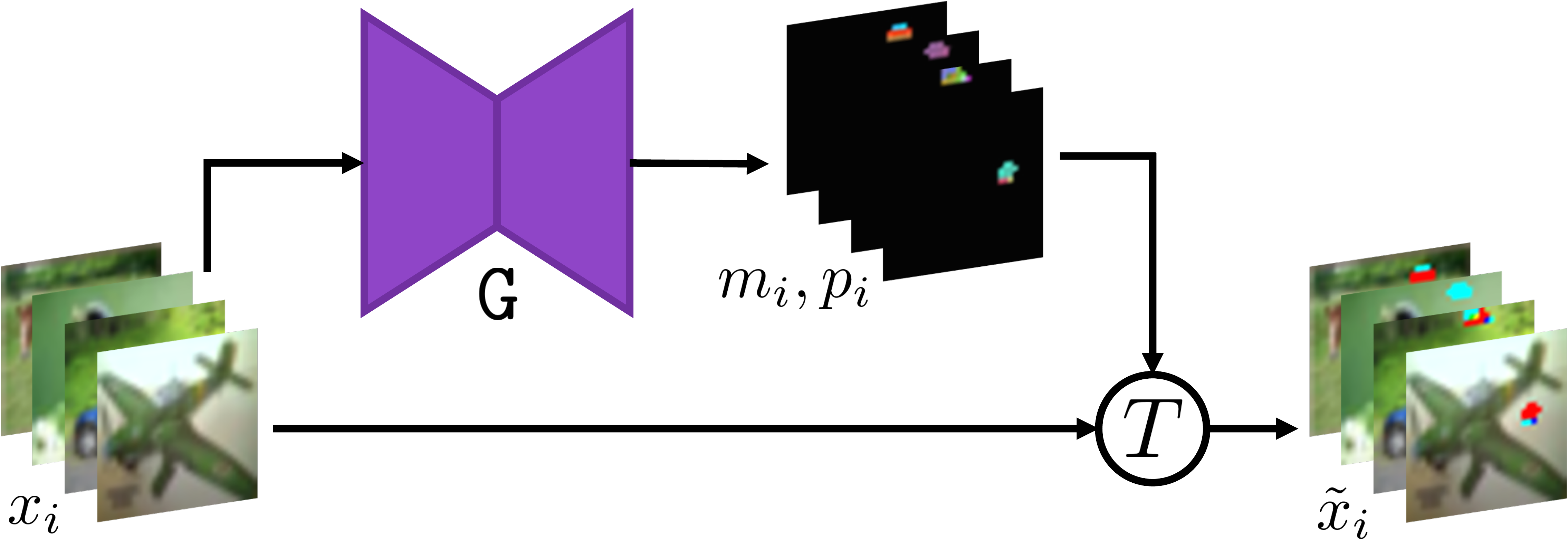}
\par\end{centering}
}\hspace{0.02\textwidth}\subfloat[\label{fig:gen_ae}]{\begin{centering}
\includegraphics[scale=0.062]{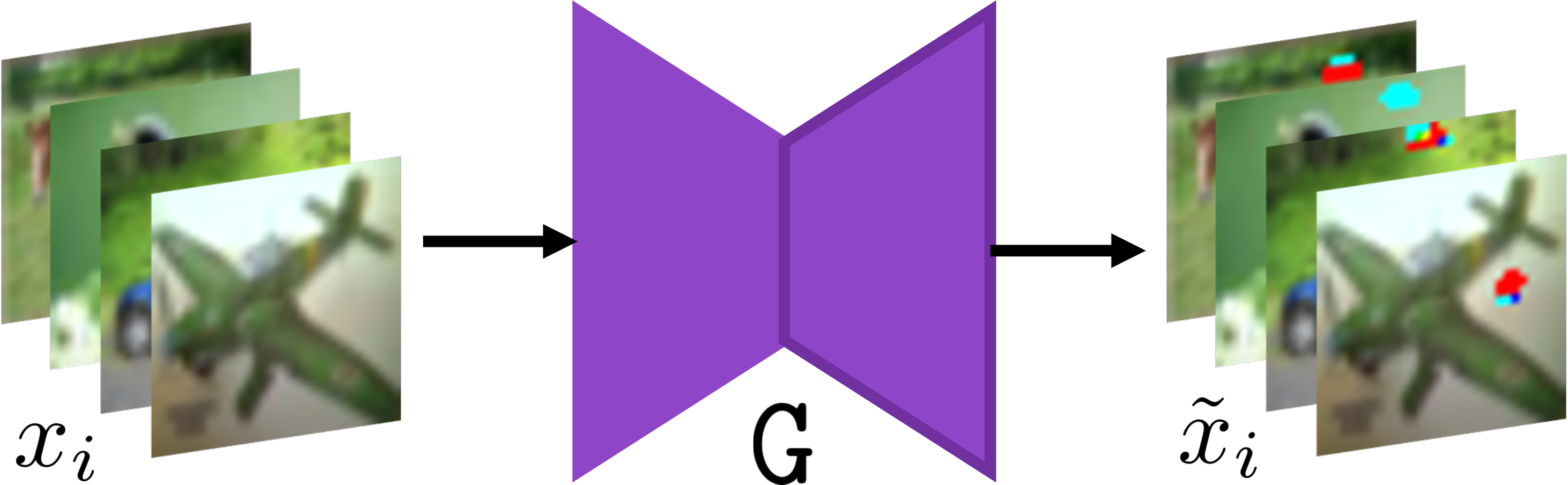}
\par\end{centering}
}
\par\end{centering}
\caption{Illustrations of three approaches to model input-specific Trojan triggers
$\psi_{i}=(m_{i},p_{i})$ w.r.t. $x_{i}$: as learnable parameters
(a), via a trigger generator (b), and via a Trojan-image autoencoder
(c).\label{fig:gen_arch}}
\end{figure}

Once trained, the Trojan-infected classifier $\Classifier$ will be
provided to victims (usually end-users) for deployment. When the victims
test $\Classifier$ with their own clean data, they do not see any
abnormalities in performance because the Trojan remains dormant for
the clean data. Thus, the victims naively believe that $\Classifier$
is normal and use $\Classifier$ as it is without any modification
or additional safeguard.

\section{Difficulty in Finding Input-Specific Triggers\label{sec:Finding-Triggers}}

In practice, we (as victims) usually have a small dataset $\mathcal{D}_{\text{val}}=\left\{ (x_{i},y_{i})\right\} _{i=1}^{N_{\text{val}}}$
containing only clean samples for evaluating the performance of $\Classifier$.
We can leverage this set to find possible Trojan triggers associated
with the target class $t$. For standard Trojan attacks \cite{chen2017targeted,gu2017badnets}
that use only a global \emph{input-agnostic} trigger $\psi=(m,p)$,
$\psi$ can be restored by minimizing the following loss w.r.t. $m$
and $p$: 
\begin{equation}
\Loss_{\text{gen}}(x,t)=-\log p_{\Classifier}(t|\tilde{x})+\lambda_{0}\max(\|m\|-\delta,0),\label{eq:NeuralCleanse_loss}
\end{equation}
where $(x,.)\sim\Data_{\text{val}}$, $\tilde{x}$ is derived from
$x$ via Eq.~\ref{eq:Trojan_img_atk}, $p_{\Classifier}(t|\tilde{x})=\frac{\exp(\Classifier_{t}(\tilde{x}))}{\sum_{k=1}^{K}\exp(\Classifier_{k}(\tilde{x}))}$
is the probability of $\tilde{x}$ belonging to the target class $t$,
$\|\cdot\|$ denotes a L1/L2 norm, $\delta\geq0$ is an upper bound
of the norm, and $\lambda_{0}\geq0$ is a coefficient. The second
term in Eq.~\ref{eq:NeuralCleanse_loss} ensures that the trigger
is small enough so that it could not be detected by human inspection.
$\Loss_{\text{gen}}$ was used by Neural Cleanse (NC) \cite{Wang_etal_19Neural}
and its variants \cite{chen2019deepinspect,guo2019tabor,harikumar2020scalable},
and was shown to work well for standard attacks. 

\begin{figure}[t]
\begin{centering}
\subfloat[]{\begin{centering}
\includegraphics[width=0.142\textwidth]{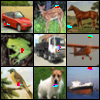}
\par\end{centering}
}\subfloat[]{\begin{centering}
\includegraphics[width=0.142\textwidth]{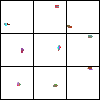}
\par\end{centering}
}\subfloat[]{\begin{centering}
\includegraphics[width=0.142\textwidth]{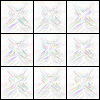}
\par\end{centering}
}\subfloat[]{\begin{centering}
\includegraphics[width=0.142\textwidth]{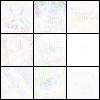}
\par\end{centering}
}\subfloat[]{\begin{centering}
\includegraphics[width=0.142\textwidth]{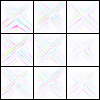}
\par\end{centering}
}\subfloat[]{\begin{centering}
\includegraphics[width=0.142\textwidth]{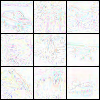}
\par\end{centering}
}
\par\end{centering}
\caption{Trojan images (a) and the corresponding triggers (b) of an Input-Aware
Attack. Triggers synthesized by Neural Cleanse (c) and by the three
approaches in Fig.~\ref{fig:gen_arch} (d, e, f). Trigger pixels
are inverted for better visualization.\label{fig:gen_imgs_4_InpAwAtk}}
\end{figure}

In this work, we however consider finding the triggers of Input-Aware
Attack (InpAwAtk) \cite{nguyen2020input}. This is a much harder problem
because InpAwAtk uses different triggers $\psi_{i}=(m_{i},p_{i})$
for different input images $x_{i}$ instead of a global one. We examine
3 different ways to model $\psi_{i}$: (i) treating $m_{i},p_{i}$
as learnable parameters for each image $x_{i}\in\Data_{\text{val}}$,
(ii) via an input-conditional trigger generator $(m_{i},p_{i})=\Generator(x_{i})$,
and (iii) generating a Trojan image $\tilde{x}_{i}$ w.r.t. $x_{i}$
via a Trojan-image generator $\tilde{x}_{i}=\Generator(x_{i})$ and
treating $\tilde{x}_{i}-x_{i}$ as $\psi_{i}$. These are illustrated
in Fig.~\ref{fig:gen_arch}. The first way does not generalize to
other images not in $\Data_{\text{val}}$ while the second and third
do. We reuse the loss $\Loss_{\text{gen}}$ in Eq.~\ref{eq:NeuralCleanse_loss}
to learn $m_{i},p_{i}$ in the first way and $\Generator$ in the
second way. The loss to train $\Generator$ in the third way is slightly
adjusted from $\Loss_{\text{gen}}$ with $\|m\|$ replaced by $\|\tilde{x}-x\|$.
As shown in Fig.~\ref{fig:gen_imgs_4_InpAwAtk}, neither NC nor the
above approaches can restore the original triggers of InpAwAtk, suggesting
new methods are on demand.

\section{Proposed Trojan Defenses\label{sec:Our-Proposed-Defenses}}

The great difficulty in finding correct input-specific triggers
(Section~\ref{sec:Finding-Triggers}) challenges a majority of existing
Trojan defenses which assume a global input-agnostic trigger is applied
to all input images \cite{chen2019deepinspect,Gao_etal_19Strip,harikumar2020scalable,liu2018fine,Liu_etal_19Abs,qiao2019defending,Wang_etal_19Neural}.
Fortunately, although we may not be able to find correct triggers,
in many cases, we can still design effective Trojan defenses by filtering
out triggers embedded in the input without concerning about the number
or the input dependence property of triggers. The fundamental idea
is learning a filter network $\Filter$ that maps the original input
image $x$ into a filtered image $x^{\circ}$, and using $x^{\circ}$
as input to the classifier $\Classifier$ instead of $x$. In order
for $\Filter$ to be considered as a good filter, $x^{\circ}$ should
satisfy the following two conditions:
\begin{itemize}
\item \emph{Condition 1}: If $x$ is clean, $x^{\circ}$ should look similar
to $x$ and should have the same label as $x$'s. This ensures a \emph{high
classification accuracy on clean images} (dubbed \emph{``clean accuracy''}).
\item \emph{Condition 2}: If $\tilde{x}$ contains triggers, $\tilde{x}^{\circ}$
should be close to $x$ and should have the same label as $x$'s where
$x$ is the clean counterpart of $\tilde{x}$. This ensures a \emph{low
attack success rate} and\emph{ }a \emph{high clean-label recovery
rate on Trojan images} (dubbed \emph{``Trojan accuracy''} and \emph{``recovery
accuracy''}, respectively).
\end{itemize}
In the next two subsections (\ref{subsec:Variational-Input-Filtering},
\ref{subsec:Adversarial-Input-Filtering}), we propose two novel filtering
defenses that leverage two different strategies to learn a good $\Filter$
which are lossy data compression and adversarial learning, respectively.

\begin{figure}[t]
\begin{centering}
\includegraphics[width=0.6\textwidth]{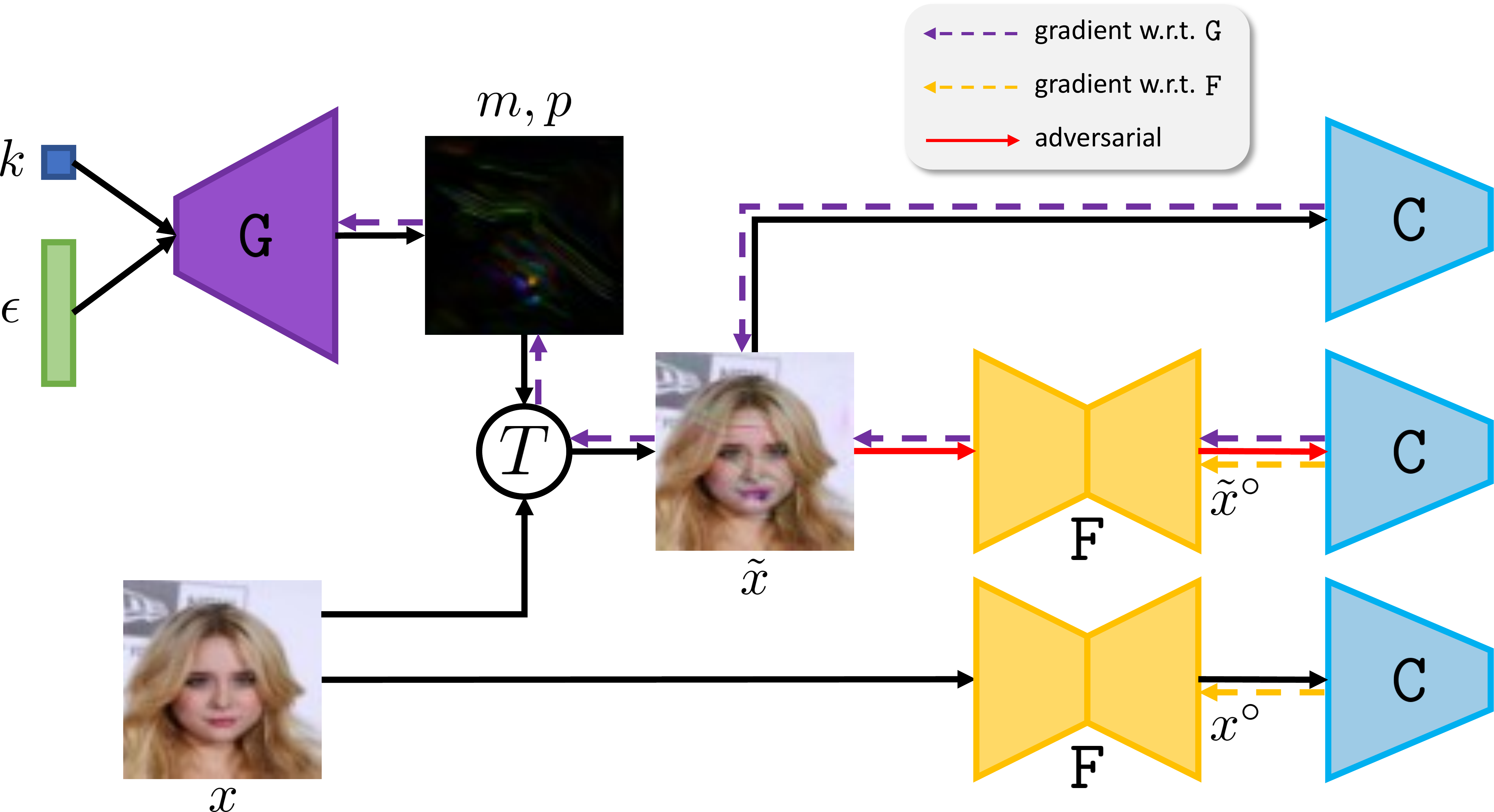}
\par\end{centering}
\caption{An illustration of Adversarial Input Filtering.\label{fig:AIF_overview}}
\end{figure}

\subsection{Variational Input Filtering\label{subsec:Variational-Input-Filtering}}

A natural choice for $\Filter$ is an autoencoder (AE) which should
be complex enough so that $\Filter$ can reconstruct clean images
well to achieve high clean accuracy. However, if $\Filter$ is too
complex, it can capture every detail of a Trojan image including the
embedded triggers, which also causes high Trojan accuracy. In general,
an optimal $\Filter$ should achieve good balance between preserving
class-related information and discarding noisy information of the
input. To reduce the dependence of $\Filter$ on architecture, we
propose to treat $\Filter$ as a variational autoencoder (VAE)\footnote{Denoising Autoencoder (DAE) \cite{vincent2008extracting} is also
a possible choice but is quite similar to VAE in terms of idea so
we do not consider it here.} \cite{kingma2013auto} and train it with the \emph{``Variational
Input Filtering''} (VIF) loss given below:
\begin{align}
\mathcal{L}_{\text{VIF}}(x,y)= & -\log p_{\Classifier}(y|x^{\circ})+\lambda_{1}\|x^{\circ}-x\|+\lambda_{2}D_{\text{KL}}(q_{\text{\ensuremath{\Filter}}}(z|x)\|p(z))\label{eq:VIF_loss}\\
= & \Loss_{\text{IF}}+\lambda_{2}D_{\text{KL}}(q_{\text{\ensuremath{\Filter}}}(z|x)\|p(z)),\label{eq:VIF_loss_2}
\end{align}
where $(x,y)\sim\mathcal{D}_{\text{val}}$, $x^{\circ}=\mathrm{\Filter}(x)$
is the filtered version of $x$, $z$ is the latent variable, $q_{\Filter}(z|x)$
denotes the variational posterior distribution and is parameterized
via the stochastic encoder of $\Filter$, $p(z)=\mathcal{N}(0,\mathrm{I})$
is the standard Gaussian distribution, $D_{\text{KL}}$ denotes the
KL divergence, $\lambda_{1},\lambda_{2}\geq0$ are coefficients. In
Eq.~\ref{eq:VIF_loss}, the first two terms force $\Filter$ to preserve
class-related information of $x$ and to reduce the visual dissimilarity
between $x^{\circ}$ and $x$ as per condition 1, 2. Meanwhile, the
last term encourages $\Filter$ (or more precisely, $q_{\Filter}(z|x)$)
to discard noisy information of $x$, which we refer to as \emph{``lossy
data compression''}. This can be explained via the following relationship
\cite{zhao2017infovae}:
\begin{equation}
\mathbb{E}_{p(x)}\left[D_{\text{KL}}(q_{\Filter}(z|x)\|p(z))\right]=D_{\text{KL}}(q_{\Filter}(z)\|p(z))+I_{\text{\ensuremath{\Filter}}}(x,z),\label{eq:VAE_info}
\end{equation}
where $q_{\Filter}(z)=\Expect_{p(x)}\left[q_{\Filter}(z|x)\right]$.
Clearly, minimizing the LHS of Eq.~\ref{eq:VAE_info} decreases the
mutual information between $z$ and $x$. And because $z$ is used
to compute $x^{\circ}$ (in decoding), this also reduces the information
between $x^{\circ}$ and $x$. 

In Eq.~\ref{eq:VIF_loss}, the first two terms alone constitute the
``Input Filtering'' (IF) loss $\Loss_{\text{IF}}$. To the best
of our knowledge, IF has not been proposed in other Trojan defense
works. Input Processing (IP) \cite{liu2017neural} is the closest
to IF but it is trained on \emph{unlabeled data} using \emph{only
the reconstruction loss} (the second term in Eq.~\ref{eq:VIF_loss}).
In Appdx.~\ref{subsec:Comparing-IF-with-IP}, we show that IP performs
worse than IF, which highlights the importance of the term $-\log p_{\Classifier}(y|x^{\circ})$.

\subsection{Adversarial Input Filtering\label{subsec:Adversarial-Input-Filtering}}

VIF, owing to its generality, do not make use of any Trojan-related
knowledge in $\mathtt{C}$ to train $\Filter$. We argue that if $\Filter$
is exposed to such knowledge, $\Filter$ could be more selective in
choosing which input information to discard, and hence, could perform
better. This motivates us to use synthetic Trojan images as additional
training data for $\Filter$ besides clean images from $\mathcal{D}_{\text{val}}$.
We synthesize a Trojan image $\tilde{x}$ from a clean image $x$
as follows:
\begin{align}
(m_{k},p_{k}) & =\Generator(\epsilon,k),\label{eq:global_gen}\\
\tilde{x} & =T(x,(m_{k},p_{k})),\label{eq:global_gen_trojan_img}
\end{align}
where $\epsilon\sim\mathcal{N}(0,\mathrm{I})$ is a standard Gaussian
noise, $k$ is a class label sampled uniformly from $\mathcal{Y}$,
$\Generator$ is a conditional generator, $T$ is the image blending
function (Eq.~\ref{eq:Trojan_img_atk}). We choose the image blending
function to craft Trojan images because it is the \emph{most general}
Trojan injection function (its output range spans the whole image
space $\mathbb{I}^{c\times h\times w}$). To make sure that the synthetic
Trojan images are useful for $\Filter$, we \emph{form an adversarial
game between $\Generator$ and $\Filter$} in which $\Generator$
attempts to generate hard Trojan images that can fool $\Filter$ into
producing the target class (sampled randomly from $\mathcal{Y}$)
while $\Filter$ becomes more robust by correcting these images. We
train $\Generator$ with the following loss:
\begin{equation}
\Loss_{\text{AIF-gen}}(x,k)=\Loss_{\text{gen}}(x,k)-\lambda_{3}\log p_{\Classifier}(k|\tilde{x}^{\circ}),\label{eq:AIF_gen_loss}
\end{equation}
where $\Loss_{\text{gen}}$ is similar to the one in Eq.~\ref{eq:NeuralCleanse_loss}
but with $m$ replaced by $m_{k}$ (Eq.~\ref{eq:global_gen}), $\tilde{x}^{\circ}=\Filter(\tilde{x})$,
$\lambda_{3}\geq0$. The loss of $\Filter$ must conform to conditions
1, 2 and is:
\begin{align}
\Loss_{\text{AIF}}(x,y)= & \Loss_{\text{IF}}(x,y)-\lambda_{4}\log p_{\mathrm{\Classifier}}(y|\tilde{x}^{\circ})+\lambda_{5}\|\tilde{x}^{\circ}-x\|\label{eq:AIF_loss}\\
= & \Loss_{\text{IF}}(x,y)+\Loss'_{\text{IF}}(\tilde{x},y),\label{eq:AIF_loss_2}
\end{align}
where AIF stands for \emph{``Adversarial Input Filtering''}, $\Loss_{\text{IF}}$
was described in Eq.~\ref{eq:VIF_loss_2}, $\tilde{x}$ is computed
from $x$ via Eq.~\ref{eq:global_gen_trojan_img}, $\lambda_{4},\lambda_{5}\geq0$.
Note that the last term in Eq.~\ref{eq:AIF_loss_2} is the reconstruction
loss between $\tilde{x}^{\circ}$ and $x$ (not $\tilde{x}$). Thus,
we denote the last two terms in Eq.~\ref{eq:AIF_loss} as $\Loss'_{\text{IF}}$
instead of $\Loss_{\text{IF}}$. AIF is depicted in Fig.~\ref{fig:AIF_overview}.

During experiment, we observed that sometimes training $\Generator$
and $\Filter$ with the above losses may not result in good performance.
The reason is that when $\Filter$ becomes better, $\Generator$ tends
to produce large-norm triggers to fool $\Filter$ despite the fact
that a regularization was applied to the norms of these triggers.
Large-norm triggers make learning $\Filter$ harder as $\tilde{x}$
is no longer close to $x$. To handle this problem, we explicitly
normalize $m_{k}$ so that its norm is always bounded by $\delta$.
We provide technical details and empirical study about this normalization
in Appdx.~\ref{subsec:Explicit-Normalization-of-Triggers-in-AIF}.

\subsection{Filtering then Contrasting}

VIF and AIF always filter $x$ even when $x$ does not contain triggers,
which often leads to the decrease in clean accuracy after filtering.
To overcome this drawback, we introduce a new defense mechanism called
\emph{``Filtering then Contrasting''} (FtC) which works as follows:
Instead of just computing the predicted label $\hat{y}^{\circ}$ of
the filtered image $x^{\circ}=\Filter(x)$ and treat it as the final
prediction, we also compute the predicted label $\hat{y}$ of $x$
without filtering and compare $\hat{y}$ with $\hat{y}^{\circ}$.
If $\hat{y}$ is different from $\hat{y}^{\circ}$, $x$ will be marked
as containing triggers and discarded. Otherwise, $x$ will be marked
as clean and $\hat{y}$ will be used as the final prediction. FtC
is especially useful for defending against attacks with large-norm
triggers (Sect.~\ref{subsec:Large-norm-triggers}) because it helps
avoid the significant drop in clean accuracy caused by the large visual
difference between $x^{\circ}$ and $x$. Under the FtC defense mechanism,
we derive two new defenses VIFtC and AIFtC from VIF and AIF, respectively.

\section{Experiments}

\subsection{Experimental Setup\label{subsec:Experimental-Setup}}

\paragraph{Datasets}

Following previous works \cite{gu2017badnets,nguyen2020input,salem2020dynamic},
we evaluate our proposed defenses on four  image datasets namely MNIST,
CIFAR10 \cite{Krizhevsky09learningmultiple}, GTSRB \cite{Stallkamp2012man},
and CelebA \cite{liu2015deep}. For CelebA, we follow Salem et al.
\cite{salem2020dynamic} and select the top 3 most balanced binary
attributes (out of 40) to form an 8-class classification problem.
The chosen attributes are \emph{``Heavy Makeup''}, \emph{``Mouth
Slightly Open''}, and \emph{``Smiling''}. Like other works \cite{Gao_etal_19Strip,Wang_etal_19Neural},
we assume that we have access to the test set of these datasets. We
use 70\% data of the test set for training our defense methods ($\Data_{\text{val}}$
in Sections~\ref{sec:Finding-Triggers}, \ref{sec:Our-Proposed-Defenses})
and 30\% for testing (denoted as $\Data_{\text{test}}$). For more
details about the datasets, please refer to Appdx.~\ref{subsec:Dataset-Description}.
Sometimes, we do not test our methods on all images in $\Data_{\text{test}}$
but on those \emph{not} belonging to the target class. This set is
denoted as $\Data'_{\text{test}}$. We also provide results with less
training data in Appdx.~\ref{subsec:Small-amounts-of-training-data}. 

\paragraph{Benchmark Attacks}

We use 5 different benchmark Trojan attacks for our defenses, which
are BadNet+, noise-BI+, image-BI+, InpAwAtk \cite{nguyen2020input},
and WaNet \cite{nguyen2021wanet}. InpAwAtk and WaNet are recent SOTA
attacks that were shown to break many strong defenses completely.
BadNet+ and noise/image-BI+ are variants of BadNet \cite{gu2017badnets}
and Blended Injection (BI) \cite{chen2017targeted} that use multiple
triggers instead of one. They are described in detail in Appdx.~\ref{subsec:BadNet-and-Noise-Image-BI}.
The training settings for the 5 attacks are given in Appdx.~\ref{subsec:Training-Settings-for-the-Attacks}.

We also consider 2 attack modes namely \emph{single-target} and \emph{all-target}
\cite{nguyen2020input,zhao2020bridging}. In the first mode, only
one class $t$ is chosen as target. Every Trojan image $\tilde{x}$
is classified as $t$ regardless of the ground-truth label of its
clean counterpart $x$. Without loss of generality, $t$ is set to
0. In the second mode, $\tilde{x}$ is classified as $(k+1)\mod K$
if $x$ belongs to the class $k$. If not clearly stated, attacks
are assumed to be \emph{single-target}.

\begin{table}[t]
\begin{centering}
\resizebox{\textwidth}{!}{%
\par\end{centering}
\begin{centering}
\begin{tabular}{cccccccccccccccccccc}
\hline 
\multirow{2}{*}{Dataset} &  & Benign &  & \multicolumn{2}{c}{BadNet+} &  & \multicolumn{2}{c}{noise-BI+} &  & \multicolumn{2}{c}{image-BI+} &  & \multicolumn{3}{c}{InpAwAtk} &  & \multicolumn{3}{c}{WaNet}\tabularnewline
 &  & Clean &  & Clean & Trojan &  & Clean & Trojan &  & Clean & Trojan &  & Clean & Trojan & Cross &  & Clean & Trojan & Noise\tabularnewline
\cline{1-1} \cline{3-3} \cline{5-6} \cline{6-6} \cline{8-9} \cline{9-9} \cline{11-12} \cline{12-12} \cline{14-16} \cline{15-16} \cline{16-16} \cline{18-20} \cline{19-20} \cline{20-20} 
MNIST &  & 99.56 &  & 99.61 & 99.96 &  & 99.46 & 100.0 &  & 99.50 & 100.0 &  & 99.47 & 99.41 & 96.05 &  & 99.48 & 98.73 & 99.38\tabularnewline
CIFAR10 &  & 94.82 &  & 94.88 & 100.0 &  & 94.69 & 100.0 &  & 95.15 & 99.96 &  & 94.58 & 99.43 & 88.68 &  & 94.32 & 99.59 & 92.58\tabularnewline
GTSRB &  & 99.72 &  & 99.34 & 100.0 &  & 99.30 & 100.0 &  & 99.18 & 100.0 &  & 98.90 & 99.54 & 95.19 &  & 99.12 & 99.54 & 99.03\tabularnewline
CelebA &  & 79.12 &  & 79.41 & 100.0 &  & 78.75 & 100.0 &  & 78.81 & 99.99 &  & 78.18 & 99.93 & 77.16 &  & 78.48 & 99.94 & 77.24\tabularnewline
\hline 
\end{tabular}}
\par\end{centering}
\caption{Test clean and Trojan accuracies of various Trojan attacks.\label{tab:main-attack-results}}
\end{table}

We report the test clean and Trojan accuracies of the benchmark attacks
(in single-target mode) in Table~\ref{tab:main-attack-results}.
It is clear that all attacks achieve very high Trojan accuracies with
little or no decrease in clean accuracy compared to the benign model's,
hence, are qualified for our experimental purpose. For results of
the attacks on $\Data_{\text{test}}$, please refer to Appdx.~\ref{sec:Additional-Results-of-Benchmark-Attacks}.

\paragraph{Baseline Defenses\label{par:Baseline-Defenses-Setup}}

We consider 5 well-known baseline defenses namely Neural Cleanse (NC)
\cite{Wang_etal_19Neural}, STRIP \cite{Gao_etal_19Strip}, Network
Pruning (NP) \cite{liu2018fine}, Neural Attention Distillation (NAD)
\cite{li2021neural}, and Februus \cite{doan2020februus}. 

Neural Cleanse (NC) assumes that attacks (i) choose only one target
class $t$ and (ii) use \emph{at least} (not exactly) one input-agnostic
trigger associated with $t$. We refer to (i) as the \emph{``single
target class''} assumption and (ii) as the \emph{``input-agnostic
trigger''} assumption. Based on these assumptions, NC finds a trigger
$\psi_{k}=(m_{k},p_{k})$ for every class $k\in\mathcal{Y}$ via reverse-engineering
(Eq.~\ref{eq:NeuralCleanse_loss}), and uses the L1 norms of the
synthetic trigger masks $\{m_{1},...,m_{K}\}$ to detect the target
class. The intuition is that if $t$ is the target class, ${\|m_{t}\|}_{1}$
will be much smaller than the rest. A $z$-value of each mask norm
is calculated via Median Absolute Deviation and the $z$-value of
the smallest mask norm (referred to as the \emph{anomaly index}) is
compared against a threshold $\zeta$ (2.0 by default). If the anomaly
index is smaller than $\zeta$, $\Classifier$ is marked as clean.
Otherwise, $\Classifier$ is marked Trojan-infected with the target
class corresponding to the smallest mask norm. In this case, the Trojans
in $\Classifier$ can be mitigated via pruning or via checking the
cleanliness of input images. Both mitigation methods make use of $\psi_{t}$
and are analyzed in Appdx.~\ref{subsec:Additional-Results-of-Baseline-Defenses}.

STRIP assumes triggers are input-agnostic and argues that if an input
image $x$ contains triggers then these triggers still have effect
if $x$ is superimposed (blended) with other images. Therefore, STRIP
superimposes $x$ with $N_{\text{s}}$ random clean images from $\Data_{\text{val}}$
and computes the \emph{average entropy} $\mathcal{H}(x)$ of $N_{\text{s}}$
predicted class probabilities corresponding to $N_{\text{s}}$ superimposed
versions of $x$. If $\mathcal{H}(x)$ is smaller than a predefined
threshold, $x$ is considered as trigger-embedded, otherwise, clean.
The threshold is set according to the false positive rate (FPR) over
the average entropies of all images in $\mathcal{D}_{\text{val}}$,
usually at FPR equal to 1/5/10\%. We evaluate the performance of STRIP
against an attack using $M_{\text{s}}$ random clean images from $\Data_{\text{test}}$
and $M_{\text{s}}$ corresponding Trojan images generated by that
attack. Following \cite{Gao_etal_19Strip}, we set $N_{\text{s}}=100$
and $M_{\text{s}}=2000$.

Network Pruning (NP) hypothesizes that idle neurons are more likely
to store Trojan-related information. Thus, it ranks neurons in the
second top layer of $\Classifier$ according to their average activation
over all data in $\Data_{\text{val}}$ and gradually prunes them until
a certain decrease in clean accuracy is reached, usually at 1/5/10\%
decrease in clean accuracy.

Neural Attention Distillation (NAD) \cite{li2021neural} is a distillation-based
Trojan defense. It first fine-tunes the pretrained classifier $\Classifier$
on clean images in $\Data_{\text{val}}$ to obtain a fine-tune classifier
$\mathtt{T}$. Then, it treats $\mathtt{T}$ and $\Classifier$ as
the teacher and student respectively, and performs attention-based
feature distillation \cite{zagoruyko2016paying} between $\mathtt{T}$
and $\Classifier$ on $\Data_{\text{val}}$ again. Since $\mathtt{T}$
is $\Classifier$ fine-tuned on clean data, $\mathtt{T}$ is expected
to have most of the Trojan in $\Classifier$ removed. Via distillation,
such Trojan-free knowledge is transferred from $\mathtt{T}$ to $\Classifier$
while performance of $\Classifier$ on clean data is still preserved.

Among the baselines, Februus is the most related to our filtering
defenses since it mitigates Trojan attacks via input purification.
It uses GradCAM \cite{selvaraju2017grad} to detect regions in an
input image $x$ that may contain triggers. Then, it removes all pixels
in the suspected regions and generates new ones via inpainiting. The
inpainted image is expected to contain no trigger and is fed to $\Classifier$
instead of $x$.

\paragraph{Model Architectures and Training Settings}

Please refer to Appdx.~\ref{subsec:Training-Settings-for-Our-Defenses}.

\paragraph{Metrics}

We evaluate VIF/AIF using 3 metrics namely \emph{decrease in clean
accuracy} ($\downarrow$C),\emph{ Trojan accuracy} (T), and \emph{decrease
in recovery accuracy} ($\downarrow$R). The first is the difference
between the classification accuracies of clean images before and after
filtering. The second is the attack success rate of Trojan images
after filtering. The last is the difference between the classification
accuracy of clean images before filtering and that of the corresponding
Trojan images after filtering. Smaller values of the metrics indicate
better results. $\downarrow$C and $\downarrow$R are computed on
$\Data_{\text{test}}$. T is computed on $\Data'_{\text{test}}$ under
single-target attacks and $\Data_{\text{test}}$ under all-target
attacks. This ensures that T can be $0$ in the best case. Otherwise,
T will be around $1/K$ where $K$ is the total number of classes.
$\downarrow$C and $\downarrow$R are upper-bounded by 1 and can be
negative. 

We evaluate VIFtC/AIFtC using FPR and FNR. FPR/FNR is defined as the
proportion of clean/Trojan images having different/similar class predictions
when the filter $\Filter$ is applied and not applied. FPR is computed
on $\Data_{\text{test}}$. FNR is computed on $\Data'_{\text{test}}$
under single-target attacks and $\Data_{\text{test}}$ under all-target
attacks. Both metrics are in {[}0, 1{]} and smaller values of them
are better. Interestingly, FPR and FNR are strongly correlated to
$\downarrow$C and T, respectively. FPR/FNR is exactly equal to $\downarrow$C/T
if $\Classifier$ achieves perfect clean/Trojan accuracy.

\subsection{Results of Baseline Defenses\label{subsec:Results-of-Baseline-Defenses}}

\begin{figure}[t]
\begin{centering}
\subfloat[Neural Cleanse\label{fig:Neural-Cleanse-result-bars}]{\begin{centering}
\includegraphics[width=0.48\textwidth]{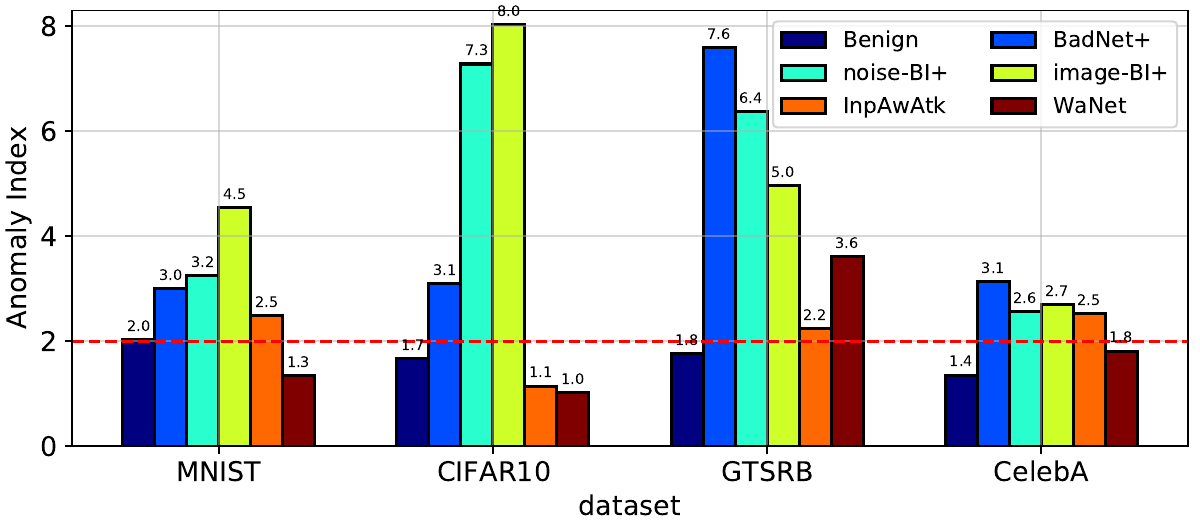}
\par\end{centering}
}\subfloat[STRIP\label{fig:STRIP-result-bars}]{\begin{centering}
\includegraphics[width=0.48\textwidth]{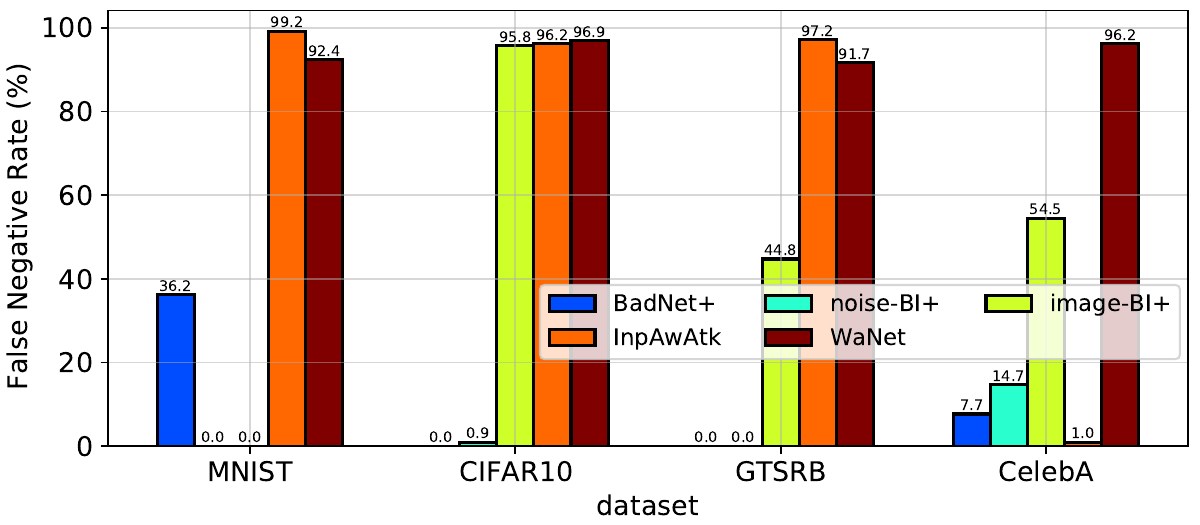}
\par\end{centering}
}
\par\end{centering}
\caption{(a) Anomaly indices of Neural Cleanse. The red dashed line indicates
the threshold. (b) FNRs of STRIP at 10\% FPR.\label{fig:Neural-Cleanse-and-STRIP-bars}}
\end{figure}

\begin{figure}[t]
\begin{centering}
\subfloat[Network Pruning\label{fig:NP-trojan-acc-bars}]{\begin{centering}
\includegraphics[width=0.47\textwidth]{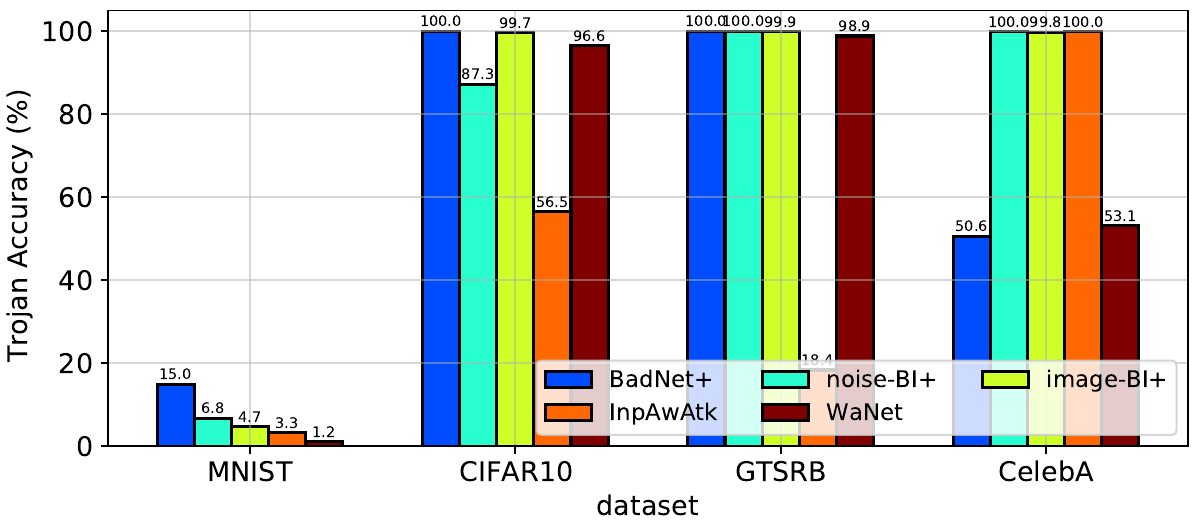}
\par\end{centering}
}\subfloat[Neural Cleanse\label{fig:NC-All-target-Anomaly-Index-bars}]{\begin{centering}
\includegraphics[width=0.235\textwidth]{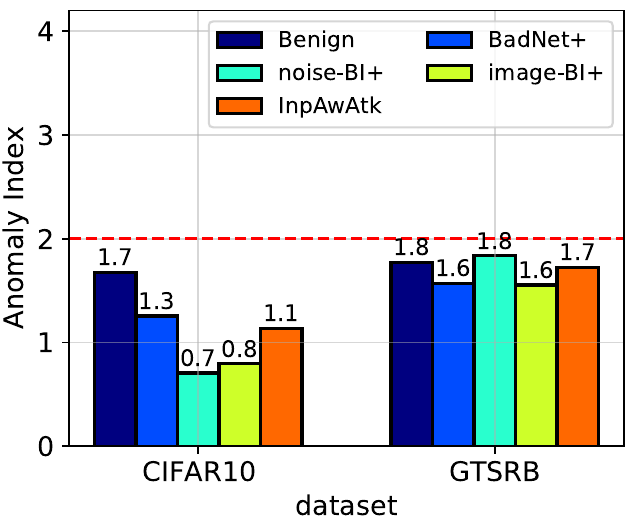}
\par\end{centering}
}\subfloat[VIF\label{fig:VIF-All-target-Trojan-Acc}]{\begin{centering}
\includegraphics[width=0.235\textwidth]{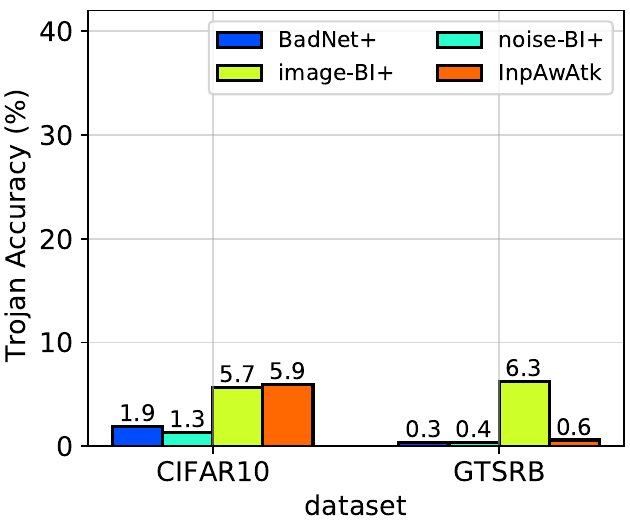}
\par\end{centering}
}
\par\end{centering}
\caption{(a) Trojan accuracies at 10\% decrease in clean accuracy of different
Trojan classifiers pruned by Network Pruning. (b) Anomaly indices
of NC and (c) Trojan accuracies of VIF against \emph{all-target} attacks
on CIFAR10 and GTSRB.}
\end{figure}

In Fig.~\ref{fig:Neural-Cleanse-and-STRIP-bars}, we show the detection
results of Neural Cleanse (NC) and STRIP w.r.t. the aforementioned
attacks. The two defenses are effective against BadNet+ and image/noise-BI+.
This is because STRIP and NC generally do not make any assumption
about the number of triggers. However, STRIP performs poorly against
InpAwAtk and WaNet (FNRs \textgreater{} 90\%) since these advanced
attacks break its ``input-agnostic trigger'' assumption. NC also
fails to detect the Trojan classifiers trained by WaNet on most datasets
for the same reason. What surprises us is that in our experiment NC
correctly detect the Trojan classifiers trained by InpAwAtk on 3/4
datasets while in the original paper \cite{nguyen2020input}, it was
shown to fail completely. We are confident that this inconsistency
does not come from our implementation of InpAwAtk since we used the
same hyperparameters and achieved the same classification results
as those in the original paper (Table~\ref{tab:main-attack-results}
versus Fig.~3b in \cite{nguyen2020input}). However, NC is still
unable to mitigate all Trojans in these correctly-detected Trojan
classifiers (Appdx.~\ref{subsec:Neural-Cleanse-supp}). In addition,
as shown in Fig.~\ref{fig:NC-All-target-Anomaly-Index-bars}, NC
is totally vulnerable to \emph{all-target attacks} since its ``single
target class'' assumption is no longer valid under these attacks.
Network Pruning (NP), despite being assumption-free, cannot mitigate
Trojans from most attacks (high Trojan accuracies in Fig.~\ref{fig:NP-trojan-acc-bars})
as it fails to prune the correct neurons containing Trojans. Februus
has certain effects on mitigating Trojans from BadNet+ while being
useless against the remaining attacks (high Ts in Table~\ref{tab:VIF-AIF-full-results}).
This is because GradCAM, the method used by Februus, is only suitable
for detecting patch-like triggers of BadNet+, not full-size noise-like
triggers of image/noise-BI+ or polymorphic triggers of InpAwAtk/WaNet.
We also observe that Februus significantly reduces the clean accuracy
(high $\downarrow$Cs in Table~\ref{tab:VIF-AIF-full-results}) as
it removes input regions that contain no Trojan trigger yet are highly
associated with the output class. This problem, however, was not discussed
in the Februus paper. NAD, thanks to its distillation-based nature,
usually achieves better clean accuracies than our filtering defenses
(Table~\ref{tab:VIF-AIF-full-results}). This defense is also effective
against InpAwAtk and WaNet. However, NAD performs poorly in recovering
Trojan samples from BadNet+ and noise/image-BI+ (high $\downarrow$Rs),
especially on MNIST. Besides, NAD is much less robust to large-norm
triggers than our filtering defenses (Sect.~\ref{subsec:Large-norm-triggers}).
For more analyses of the baseline defenses, please refer to Appdx.~\ref{subsec:Additional-Results-of-Baseline-Defenses}.

\subsection{Results of Proposed Defenses}

\begin{table}[t]
\begin{centering}
\resizebox{\textwidth}{!}{%
\par\end{centering}
\begin{centering}
\begin{tabular}{|c|c|d|dab|dab|dab|dab|dab|}
\hline 
\multirow{2}{*}{Dataset} & \multirow{2}{*}{Defense} & \multicolumn{1}{c|}{Benign} & \multicolumn{3}{c|}{BadNet+} & \multicolumn{3}{c|}{noise-BI+} & \multicolumn{3}{c|}{image-BI+} & \multicolumn{3}{c|}{InpAwAtk} & \multicolumn{3}{c|}{WaNet}\tabularnewline
\cline{3-18} \cline{4-18} \cline{5-18} \cline{6-18} \cline{7-18} \cline{8-18} \cline{9-18} \cline{10-18} \cline{11-18} \cline{12-18} \cline{13-18} \cline{14-18} \cline{15-18} \cline{16-18} \cline{17-18} \cline{18-18} 
 &  & \multicolumn{1}{c|}{$\downarrow$C} & \multicolumn{1}{c}{$\downarrow$C} & \multicolumn{1}{c}{T} & \multicolumn{1}{c|}{$\downarrow$R} & \multicolumn{1}{c}{$\downarrow$C} & \multicolumn{1}{c}{T} & \multicolumn{1}{c|}{$\downarrow$R} & \multicolumn{1}{c}{$\downarrow$C} & \multicolumn{1}{c}{T} & \multicolumn{1}{c|}{$\downarrow$R} & \multicolumn{1}{c}{$\downarrow$C} & \multicolumn{1}{c}{T} & \multicolumn{1}{c|}{$\downarrow$R} & \multicolumn{1}{c}{$\downarrow$C} & \multicolumn{1}{c}{T} & \multicolumn{1}{c|}{$\downarrow$R}\tabularnewline
\hline 
\hline 
\multirow{5}{*}{MNIST} & Feb. & 5.96 & 39.08 & 96.24 & 86.32 & 2.30 & 100.0 & 89.58 & 8.19 & 100.0 & 89.58 & 9.90 & 92.40 & 83.32 & 25.43 & 80.46 & 88.75\tabularnewline
 & NAD & 0.45 & 0.82 & 35.72 & 36.41 & 0.75 & 84.83 & 76.22 & 0.78 & 88.34 & 79.18 & 0.80 & 4.46 & 5.29 & 0.42 & 0.44 & 0.98\tabularnewline
 & IF & \textbf{0.10} & 0.27 & 2.47 & 4.99 & \textbf{0.10} & 0.16 & 13.52 & 0.13 & 1.29 & 12.02 & 0.21 & \textbf{0.96} & 2.08 & 0.23 & 0.34 & 0.61\tabularnewline
 & VIF & 0.13 & \textbf{0.17} & \textbf{2.36} & \textbf{3.63} & 0.12 & \textbf{0.04} & 0.63 & \textbf{0.03} & \textbf{0.11} & 0.40 & 0.20 & 1.25 & 1.83 & \textbf{0.10} & 0.48 & 0.53\tabularnewline
 & AIF & \textbf{0.10} & \textbf{0.17} & 3.80 & 4.86 & 0.13 & 0.15 & \textbf{0.11} & 0.10 & \textbf{0.11} & \textbf{0.10} & \textbf{0.03} & 1.14 & \textbf{1.66} & 0.13 & \textbf{0.15} & \textbf{0.20}\tabularnewline
\hline 
\hline 
\multirow{5}{*}{CIFAR10} & Feb. & 32.67 & 49.17 & 12.63 & 19.57 & 26.73 & 43.59 & 78.90 & 39.70 & 92.67 & 81.00 & 53.43 & 49.52 & 66.50 & 55.80 & 98.70 & 83.30\tabularnewline
 & NAD & \textbf{3.16} & \textbf{3.81} & 35.71 & 41.68 & \textbf{2.52} & 1.81 & 28.89 & \textbf{3.87} & \textbf{1.63} & 18.92 & \textbf{2.98} & \textbf{1.81} & \textbf{4.75} & \textbf{2.95} & \textbf{0.93} & \textbf{5.42}\tabularnewline
 & IF & 3.34 & 4.15 & \textbf{2.30} & \textbf{7.79} & 3.32 & \textbf{1.01} & \textbf{4.43} & 4.76 & 37.48 & 34.30 & 4.47 & 16.35 & 18.96 & 3.21 & 4.82 & 6.80\tabularnewline
 & VIF & 7.81 & 7.70 & 2.52 & 11.27 & 6.43 & 1.22 & 7.10 & 7.53 & 10.52 & 16.50 & 7.67 & 3.07 & 12.38 & 7.97 & 3.96 & 10.67\tabularnewline
 & AIF & 4.67 & 5.60 & 2.37 & 9.03 & 4.87 & 1.14 & 6.02 & 5.23 & 1.96 & \textbf{7.10} & 5.28 & 5.30 & 11.87 & 4.30 & 1.22 & 5.67\tabularnewline
\hline 
\hline 
\multirow{5}{*}{GTSRB} & Feb. & 42.01 & 35.30 & 21.02 & 44.11 & 43.40 & 75.75 & 95.90 & 32.18 & 97.83 & 97.37 & 21.27 & 70.02 & 72.71 & 33.18 & 70.10 & 71.69\tabularnewline
 & NAD & -0.13 & \textbf{-0.35} & \textbf{0.00} & 8.20 & \textbf{-0.32} & \textbf{0.00} & 4.06 & \textbf{-0.42} & \textbf{0.05} & \textbf{8.33} & \textbf{-0.28} & 0.05 & \textbf{0.56} & \textbf{-0.40} & \textbf{0.00} & \textbf{0.11}\tabularnewline
 & IF & 0.12 & 0.13 & \textbf{0.00} & 2.55 & 0.13 & 0.03 & 1.52 & 0.37 & 52.27 & 51.95 & 0.03 & 0.66 & 3.60 & 0.08 & 9.83 & 9.62\tabularnewline
 & VIF & 0.18 & 0.45 & \textbf{0.00} & 3.55 & 0.18 & \textbf{0.00} & 1.12 & 0.37 & 12.12 & 16.56 & 0.11 & \textbf{0.03} & 1.87 & 0.55 & 3.67 & 3.89\tabularnewline
 & AIF & \textbf{0.05} & -0.16 & \textbf{0.00} & \textbf{1.87} & 0.05 & \textbf{0.00} & \textbf{0.81} & 0.13 & 7.47 & 9.54 & -0.03 & 0.05 & 1.37 & -0.05 & 0.50 & 0.42\tabularnewline
\hline 
\hline 
\multirow{5}{*}{CelebA} & Feb. & 12.71 & 18.80 & 42.96 & 21.33 & 11.76 & 93.27 & 49.05 & 13.30 & 98.59 & 49.84 & 5.60 & 99.98 & 49.71 & 9.16 & 97.30 & 48.53\tabularnewline
 & NAD & 3.06 & \textbf{3.19} & 12.14 & 9.98 & 3.56 & 25.31 & 23.07 & 3.51 & 16.97 & 9.46 & 3.14 & 13.85 & 11.24 & 2.51 & 11.48 & \textbf{3.21}\tabularnewline
 & IF & \textbf{2.23} & 4.21 & 8.62 & \textbf{4.75} & \textbf{2.57} & 13.83 & 6.00 & \textbf{2.25} & 59.39 & 27.94 & \textbf{2.86} & 11.95 & \textbf{6.07} & \textbf{2.43} & 15.21 & 4.75\tabularnewline
 & VIF & 3.74 & 4.63 & 9.28 & 4.90 & 3.20 & \textbf{11.51} & \textbf{4.08} & 3.54 & \textbf{14.32} & \textbf{5.62} & 3.89 & 11.55 & 6.27 & 3.96 & \textbf{8.30} & 4.19\tabularnewline
 & AIF & 4.95 & 6.46 & \textbf{7.85} & 6.49 & 4.18 & 12.56 & 6.52 & 4.37 & 18.40 & 9.23 & 3.71 & \textbf{10.43} & 7.65 & 4.02 & 12.82 & 5.74\tabularnewline
\hline 
\end{tabular}}
\par\end{centering}
\caption{Trojan filtering results (in \%) of Februus, NAD, and our filtering
defenses against different attacks. \emph{Smaller values are better}.
For a particular dataset, attack, and metric, the best defense is
highlighted in bold.\label{tab:VIF-AIF-full-results}}
\end{table}

From Table~\ref{tab:VIF-AIF-full-results}, it is clear that VIF
and AIF achieve superior performances in mitigating Trojans of all
the single-target attacks compared to most of the baseline defenses.
For example, on MNIST and GTSRB, our filtering defenses impressively
reduce T from about 100\% (Table~\ref{tab:main-attack-results})
to less than 2\% for most attacks yet only cause less than 1\% drop
of clean accuracy ($\downarrow$C \textless{} 1\%). On more diverse
datasets such as CIFAR10 and CelebA, VIF and AIF still achieve T less
than 6\% and 12\% for most attacks while maintaining $\downarrow$C
below 8\% and 5\%, respectively. We note that on CelebA, the nonoptimal
performance of $\Classifier$ (accuracy $\approx$ 79\%) makes T higher
than normal because T may contains the error of samples from non-target
classes misclassified as the target class. However, it is not trivial
to disentangle the two quantities so we leave this problem for future
work. As there is no free lunch, our filtering defenses may be not
as good as some baselines in some specific cases. For example, on
CIFAR10, STRIP achieves FNRs $\approx$ 0\% against BadNet+/noise-BI+
(Fig.~\ref{fig:STRIP-result-bars}) while VIF/AIF achieves Ts $\approx$
1-3\%. However, the gaps are very small and in general, our filtering
defenses are still much more effective than the baseline defenses
against all the single-target attacks. Our filtering defenses also
perform well against all-target attacks (Fig.~\ref{fig:VIF-All-target-Trojan-Acc}
and Appdx.~\ref{subsec:Results-of-Our-Defenses-All-target}) as ours
are not sensitive to the number of target classes. To gain a better
insight into the performance of VIF/AIF, we visualize the filtered
images produced by VIF/AIF and their corresponding \emph{``counter-triggers''}
in Appdx.~\ref{subsec:Qualitative-Results-Against-Benchmark-Attacks}.

\begin{table}[t]
\begin{centering}
\resizebox{.92\textwidth}{!}{%
\par\end{centering}
\begin{centering}
\begin{tabular}{|c|c|d|da|da|da|da|da|}
\hline 
\multirow{2}{*}{Dataset} & \multirow{2}{*}{Defense} & \multicolumn{1}{c|}{Benign} & \multicolumn{2}{c|}{BadNet+} & \multicolumn{2}{c|}{noise-BI+} & \multicolumn{2}{c|}{image-BI+} & \multicolumn{2}{c|}{InpAwAtk} & \multicolumn{2}{c|}{WaNet}\tabularnewline
\cline{3-13} \cline{4-13} \cline{5-13} \cline{6-13} \cline{7-13} \cline{8-13} \cline{9-13} \cline{10-13} \cline{11-13} \cline{12-13} \cline{13-13} 
 &  & \multicolumn{1}{c|}{FPR} & \multicolumn{1}{c}{FPR} & \multicolumn{1}{c|}{FNR} & \multicolumn{1}{c}{FPR} & \multicolumn{1}{c|}{FNR} & \multicolumn{1}{c}{FPR} & \multicolumn{1}{c|}{FNR} & \multicolumn{1}{c}{FPR} & \multicolumn{1}{c|}{FNR} & \multicolumn{1}{c}{FPR} & \multicolumn{1}{c|}{FNR}\tabularnewline
\hline 
\hline 
\multirow{3}{*}{MNIST} & IFtC & 0.40 & 0.50 & \textbf{2.21} & 0.37 & 0.22 & 0.37 & 1.51 & 0.53 & \textbf{1.71} & 0.60 & 1.33\tabularnewline
 & VIFtC & 0.27 & \textbf{0.30} & 2.84 & 0.23 & \textbf{0.07} & 0.40 & 0.15 & 0.47 & 1.99 & 0.30 & 1.51\tabularnewline
 & AIFtC & \textbf{0.23} & 0.32 & 4.17 & \textbf{0.17} & 0.15 & \textbf{0.17} & \textbf{0.11} & \textbf{0.33} & 1.87 & \textbf{0.13} & \textbf{1.18}\tabularnewline
\hline 
\hline 
\multirow{3}{*}{CIFAR10} & IFtC & \textbf{6.83} & \textbf{7.47} & \textbf{1.70} & \textbf{6.57} & 0.93 & \textbf{7.83} & 36.56 & \textbf{7.25} & 16.89 & \textbf{7.17} & 5.15\tabularnewline
 & VIFtC & 12.30 & 11.00 & 2.63 & 10.67 & 1.26 & 11.03 & 10.89 & 10.63 & \textbf{3.67} & 11.40 & 4.26\tabularnewline
 & AIFtC & 8.63 & 8.87 & 1.96 & 8.73 & \textbf{0.89} & 8.77 & \textbf{2.15} & 8.27 & 5.93 & 7.93 & \textbf{1.56}\tabularnewline
\hline 
\hline 
\multirow{3}{*}{GTSRB} & IFtC & \textbf{0.38} & 0.45 & \textbf{0.00} & 0.24 & 0.03 & 0.66 & 52.91 & \textbf{0.29} & 1.00 & 0.66 & 10.41\tabularnewline
 & VIFtC & 0.45 & 0.74 & 0.03 & 0.47 & \textbf{0.00} & 0.87 & 12.63 & 0.53 & \textbf{0.37} & 1.31 & 4.25\tabularnewline
 & AIFtC & 0.50 & \textbf{0.37} & 0.03 & 0.39 & \textbf{0.00} & \textbf{0.63} & \textbf{7.87} & 0.47 & 0.40 & \textbf{0.60} & \textbf{1.08}\tabularnewline
\hline 
\hline 
\multirow{3}{*}{CelebA} & IFtC & \textbf{14.24} & \textbf{15.78} & 8.36 & \textbf{14.94} & 14.25 & \textbf{13.99} & 59.43 & \textbf{12.84} & 11.95 & \textbf{13.08} & 15.27\tabularnewline
 & VIFtC & 17.74 & 18.90 & 9.09 & 18.50 & \textbf{11.67} & 18.09 & \textbf{14.30} & 16.37 & 11.54 & 17.22 & \textbf{8.34}\tabularnewline
 & AIFtC & 20.24 & 20.95 & \textbf{7.71} & 19.08 & 12.82 & 19.29 & 18.65 & 16.54 & \textbf{10.43} & 16.55 & 12.87\tabularnewline
\hline 
\end{tabular}}
\par\end{centering}
\caption{Trojan mitigation results (in \%) of our FtC defenses against different
attacks. \emph{Smaller values are better}. For a particular attack,
dataset, and metric, the best defense is highlighted in bold.\label{tab:VIFtC-AIFtC-full-results}}
\end{table}

Among the filtering defenses, IF usually achieves the smallest $\downarrow$Cs
because its loss does not have any term that encourages information
removal like VIF's and AIF's. The gaps in $\downarrow$C between IF
and AIF/VIF are the largest on CIFAR10 but do not exceed 5\%. However,
IF usually performs much worse than VIF/AIF in mitigating Trojans,
especially those from image-BI+, InpAwAtk, and WaNet. For example,
on CIFAR10, GTSRB, and CelebA, IF reduces the attack success rate
(T) of image-BI+ to 37.48\%, 52.27\%, and 59.39\% respectively. These
numbers are only 1.96\%, 7.47\%, and 18.40\% for AIF and 10.52\%,
12.12\%, and 14.32\% for VIF. Therefore, when considering the trade-off
between $\downarrow$C and T, VIF and AIF are clearly better than
IF. We also observe that AIF usually achieves lower $\downarrow$Cs
and $\downarrow$Rs than VIF. It is because AIF discards only potential
malicious information instead of all noisy information like VIF. However,
VIF is simpler and easier to train than AIF.

From Table~\ref{tab:VIFtC-AIFtC-full-results}, we see that the FPRs
and FNRs of VIFtC/AIFtC are close to the $\downarrow$Cs and Ts of
VIF/AIF respectively on MNIST, GTSRB, and CIFAR10. This is because
$\Classifier$ achieves nearly 100\% clean and Trojan accuracies on
these datasets. Thus, we can interpret the results of VIFtC/AIFtC
in the same way as what we have done for VIF/AIF. Since FPR only affects
the classification throughput not (clean) accuracy, VIFtC/AIFtC are
preferred to VIF/AIF in applications that favor (clean) accuracy (e.g.,
defending against attacks with large-norm triggers in Sect.~\ref{subsec:Large-norm-triggers}).

\subsection{Ablation Studies}

It is undoubted that our defenses require some settings to work well.
However, these settings \emph{cannot be managed by attackers} unlike
the assumptions of most existing defenses \cite{Gao_etal_19Strip,Wang_etal_19Neural}.
Below, we examine the contribution of lossy data compression to the
performance of VIF, and the robustness of our proposed defenses to
small amounts of training data and to large-norm triggers. For other
ablation studies, please refer to Appdx.~\ref{sec:Additional-Results-of-Our-Defenses}.

\begin{figure}[t]
\centering{}\subfloat[Clean Accuracy]{\begin{centering}
\includegraphics[width=0.315\textwidth]{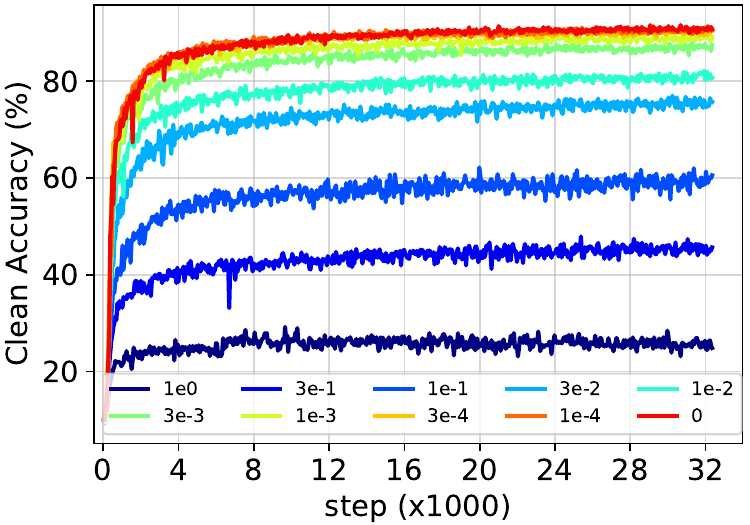}
\par\end{centering}
}\subfloat[Trojan Accuracy]{\begin{centering}
\includegraphics[width=0.315\textwidth]{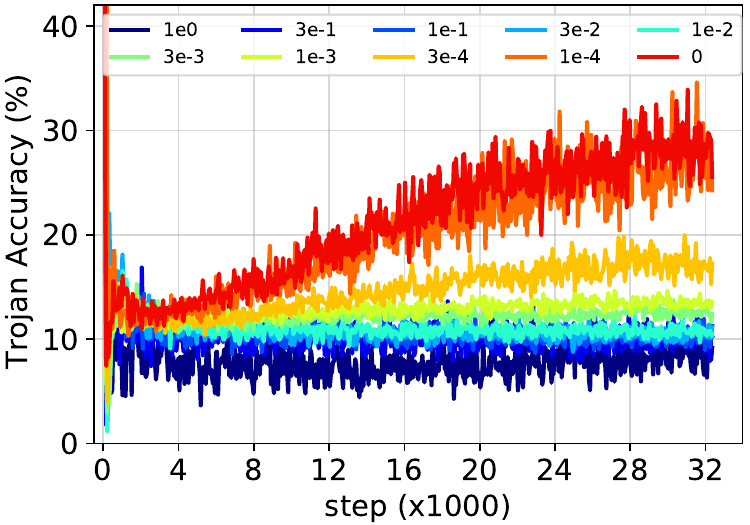}
\par\end{centering}
}\subfloat[Recovery Accuracy]{\begin{centering}
\includegraphics[width=0.315\textwidth]{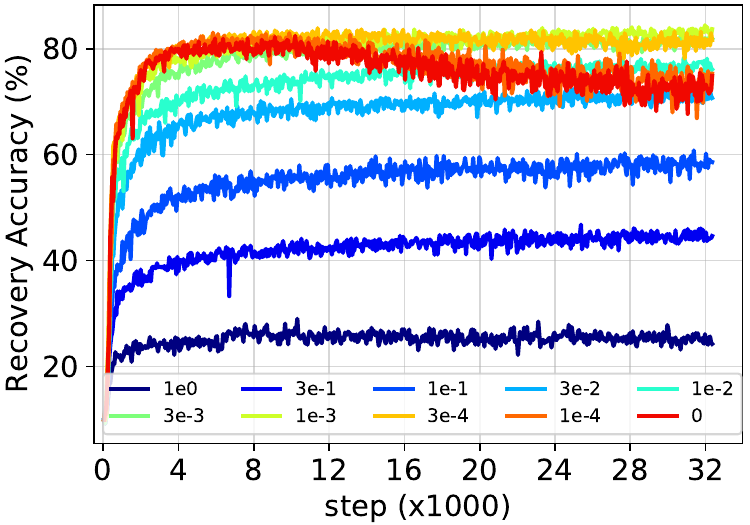}
\par\end{centering}
}\caption{Clean, Trojan, and recovery accuracy curves of VIF against InpAwAtk
on CIFAR10 w.r.t. different values of $\lambda_{2}$ in Eq.~\ref{eq:VIF_loss}.
The Trojan accuracy curves in (b) fluctuate around 10\% since they
are computed on $\protect\Data_{\text{test}}$ instead of $\protect\Data'_{\text{test}}$.\label{fig:KLD-coeff-curves-CIFAR10}}
\end{figure}

\subsubsection{Different data compression rates in VIF\label{subsec:Different-coefficients-of-DKL}}

The lossy data compression in VIF can be managed via changing the
coefficients of $D_{\text{KL}}$ in $\Loss_{\text{VIF}}$ ($\lambda_{2}$
in Eq.~\ref{eq:VIF_loss}). A smaller values of $\lambda_{2}$ means
a lower data compression rate and vice versa. From Fig.~\ref{fig:KLD-coeff-curves-CIFAR10},
it is clear that when $\lambda_{2}$ is small, most information in
the input including both semantic information and embedded triggers
is retained, thus, the clean accuracy (C) and the Trojan accuracy
(T) are both high. To decide the optimal value of $\lambda_{2}$,
we base on recovery accuracy (R) since R can be seen as a combination
of C and T to some extent. From the results on CIFAR10 (Fig.~\ref{fig:KLD-coeff-curves-CIFAR10})
and on other datasets, we found $\lambda_{2}=0.003$ to be the best.

\begin{table}[t]
\begin{centering}
\resizebox{\textwidth}{!}{%
\par\end{centering}
\begin{centering}
\begin{tabular}{|c|c|c|cccccc|cccccc|}
\hline 
\multirow{3}{*}{Def.} & \multirow{3}{*}{Metric} & \multicolumn{13}{c|}{InpAwAtk}\tabularnewline
\cline{3-15} \cline{4-15} \cline{5-15} \cline{6-15} \cline{7-15} \cline{8-15} \cline{9-15} \cline{10-15} \cline{11-15} \cline{12-15} \cline{13-15} \cline{14-15} \cline{15-15} 
 &  & Default & \multicolumn{6}{c|}{Fixed \#epochs} & \multicolumn{6}{c|}{Fixed total \#iterations}\tabularnewline
\cline{3-15} \cline{4-15} \cline{5-15} \cline{6-15} \cline{7-15} \cline{8-15} \cline{9-15} \cline{10-15} \cline{11-15} \cline{12-15} \cline{13-15} \cline{14-15} \cline{15-15} 
 &  & 1.0 & 0.8 & 0.6 & 0.4 & 0.2 & 0.1 & 0.05 & 0.8 & 0.6 & 0.4 & 0.2 & 0.1 & 0.05\tabularnewline
\hline 
\hline 
\rowcolor{gray0}\cellcolor{white} & \cellcolor{white}$\downarrow$C & \emph{4.47} & 4.73 & 4.43 & 5.37 & 7.90 & 11.60 & 25.93 & 3.53 & 3.87 & 3.83 & 4.20 & 4.23 & 4.87\tabularnewline
\rowcolor{gray1}\cellcolor{white}IF & \cellcolor{white}T & \emph{16.35} & 15.11 & 11.56 & 9.48 & 6.70 & 4.07 & 6.70 & 16.11 & 15.41 & 20.22 & 18.37 & 26.33 & 21.00\tabularnewline
\rowcolor{gray2}\cellcolor{white} & \cellcolor{white}$\downarrow$R & \emph{18.96} & 19.00 & 15.10 & 15.47 & 15.07 & 17.47 & 32.23 & 18.47 & 18.03 & 21.87 & 20.57 & 26.63 & 23.17\tabularnewline
\hhline{*{15}{-}}\rowcolor{gray0}\cellcolor{white} & \cellcolor{white}$\downarrow$C & \emph{7.67} & 9.17 & 8.57 & 9.83 & 12.13 & 17.10 & 27.80 & 8.53 & 7.93 & 7.87 & 8.97 & 9.37 & 11.27\tabularnewline
\rowcolor{gray1}\cellcolor{white}VIF & \cellcolor{white}T & \emph{3.07} & 3.81 & 2.81 & 4.30 & 3.30 & 2.70 & 5.67 & 3.26 & 4.19 & 4.26 & 3.70 & 3.44 & 4.67\tabularnewline
\rowcolor{gray2}\cellcolor{white} & \cellcolor{white}$\downarrow$R & \emph{12.38} & 14.90 & 13.17 & 16.30 & 17.37 & 22.10 & 32.30 & 13.73 & 13.87 & 13.67 & 14.73 & 15.03 & 17.87\tabularnewline
\hhline{*{15}{-}}\rowcolor{gray0}\cellcolor{white} & \cellcolor{white}$\downarrow$C & \emph{5.28} & 5.47 & 6.53 & 7.90 & 11.70 & 16.67 & 26.13 & 5.07 & 5.23 & 5.17 & 6.17 & 5.77 & 7.83\tabularnewline
\rowcolor{gray1}\cellcolor{white}AIF & \cellcolor{white}T & \emph{5.30} & 4.96 & 3.59 & 4.89 & 3.15 & 2.89 & 4.00 & 6.63 & 9.04 & 5.70 & 4.56 & 8.48 & 12.96\tabularnewline
\rowcolor{gray2}\cellcolor{white} & \cellcolor{white}$\downarrow$R & \emph{11.87} & 11.60 & 11.90 & 13.87 & 17.03 & 21.80 & 30.40 & 11.73 & 13.83 & 11.77 & 12.57 & 14.50 & 19.93\tabularnewline
\hline 
\end{tabular}}
\par\end{centering}
\caption{Trojan filtering results (in \%) of IF, VIF, and AIF against InpAwAtk
on CIFAR10 w.r.t. different proportions of training data (the third
row) and two broader training settings (the second row): i) fixed
number of epochs, and ii) fixed total number of iterations. Results
taken from Table~\ref{tab:VIF-AIF-full-results} are shown in italic.\label{tab:Diff-train-data-main-results}}
\end{table}

\begin{figure}[t]
\begin{centering}
\resizebox{\textwidth}{!}{%
\par\end{centering}
\begin{centering}
\begin{tabular}{ccc?cc}
 & \multicolumn{2}{c}{Fixed \#epochs} & \multicolumn{2}{c}{Fixed total \#iterations}\tabularnewline
 &  &  &  & \tabularnewline
IF & \includegraphics[width=0.32\textwidth,valign=m]{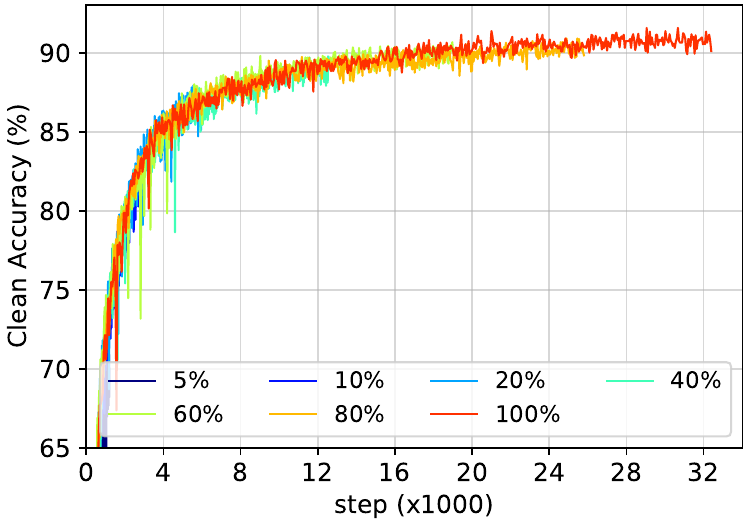} & \includegraphics[width=0.32\textwidth,valign=m]{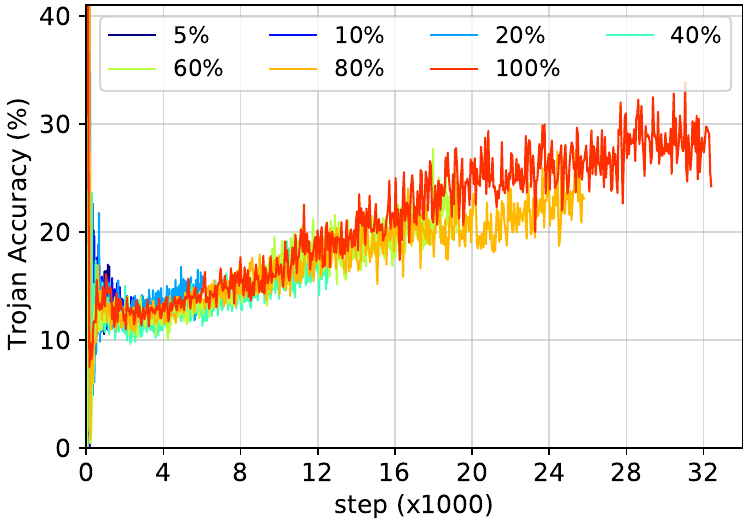} & \includegraphics[width=0.32\textwidth,valign=m]{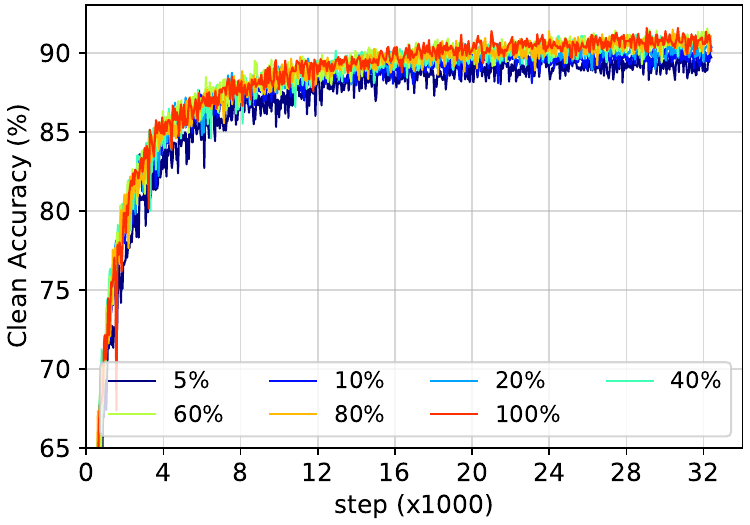} & \includegraphics[width=0.32\textwidth,valign=m]{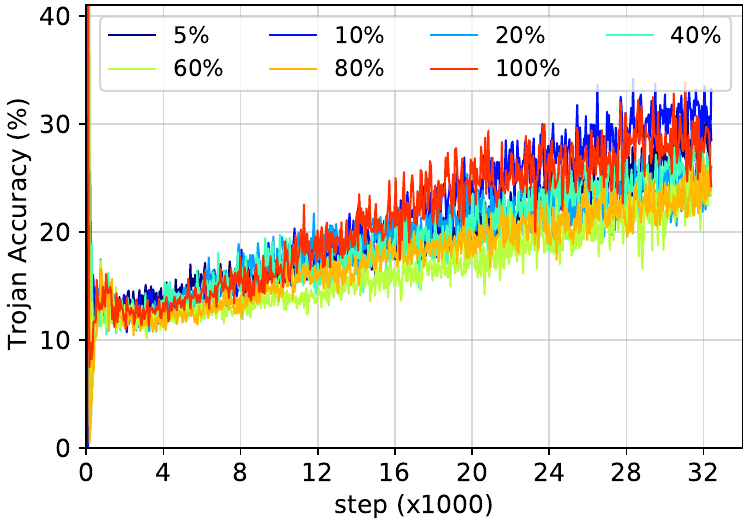}\tabularnewline
 &  &  &  & \tabularnewline
VIF & \includegraphics[width=0.32\textwidth,valign=m]{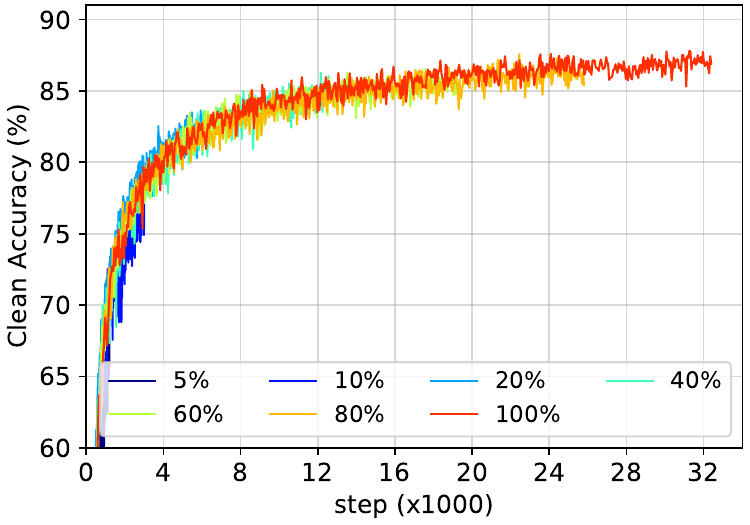} & \includegraphics[width=0.32\textwidth,valign=m]{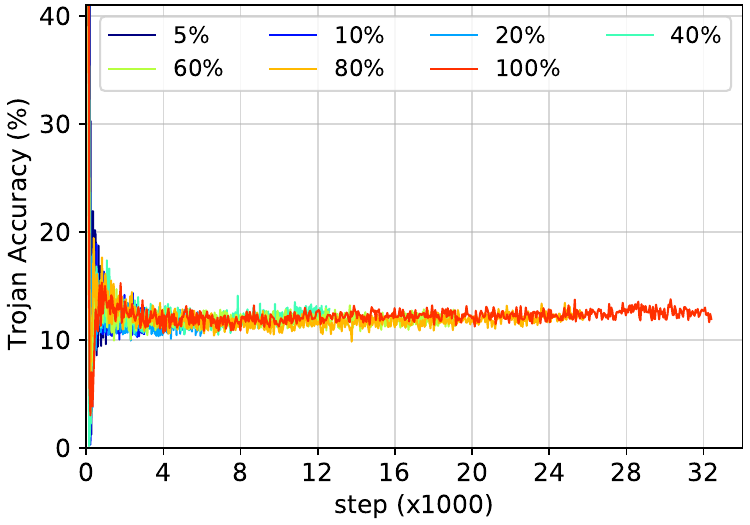} & \includegraphics[width=0.32\textwidth,valign=m]{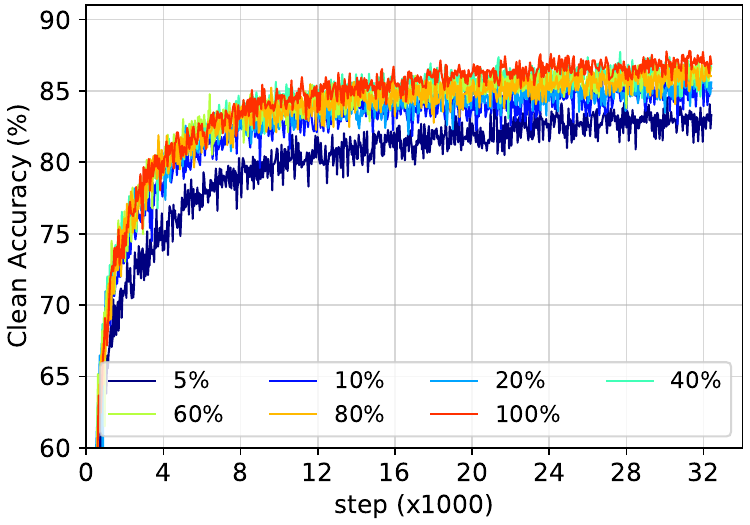} & \includegraphics[width=0.32\textwidth,valign=m]{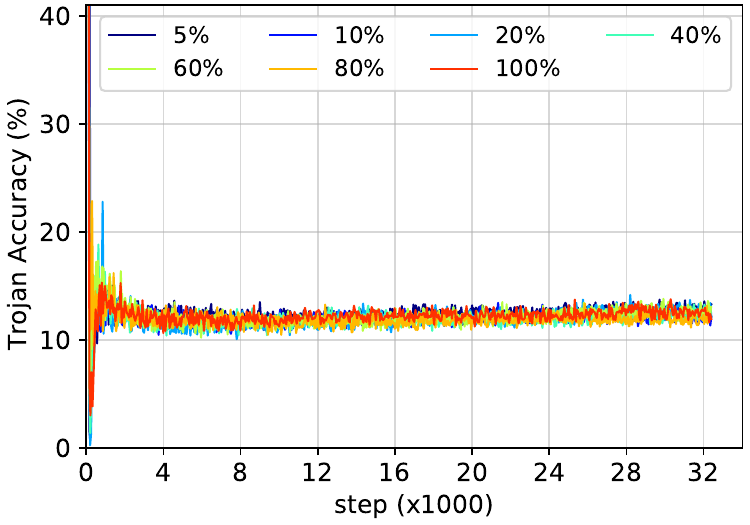}\tabularnewline
 &  &  &  & \tabularnewline
AIF & \includegraphics[width=0.32\textwidth,valign=m]{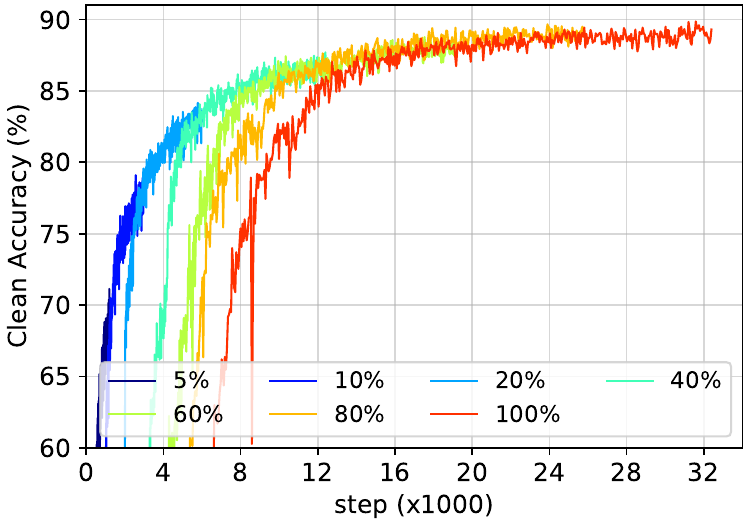} & \includegraphics[width=0.32\textwidth,valign=m]{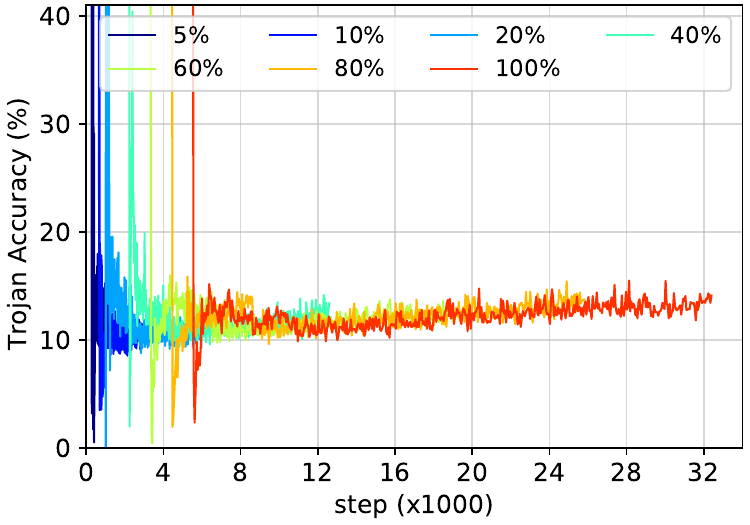} & \includegraphics[width=0.32\textwidth,valign=m]{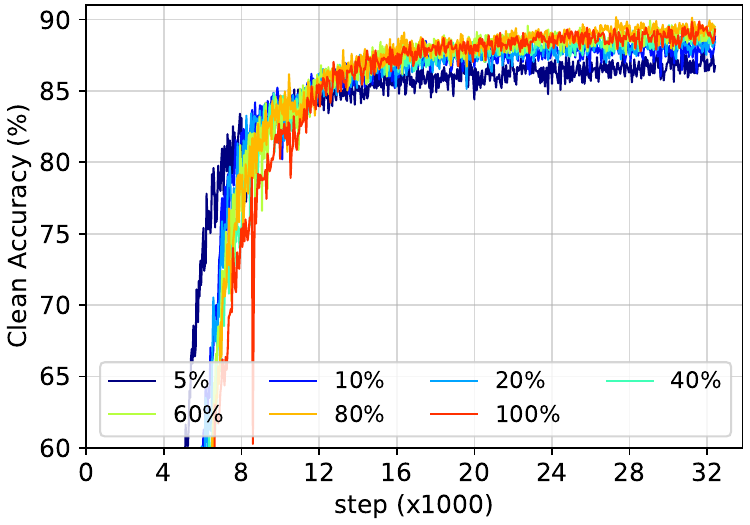} & \includegraphics[width=0.32\textwidth,valign=m]{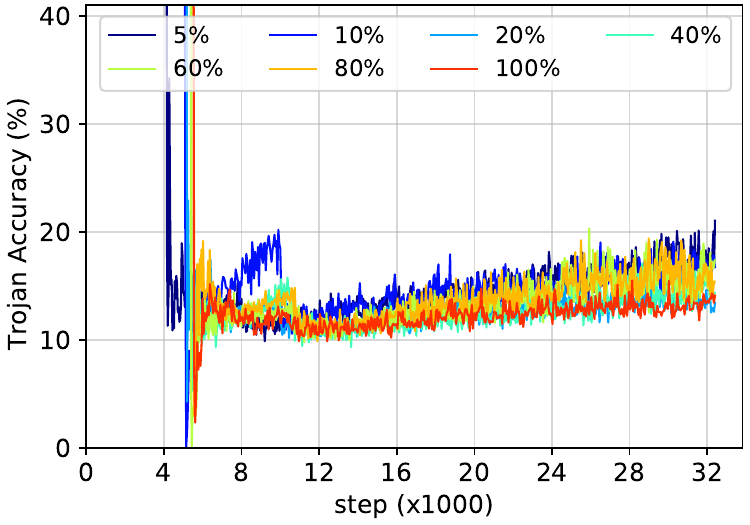}\tabularnewline
\end{tabular}}
\par\end{centering}
\caption{Test clean accuracy and Trojan accuracy curves of our filtering defenses
(IF, VIF, AIF) against InpAwAtk on CIFAR10 w.r.t. different proportions
of training data and 2 broader training settings: i) fixed number
of epochs and ii) fixed total number of iterations.\label{fig:Learning-curves-diff-num-data}}
\end{figure}

\subsubsection{Robustness to small amounts of training data\label{subsec:Small-amounts-of-training-data}}

We are curious to know how well our proposed defenses will perform
if we reduce the amount of training data. We select the proportion
of training data from \{0.8, 0.6, 0.4, 0.2, 0.1, 0.05\}. In addition,
we consider two broader training settings. In the first setting, the
number of training epochs is fixed at 600 (Section~\ref{subsec:Training-Settings-for-Our-Defenses})
regardless of the amount of training data. Because less training data
results in fewer iterations per epoch, fixing the number of training
epochs means \emph{smaller total number of training iterations for
less training data}. In the second setting, we adjust the number of
training epochs based on the proportion of training data so that the
total number of training iterations is fixed and similar to that when
full data is used. Table~\ref{tab:Diff-train-data-main-results}
shows the results of our filtering defenses w.r.t. the above settings.
When the number of epochs is fixed, we see that our defenses often
achieve lower Ts yet larger $\downarrow$Cs (and $\downarrow$Rs)
for less training data. This is because the filter $\Filter$ has
not been fully trained to reconstruct the input image well enough
(the first column in Fig.~\ref{fig:Learning-curves-diff-num-data}).
On the other hand, when the total number of iterations is fixed, $\Filter$
has been fully trained and we do not see much difference in Trojan
accuracy of VIF for different amounts of training data. The Trojan
accuracy of AIF slightly increase for less training data but is still
acceptable (the last column in Fig.~\ref{fig:Learning-curves-diff-num-data}).
Changes in clean accuracy of our filtering defenses are small (the
third column in Fig.~\ref{fig:Learning-curves-diff-num-data}). In
summary, these results suggest that our filtering defenses are quite
robust to the limited amount of training data.

\subsubsection{Robustness to large-norm triggers\label{subsec:Large-norm-triggers}}

\begin{figure}[t]
\begin{centering}
\resizebox{\textwidth}{!}{%
\par\end{centering}
\begin{centering}
\begin{tabular}{cccccc}
 & STRIP & Neural Cleanse & NAD & AIF & VIF\tabularnewline
 &  &  &  &  & \tabularnewline
\begin{turn}{90}
BadNet+
\end{turn} & \includegraphics[width=0.25\textwidth,valign=m]{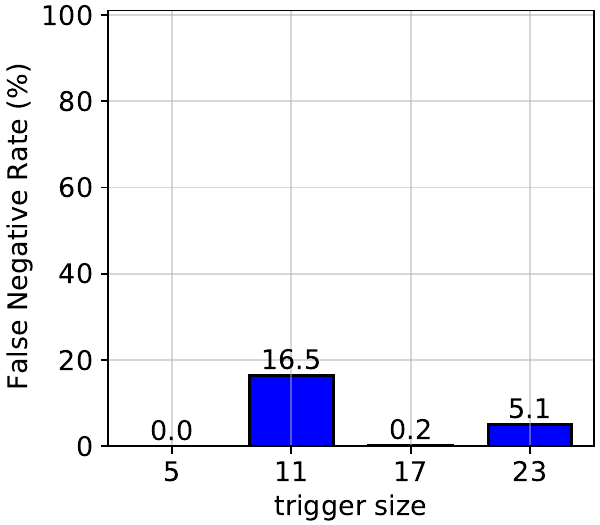} & \includegraphics[width=0.25\textwidth,valign=m]{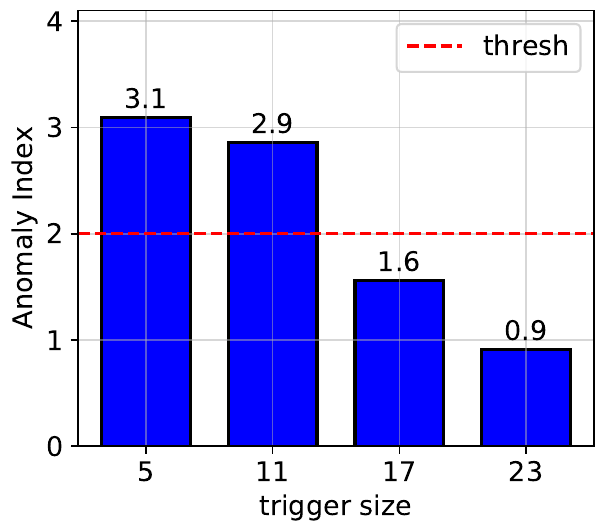} & \includegraphics[width=0.25\textwidth,valign=m]{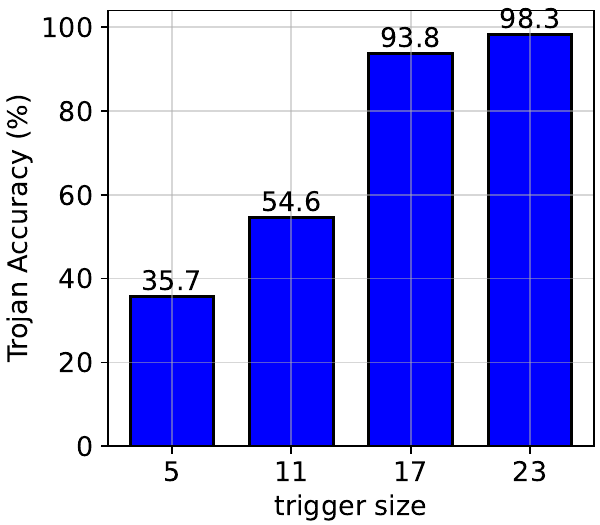} & \includegraphics[width=0.25\textwidth,valign=m]{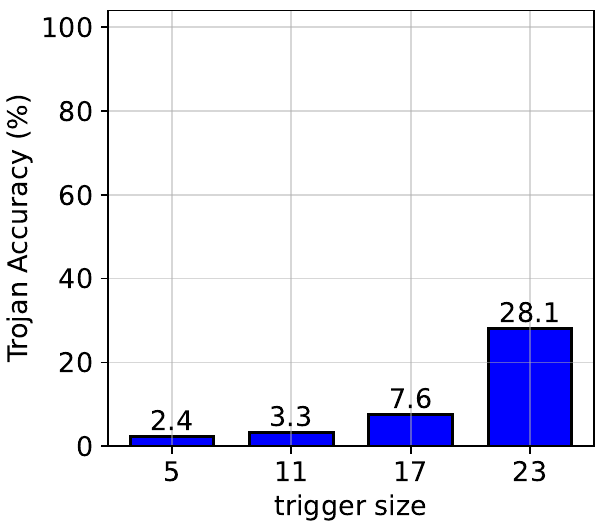} & \includegraphics[width=0.25\textwidth,valign=m]{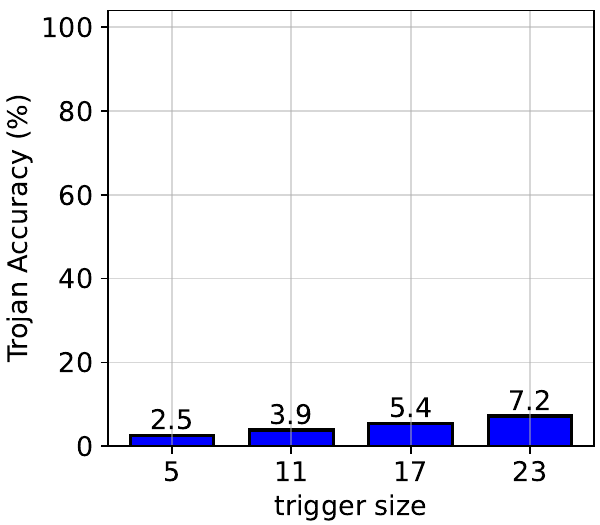}\tabularnewline
 &  &  &  &  & \tabularnewline
\begin{turn}{90}
noise-BI+
\end{turn} & \includegraphics[width=0.25\textwidth,valign=m]{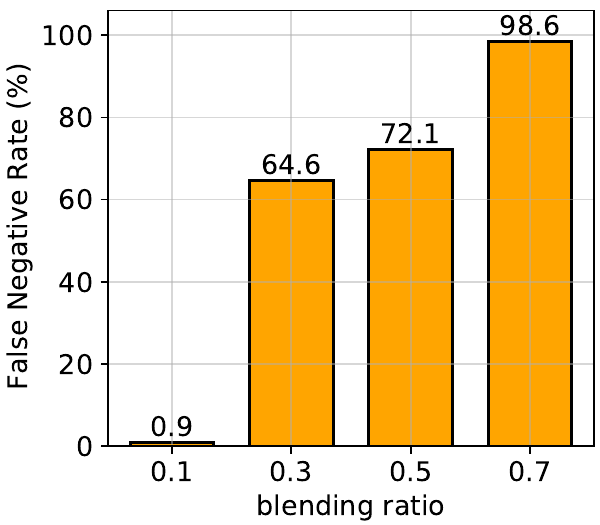} & \includegraphics[width=0.25\textwidth,valign=m]{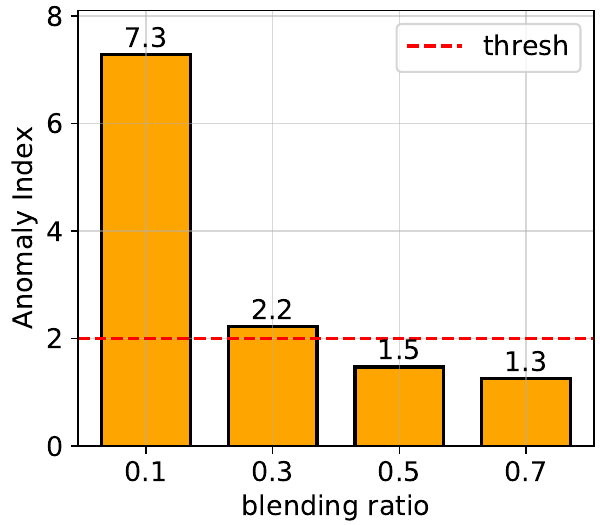} & \includegraphics[width=0.25\textwidth,valign=m]{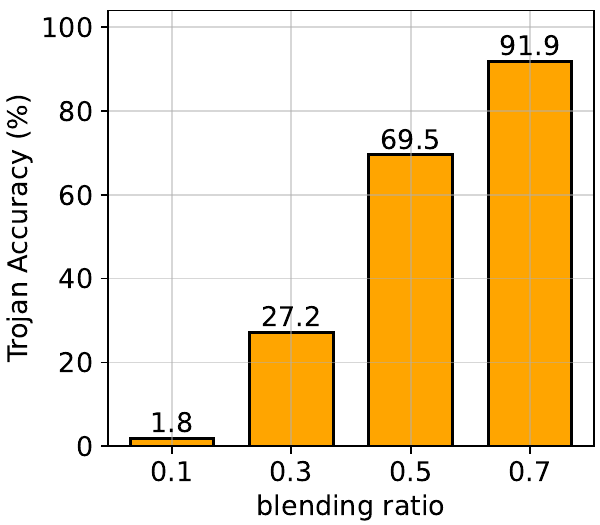} & \includegraphics[width=0.25\textwidth,valign=m]{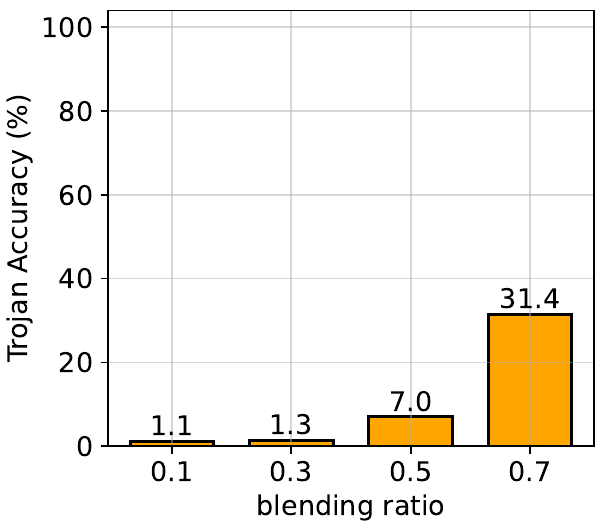} & \includegraphics[width=0.25\textwidth,valign=m]{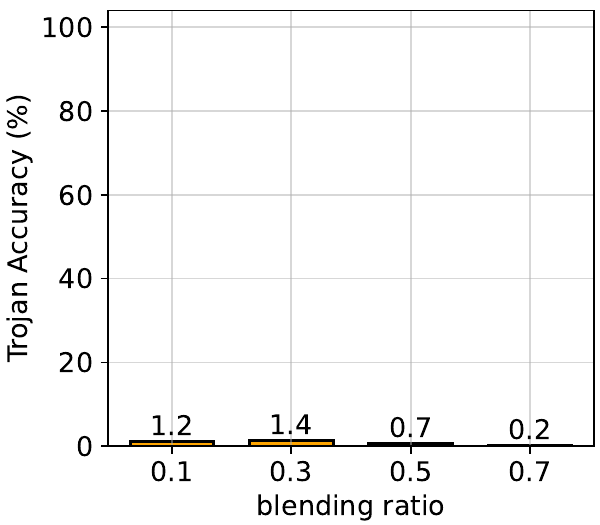}\tabularnewline
\end{tabular}}
\par\end{centering}
\caption{FNRs at 10\% FPR of STRIP, anomaly indices of Neural Cleanse, Trojan
accuracies of NAD, and Trojan accuracies of our filtering defenses
VIF, AIF against BadNet+ with different trigger sizes (top) and noise-BI+
with different blending ratios (bottom).\label{fig:Large-norm-triggers-comparison}}
\end{figure}

In this section, we examine the performances of our defenses against
attacks that use large-norm triggers. We consider BadNet+ and noise-BI+
for this study. For BadNet+, we increase the trigger norm by increasing
the trigger size $s$. The results for BadNet+ are shown in Table~\ref{tab:Large-norm-results-BadNet}.
For noise-BI+, we increase the trigger norm by increasing the blending
ratio $\alpha$. The results for noise-BI+ are shown in Table~\ref{fig:Large-norm-results-noiseBI}.
It is clear that even when triggers have large norms, our filtering
defenses, especially VIF, still effectively erase most of the trigger
pixels that could activate the Trojans in $\Classifier$ (Fig.~\ref{fig:Large-norm-triggers-visualization})
and achieve \emph{low Trojan accuracies} (Tables.~\ref{tab:Large-norm-results-BadNet},
\ref{fig:Large-norm-results-noiseBI}). However, large-norm triggers
cause a lot of difficulty in reconstructing the original clean images
from Trojan images (Fig.~\ref{fig:Large-norm-triggers-visualization}),
which leads to large decreases in recovery accuracy of our methods.
Note that such poor performance is inevitable for filtering defenses
like VIF and AIF. A solution to this problem is using other types
of defenses. STRIP \cite{Gao_etal_19Strip}, Neural Cleanse \cite{Wang_etal_19Neural},
and NAD \cite{li2021neural} are possible yet not good options. As
shown in Fig.~\ref{fig:Large-norm-triggers-comparison}, STRIP works
well against BadNet+ with large trigger sizes but poorly against noise-BI+
with large blending ratios. We guess the reason is that with large
blending ratios, Trojan images of noise-BI+ will look like noises,
and superimposing a noise-like Trojan image with a clean image is
like adding noise to the clean image, hence, won't affect of the class
prediction of the clean image. Neural Cleanse tends to wrongly identify
Trojan models as benign if the behind attacks use triggers with large
enough norms. This is because the reverse-engineered trigger w.r.t.
the true target class also has large norm which is not very different
from the norms of the reverse-engineered triggers w.r.t. other classes.
NAD achieves impracticably high Trojan accuracies when trigger norms
are large for both BadNet+ and noise-BI+. By contrast, our FtC defenses,
especially VIFtC, are \emph{good alternatives}. VIFtC achieves very
low FNRs (\textless 7\% in case of BadNet+ and \textless 2\% in
case of noise-BI+) while still keeping FPRs in an acceptable range
between 10\% and 15\% (Tables.~\ref{tab:Large-norm-results-BadNet},
\ref{fig:Large-norm-results-noiseBI}). 

\begin{table}[t]
\begin{centering}
\resizebox{\textwidth}{!}{%
\par\end{centering}
\begin{centering}
\begin{tabular}{|c|dab|da|dab|da|dab|da|cab|da|}
\hline 
\multirow{3}{*}{Defense} & \multicolumn{20}{c|}{BadNet+}\tabularnewline
\cline{2-21} \cline{3-21} \cline{4-21} \cline{5-21} \cline{6-21} \cline{7-21} \cline{8-21} \cline{9-21} \cline{10-21} \cline{11-21} \cline{12-21} \cline{13-21} \cline{14-21} \cline{15-21} \cline{16-21} \cline{17-21} \cline{18-21} \cline{19-21} \cline{20-21} \cline{21-21} 
 & \multicolumn{5}{c|}{$s=5$} & \multicolumn{5}{c|}{$s=11$} & \multicolumn{5}{c|}{$s=17$} & \multicolumn{5}{c|}{$s=23$}\tabularnewline
\cline{2-21} \cline{3-21} \cline{4-21} \cline{5-21} \cline{6-21} \cline{7-21} \cline{8-21} \cline{9-21} \cline{10-21} \cline{11-21} \cline{12-21} \cline{13-21} \cline{14-21} \cline{15-21} \cline{16-21} \cline{17-21} \cline{18-21} \cline{19-21} \cline{20-21} \cline{21-21} 
 & \multicolumn{1}{c}{$\downarrow$C} & \multicolumn{1}{c}{T} & \multicolumn{1}{c|}{$\downarrow$R} & \multicolumn{1}{c}{FPR} & \multicolumn{1}{c|}{FNR} & \multicolumn{1}{c}{$\downarrow$C} & \multicolumn{1}{c}{T} & \multicolumn{1}{c|}{$\downarrow$R} & \multicolumn{1}{c}{FPR} & \multicolumn{1}{c|}{FNR} & \multicolumn{1}{c}{$\downarrow$C} & \multicolumn{1}{c}{T} & \multicolumn{1}{c|}{$\downarrow$R} & \multicolumn{1}{c}{FPR} & \multicolumn{1}{c|}{FNR} & \multicolumn{1}{c}{$\downarrow$C} & \multicolumn{1}{c}{T} & \multicolumn{1}{c|}{$\downarrow$R} & \multicolumn{1}{c}{FPR} & \multicolumn{1}{c|}{FNR}\tabularnewline
\hline 
\hline 
IF \textbar{} IFtC & \textbf{\emph{4.15}} & \emph{2.30} & \textbf{\emph{7.79}} & \textbf{\emph{7.47}} & \textbf{\emph{1.70}} & 5.83 & 6.33 & 29.27 & \textbf{8.83} & 6.81 & \textbf{4.17} & 22.52 & 52.50 & \textbf{6.83} & 22.37 & \textbf{4.30} & 64.41 & 78.77 & \textbf{7.47} & 63.74\tabularnewline
VIF \textbar{} VIFtC & \emph{7.70} & \emph{2.52} & \emph{11.27} & \emph{11.00} & \emph{2.63} & 11.07 & 3.89 & 31.07 & 14.43 & 3.85 & 9.13 & \textbf{5.41} & 55.60 & 12.80 & \textbf{4.89} & 9.77 & \textbf{7.22} & 77.73 & 14.07 & \textbf{6.81}\tabularnewline
AIF \textbar{} AIFtC & \emph{5.60} & \emph{2.37} & \emph{9.03} & \emph{8.87} & \emph{1.96} & \textbf{5.53} & \textbf{3.33} & \textbf{23.97} & 9.03 & \textbf{3.15} & 4.67 & 7.56 & \textbf{48.40} & 8.30 & 8.04 & 5.10 & 28.07 & \textbf{77.70} & 8.30 & 28.0\tabularnewline
AIF$^{*}$ \textbar{} AIFtC$^{*}$ & 5.90 & \textbf{2.15} & 10.00 & 9.30 & 2.15 & 6.50 & 5.00 & 34.33 & 9.97 & 5.30 & 5.77 & 14.63 & 52.67 & 9.33 & 13.70 & 6.97 & 29.96 & 77.93 & 10.60 & 30.19\tabularnewline
\hline 
\end{tabular}}
\par\end{centering}
\caption{Trojan filtering results (in \%) of IF, VIF, AIF, and AIF without
explicit trigger normalization (AIF$^{*}$) against BadNet+ with different
trigger sizes ($s$) on CIFAR10. For a particular trigger size and
metric, the best result is highlighted in bold. Results taken from
Tables~\ref{tab:VIF-AIF-full-results}, \ref{tab:VIFtC-AIFtC-full-results}
are shown in italic.\label{tab:Large-norm-results-BadNet}}
\end{table}

\begin{table}[t]
\begin{centering}
\resizebox{\textwidth}{!}{%
\par\end{centering}
\begin{centering}
\begin{tabular}{|c|dabda|dabda|daada|cabda|}
\hline 
\multirow{3}{*}{Defense} & \multicolumn{20}{c|}{noise-BI+}\tabularnewline
\cline{2-21} \cline{3-21} \cline{4-21} \cline{5-21} \cline{6-21} \cline{7-21} \cline{8-21} \cline{9-21} \cline{10-21} \cline{11-21} \cline{12-21} \cline{13-21} \cline{14-21} \cline{15-21} \cline{16-21} \cline{17-21} \cline{18-21} \cline{19-21} \cline{20-21} \cline{21-21} 
 & \multicolumn{5}{c|}{$\alpha=0.1$} & \multicolumn{5}{c|}{$\alpha=0.3$} & \multicolumn{5}{c|}{$\alpha=0.5$} & \multicolumn{5}{c|}{$\alpha=0.7$}\tabularnewline
\cline{2-21} \cline{3-21} \cline{4-21} \cline{5-21} \cline{6-21} \cline{7-21} \cline{8-21} \cline{9-21} \cline{10-21} \cline{11-21} \cline{12-21} \cline{13-21} \cline{14-21} \cline{15-21} \cline{16-21} \cline{17-21} \cline{18-21} \cline{19-21} \cline{20-21} \cline{21-21} 
 & \multicolumn{1}{c}{$\downarrow$C} & \multicolumn{1}{c}{T} & \multicolumn{1}{c}{$\downarrow$R} & \multicolumn{1}{c}{FNR} & \multicolumn{1}{c|}{FPR} & \multicolumn{1}{c}{$\downarrow$C} & \multicolumn{1}{c}{T} & \multicolumn{1}{c}{$\downarrow$R} & \multicolumn{1}{c}{FNR} & \multicolumn{1}{c|}{FPR} & \multicolumn{1}{c}{$\downarrow$C} & \multicolumn{1}{c}{T} & \multicolumn{1}{c}{$\downarrow$R} & \multicolumn{1}{c}{FNR} & \multicolumn{1}{c|}{FPR} & \multicolumn{1}{c}{$\downarrow$C} & \multicolumn{1}{c}{T} & \multicolumn{1}{c}{$\downarrow$R} & \multicolumn{1}{c}{FNR} & \multicolumn{1}{c|}{FPR}\tabularnewline
\hline 
\hline 
IF/IFtC & \textbf{\emph{3.32}} & \emph{1.01} & \textbf{\emph{4.43}} & \textbf{\emph{6.57}} & \emph{0.93} & \textbf{4.30} & 1.26 & 36.73 & \textbf{7.80} & 1.19 & \textbf{3.53} & 27.59 & 73.03 & \textbf{7.10} & 26.74 & \textbf{3.87} & 49.37 & 83.10 & \textbf{7.13} & 49.00\tabularnewline
VIF/VIFtC & \emph{6.43} & \emph{1.22} & \emph{7.10} & \emph{10.67} & \emph{1.26} & 7.53 & 1.44 & 27.57 & 11.80 & 1.56 & 7.67 & \textbf{0.74} & \textbf{67.33} & 11.63 & \textbf{0.74} & 7.93 & \textbf{0.22} & 80.80 & 11.57 & \textbf{0.26}\tabularnewline
AIF/AIFtC & \emph{4.87} & \emph{1.14} & \emph{6.02} & \emph{8.73} & \emph{0.89} & 4.90 & 1.33 & 39.50 & 8.60 & 1.37 & 5.30 & 7.00 & 72.30 & 9.10 & 7.11 & 6.07 & 31.41 & \textbf{80.70} & 8.87 & 31.67\tabularnewline
AIF$^{*}$/AIFtC$^{*}$ & 5.60 & \textbf{0.78} & 7.50 & 9.53 & \textbf{0.74} & 6.07 & \textbf{0.48} & 45.13 & 10.10 & \textbf{0.74} & 6.27 & 1.48 & 74.83 & 10.67 & 1.78 & 6.47 & 38.74 & 80.97 & 10.00 & 39.56\tabularnewline
\hline 
\end{tabular}}
\par\end{centering}
\caption{Trojan filtering results (in \%) of IF, VIF, AIF, and AIF without
explicit trigger normalization (AIF$^{*}$) against noise-BI+ with
different blending ratios ($\alpha$) on CIFAR10. For a particular
blending ratio and metric, the best result is highlighted in bold.
Results taken from Tables~\ref{tab:VIF-AIF-full-results}, \ref{tab:VIFtC-AIFtC-full-results}
are shown in italic.\label{fig:Large-norm-results-noiseBI}}
\end{table}

\section{Related Work\label{sec:Related-Work}}

Due to space limit, in this section we only discuss related work about
Trojan defenses. Related work about Trojan attacks are provided in
Appdx.~\ref{sec:Related-Work-about-Attacks}. A large number of Trojan
defenses have been proposed so far, among which Neural Cleanse (NC)
\cite{Wang_etal_19Neural}, Network Pruning (NP) \cite{liu2018fine},
STRIP \cite{Gao_etal_19Strip}, Neural Attention Distillation (NAD)
\cite{li2021neural}, and Februus \cite{doan2020februus} are representative
for five different types of defenses and are carefully analyzed in
Section~\ref{par:Baseline-Defenses-Setup}. DeepInspect \cite{chen2019deepinspect},
MESA \cite{qiao2019defending} improve upon NC by synthesizing a distribution
of triggers for each class instead of just a single one. TABOR \cite{guo2019tabor}
adds more regularization losses to NC to better handle large and scattered
triggers. STS \cite{harikumar2020scalable} restores triggers by minimizing
a novel loss function which is the pairwise difference between the
class probabilities of two random synthetic Trojan images. This makes
STS independent of the number of classes and more efficient than NC
on datasets with many classes. ABS \cite{Liu_etal_19Abs} is a quite
complicated defense inspired by brain stimulation. It analyzes all
neurons in the classifier $\Classifier$ to find ``compromised''
ones and use these neurons to validate whether $\Classifier$ is attacked
or not. DL-TND \cite{wang2020practical}, B3D \cite{dong2021black}
focus on detecting Trojan-infected models in case validation data
are limited. However, all the aforementioned defenses derived from
NC still make the same ``input-agnostic trigger'' and ``single
target class'' assumptions as NC, and hence, are supposed to be ineffective
against attacks that break these assumptions such as input-specific
\cite{nguyen2020input,nguyen2021wanet} and all-target attacks. Activation
Clustering \cite{chen2018detecting} and Spectral Signatures \cite{Tran_etal_18Spectral}
regard hidden activations as a clue to detect Trojan samples from
BadNet \cite{gu2017badnets}. They base on an empirical observation
that the hidden activations of Trojan samples and clean samples of
the target class usually form distinct clusters in the hidden activation
space. These defenses are of the same kind as STRIP and are not applicable
to all-target attacks. Mode Connectivity Repair (MCR) \cite{zhao2020bridging}
mitigates Trojans by choosing an interpolated model near the two end
points of a parametric path connecting a Trojan model and its fine-tuned
version. MCR was shown to be defeated by InpAwAtk in \cite{nguyen2020input}.
Adversarial Neuron Pruning (ANP) \cite{wu2021adversarial} leverages
adversarial learning to find compromised neurons for pruning. Our
AIF is different from ANP in the sense that we use adversarial learning
to train an entire generator $\Generator$ for filtering input instead
of pruning neurons.

\section{Conclusion}

We have proposed two novel ``filtering'' Trojan defenses dubbed
VIF and AIF that leverage lossy data compression and adversarial learning
respectively to effectively remove all potential Trojan triggers embedded
in the input. We have also introduced a new defense mechanism called
``Filtering-then-Contrasting'' (FtC) that circumvents the loss in
clean accuracy caused by ``filtering''. Unlike most existing defenses,
our proposed filtering and FtC defenses make no assumption about the
number of triggers/target classes or the input dependency property
of triggers. Through extensive experiments, we have demonstrated that
our proposed defenses significantly outperform many well-known defenses
in mitigating various strong attacks. In the future, we would like
to extend our proposed defenses to other domains (e.g., texts, graphs)
and other tasks (e.g., object detection, visual reasoning) which we
believe are more challenging than those considered in this work.

\section*{Acknowledgement}

This research was partially supported by the Australian Government
through the Australian Research Council's Discovery Projects funding
scheme (project DP210102798). The views expressed herein are those
of the authors and are not necessarily those of the Australian Government
or Australian Research Council.

\bibliographystyle{splncs04}
\bibliography{Trojan}

\addtocontents{toc}{\protect\setcounter{tocdepth}{2}}

\appendix
\tableofcontents{}

\section{Related Work about Trojan Attacks\label{sec:Related-Work-about-Attacks}}

In this paper, we mainly consider a class of Trojan attacks in which
attackers fully control the training processes of a classifier. We
refer to these attacks as \emph{``full-control''} attacks. There
is another less common type of Trojan attacks called \emph{``clean-label''}
attacks \cite{saha2020hidden,shafahi2018poison,zhu2019transferable}.
These attacks assume a scenario in which people want to adapt a popular
pretrained classifier $\Classifier$ (e.g., ResNet \cite{he2016deep})
for their tasks by retraining the top layers of $\Classifier$ with
additional data collected from the web. The goal of an attacker is
to craft a poisoning image $\tilde{x}$ that looks visually indistinguishable
from an image $x_{t}$ of the target class $t$ while being close
to some source image $x_{s}$ in the feature space by optimizing the
following objective:
\[
\tilde{x}=\argmin x{\|\Classifier_{\text{f}}(x)-\Classifier_{\text{f}}(x_{s})\|}_{2}^{2}+\lambda{\|x-x_{t}\|}_{2}^{2}
\]
where $\Classifier_{\text{f}}(\cdot)$ denotes the output of the penultimate
layer of $\Classifier$. The attacker then puts $\tilde{x}$ on the
web so that it can be collected and labeled by victims. Since $\tilde{x}$
looks like an image of the target class $t$, $\tilde{x}$ will be
labeled as $t$. When the victims retrain $\Classifier$ using a dataset
containing $\tilde{x}$, $\tilde{x}$ will create a \emph{``backdoor''}
in $\Classifier$. The attacker can use $x_{s}$ to access this \emph{``backdoor''}
and forces $\Classifier$ to output $t$. Apart from the advantage
that the attacker does not need to control the labeling process (while
in fact, he can't), the clean-label attack has several drawbacks due
to its impractical assumptions. For example, the victims may retrain
the whole $\Classifier$ instead of just the last softmax layer of
$\Classifier$; the victims may use their own training data to which
the attacker can't access; the victims may use $\Classifier$ for
a completely new task that the attacker does not know; the attacker
doesn't even know who are the victims. Moreover, since $x_{s}$ often
looks very different from images of the target class $t$, $x_{s}$
can be easily detected by human inspection at test time.

Compared to clean-label attacks, full-control attacks are much harder
to defend against because attackers have all freedom to do whatever
they want with $\Classifier$ before sending it to the victims. BadNet
\cite{gu2017badnets}, Blended Injection \cite{chen2017targeted}
are among the earliest attacks \cite{Ji_etal_17Backdoor,liu2017neural}
of this type that use only one global trigger and use image blending
as an injection function. These attacks can be mitigated by well-known
defenses like Neural Cleanse \cite{Wang_etal_19Neural} or STRIP.
Besides, their triggers also look unnatural. Therefore, subsequent
attacks focus mainly on improving the robustness and stealthiness
of triggers at test time. Some attacks use dynamic and/or input-specific
triggers \cite{li2020backdoor,nguyen2020input,nguyen2021wanet,salem2020dynamic}.
Others use more advanced injection functions \cite{liu2020reflection,nguyen2021wanet}
or GANs \cite{munoz2019poisoning} to create hidden triggers or use
physical objects as triggers \cite{chen2017targeted,wenger2021backdoor}.

\section{Experimental Setup}

\begin{table}[t]
\begin{centering}
\begin{tabular}{ccccccc}
\hline 
\multirow{2}{*}{Dataset} & \multirow{2}{*}{\#Classes} & \multirow{2}{*}{Image size} & \multicolumn{2}{c}{Attack} & \multicolumn{2}{c}{Defense}\tabularnewline
 &  &  & \#Train & \#Test & \#Train & \#Test\tabularnewline
\hline 
MNIST & 10 & 28$\times$28$\times$1 & 60000 & 10000 & 7000 & 3000\tabularnewline
CIFAR10 & 10 & 32$\times$32$\times$3 & 50000 & 10000 & 7000 & 3000\tabularnewline
GTSRB & 43 & 32$\times$32$\times$3 & 39209 & 12630 & 8826 & 3804\tabularnewline
CelebA & 8 & 64$\times$64$\times$3 & 162770 & 19867 & 13904 & 5963\tabularnewline
\hline 
\end{tabular}
\par\end{centering}
\caption{Datasets used in our experiments.\label{tab:Datasets}}
\end{table}

\subsection{Datasets\label{subsec:Dataset-Description}}

We provide details of the datasets used in our experiments in Table~\ref{tab:Datasets}.
The training and test sets for defense ($\Data_{\text{val}}$ and
$\Data_{\text{test}}$) are taken from the test set for attack with
the \#training/\#test ratio of 0.7/0.3.

\begin{table}[t]
\begin{centering}
\begin{tabular}{>{\centering}p{0.32\textwidth}>{\centering}p{0.32\textwidth}}
\hline 
Encoder & Encoder\tabularnewline
\hline 
ConvBlockX(1, 16) & ConvBlockY(3, 64)\tabularnewline
ConvBlockX(16, 32) & ConvBlockY(64, 128)\tabularnewline
ConvBlockX(32, 64) & ConvBlockY(128, 256)\tabularnewline
Reshape {[}64, 3, 3{]} to {[}576{]} & ConvBlockY(256, 512)\tabularnewline
LinearBlockX(576, 256) & \tabularnewline
 & \tabularnewline
\hline 
Decoder & Decoder\tabularnewline
\hline 
LinearBlockX(256, 576) & DeconvBlockY(512, 512)\tabularnewline
Reshape {[}576{]} to {[}64, 3, 3{]} & DeconvBlockY(512, 256)\tabularnewline
DeconvBlockX(64, 32) & DeconvBlockY(256, 128)\tabularnewline
DeconvBlockX(32, 16) & DeconvBlockY(128, 3)\tabularnewline
DeconvBlockX(16, 1) & \tabularnewline
 & \tabularnewline
MNIST & CelebA\tabularnewline
\end{tabular}
\par\end{centering}
\centering{}\caption{Architectures of $\protect\Filter$ (in AIF) for MNIST and CelebA.\label{tab:Architectures-MNIST-CelebA}}
\end{table}

\begin{table}[t]
\begin{centering}
\begin{tabular}{>{\centering}p{0.35\textwidth}>{\centering}p{0.4\textwidth}}
\cline{1-1} 
LinearBlockX($d_{i}$ ,$d_{o}$) & \tabularnewline
\cline{1-1} 
Linear($d_{i}$, $d_{o}$, $b$=False) & \tabularnewline
BatchNorm2d($d_{o}$, $m$=0.01) & \tabularnewline
ReLU() & \tabularnewline
 & \tabularnewline
\hline 
ConvBlockX($c_{i}$, $c_{o}$) & DeconvBlockX($c_{i}$, $c_{o}$)\tabularnewline
\hline 
Conv2d($c_{i}$, $c_{o}$, $k$=4, $s$=2, \\
$p$=1, $b$=False) & ConvTranspose2d($c_{i}$, $c_{o}$, $k$=4,

$s$=2, $p$=1, $p_{o}$=1, $b$=False)\tabularnewline
BatchNorm2d($c_{o}$, $m$=0.01) & BatchNorm2d($c_{o}$, $m$=0.01)\tabularnewline
ReLU() & ReLU()\tabularnewline
 & \tabularnewline
\hline 
ConvBlockY($c_{i}$, $c_{o}$) & DeconvBlockY($c_{i}$, $c_{o}$)\tabularnewline
\hline 
Conv2d($c_{i}$, $c_{o}$, $k$=4, $s$=2, \\
$p$=1, $b$=False) & ConvTranspose2d($c_{i}$, $c_{o}$, $k$=4,

$s$=2, $p$=1, $b$=False)\tabularnewline
BatchNorm2d($c_{o}$, $m$=0.01) & BatchNorm2d($c_{o}$, $m$=0.01)\tabularnewline
LeakyReLU(0.2) & ReLU()\tabularnewline
\end{tabular}
\par\end{centering}
\caption{Linear, convolutional, and deconvolutional blocks of the architectures
in Table~\ref{tab:Architectures-MNIST-CelebA}.\label{tab:Architectures-MNIST-CelebA-Blocks}}
\end{table}

\subsection{Model Architectures and Training Settings for Benchmark Attacks\label{subsec:Training-Settings-for-the-Attacks}}

\paragraph{Model architectures}

Architectures of the classifier $\Classifier$ in our work follow
exactly those in \cite{nguyen2020input,nguyen2021wanet}. Specifically,
we use PreactResNet18 \cite{he2016identity} for CIFAR10 and GTSRB,
ResNet18 \cite{he2016deep} for CelebA, and the convolutional network
described in \cite{nguyen2020input} for MNIST.

\paragraph{Training settings}

We train Input-Aware Attack and WaNet based on the official implementations
of the two attacks\footnote{Input-Aware Attack: \url{https://github.com/VinAIResearch/input-aware-backdoor-attack-release}}\footnote{WaNet: \url{https://github.com/VinAIResearch/Warping-based_Backdoor_Attack-release}}
with the same settings as in the original papers \cite{nguyen2020input,nguyen2021wanet}.
We implement and train BadNet+, noise/image-BI+ ourselves. The settings
of these attacks are given in Appdx.~\ref{subsec:BadNet-and-Noise-Image-BI}.
We set the poisoning rate of 0.1 for all the benchmark attacks.

\begin{table}[t]
\begin{centering}
\resizebox{\textwidth}{!}{%
\par\end{centering}
\begin{centering}
\begin{tabular}{cccc}
\hline 
{\small{}Encoder} & {\small{}Encoder} & {\small{}Encoder} & {\small{}Encoder}\tabularnewline
\hline 
{\small{}ConvBlockA(3, 32, $k$=5, $s$=1)} & {\small{}ConvBlockB(3, 32)} & {\small{}ConvBlockC(3, 32)} & {\small{}ConvBlockC(3, 32) $\rightarrow$ $z_{1}$}\tabularnewline
{\small{}ConvBlockA(32, 64, $k$=4, $s$=2)} & {\small{}ConvBlockB(32, 32)} & {\small{}MaxPool2d(2, 2)} & {\small{}MaxPool2d(2, 2)}\tabularnewline
{\small{}ConvBlockA(64, 128, $k$=4, $s$=1)} & {\small{}MaxPool2d(2, 2)} & {\small{}ConvBlockC(32, 64)} & {\small{}ConvBlockC(32, 64) $\rightarrow$ $z_{2}$}\tabularnewline
{\small{}ConvBlockA(128, 256, $k$=4, $s$=2)} & {\small{}ConvBlockB(32, 64)} & {\small{}MaxPool2d(2, 2)} & {\small{}MaxPool2d(2, 2)}\tabularnewline
{\small{}ConvBlockA(256, 512, $k$=4, $s$=1)} & {\small{}ConvBlockB(64, 64)} & {\small{}ConvBlockC(64, 128)} & {\small{}ConvBlockC(64, 128) $\rightarrow$ $z$}\tabularnewline
{\small{}ConvBlockA(512, 512, $k$=1, $s$=1)} & {\small{}MaxPool2d(2, 2)} &  & \tabularnewline
{\small{}Linear(512, 256)} & {\small{}ConvBlockB(64, 128)} &  & \tabularnewline
 & {\small{}ConvBlockB(128, 128)} &  & \tabularnewline
 & {\small{}MaxPool2d(2, 2)} &  & \tabularnewline
 & {\small{}ConvBlockB(128, 128)} &  & \tabularnewline
 &  &  & \tabularnewline
\hline 
{\small{}Decoder} & {\small{}Decoder} & {\small{}Decoder} & {\small{}Decoder}\tabularnewline
\hline 
{\small{}DeconvBlockA(256, 256, $k$=4, $s$=1)} & {\small{}UpsamplingBilinear2d(2)} & {\small{}UpsamplingBilinear2d(2)} & {\small{}ConvTranspose2d(128, 64, $k$=2, $s$=2) $\rightarrow$ $y_{2}$}\tabularnewline
{\small{}DeconvBlockA(256, 128, $k$=4, $s$=2)} & {\small{}ConvBlockB(128, 128)} & {\small{}ConvBlockC(128, 64)} & {\small{}Concat({[}$y_{2}$, $z_{2}${]})}\tabularnewline
{\small{}DeconvBlockA(128, 64, $k$=4, $s$=1)} & {\small{}ConvBlockB(128, 64)} & {\small{}UpsamplingBilinear2d(2)} & {\small{}ConvBlockC(128,64)}\tabularnewline
{\small{}DeconvBlockA(64, 32, $k$=4, $s$=2)} & {\small{}UpsamplingBilinear2d(2)} & {\small{}ConvBlockC(64, 32)} & {\small{}ConvTranspose2d(64, 32, $k$=2, $s$=2) $\rightarrow$ $y_{1}$}\tabularnewline
{\small{}DeconvBlockA(32, 32, $k$=5, $s$=1)} & {\small{}ConvBlockB(64, 64)} & {\small{}Conv2d(32, 3, $k$=1)} & {\small{}Concat({[}$y_{1}$, $z_{1}${]})}\tabularnewline
{\small{}DeconvBlockA(32, 32, $k$=1, $s$=1)} & {\small{}ConvBlockB(64, 32)} &  & {\small{}ConvBlockC(64,32)}\tabularnewline
{\small{}ConvTranspose2d(32, 3, $k$=1, $s$=1)} & {\small{}UpsamplingBilinear2d(2)} &  & {\small{}Conv2d(32, 3, $k$=1)}\tabularnewline
 & {\small{}ConvBlockB(32, 32)} &  & \tabularnewline
 & {\small{}Conv2d(32, 3, $k$=3, $p$=1)} &  & \tabularnewline
 &  &  & \tabularnewline
{\small{}(A)} & {\small{}(B)} & {\small{}(C)} & {\small{}(D)}\tabularnewline
\end{tabular}
\par\end{centering}
\begin{centering}
}
\par\end{centering}
\centering{}\caption{Architectures of $\protect\Filter$ (in AIF) for CIFAR10 and GTSRB.\label{tab:4-Architectures-CIFAR10}}
\end{table}

\begin{table}[t]
\begin{centering}
\resizebox{\textwidth}{!}{%
\par\end{centering}
\begin{centering}
\begin{tabular}{cccc}
\hline 
ConvBlockA($c_{i}$, $c_{o}$, $k$, $s$) & DeconvBlockA($c_{i}$, $c_{o}$, $k$, $s$) & ConvBlockB($c_{i}$ ,$c_{o}$) & ConvBlockC($c_{i}$, $c_{o}$)\tabularnewline
\hline 
Conv2d($c_{i}$, $c_{o}$, $k$, $s$) & ConvTranspose2d($c_{i}$, $c_{o}$, $k$, $s$) & Conv2d($c_{i}$, $c_{o}$, $k$=3, $p$=1) & Conv2d($c_{i}$, $c_{o}$, $k$=3, $p$=1)\tabularnewline
BatchNorm2d($c_{o}$, $m$=0.1) & BatchNorm2d($c_{o}$, $m$=0.1) & BatchNorm2d($c_{o}$, $m$=0.05) & ReLU()\tabularnewline
LeakyReLU() & LeakyReLU() & ReLU() & BatchNorm2d($c_{\ensuremath{o}}$, $m$=0.01)\tabularnewline
 &  &  & Conv2d($c_{o}$, $c_{o}$, $k$=3, $p$=1)\tabularnewline
 &  &  & ReLU()\tabularnewline
 &  &  & BatchNorm2d($c_{\ensuremath{o}}$, $m$=0.01)\tabularnewline
\end{tabular}
\par\end{centering}
\begin{centering}
}
\par\end{centering}
\caption{Convolutional and deconvolutional blocks of the architectures in Table~\ref{tab:4-Architectures-CIFAR10}.\label{tab:4-Architectures-CIFAR10-Blocks}}
\end{table}

\subsection{Model Architectures and Training Settings for Our Defenses\label{subsec:Training-Settings-for-Our-Defenses}}

\paragraph{Model architectures}

In AIF, $\Filter$ is a plain autoencoder. We use the two architectures
in Table~\ref{tab:Architectures-MNIST-CelebA} for $\Filter$ when
working on MNIST and CelebA and the architecture (C) in Table~\ref{tab:4-Architectures-CIFAR10}
when working on CIFAR10 and GTSRB. The remaining architectures in
Table~\ref{tab:4-Architectures-CIFAR10} (A, B, D) are for our ablation
study in Appdx.~\ref{subsec:Different-architectures-of-F}. The architecture
of $\Generator$ is derived from the decoder of $\Filter$ with additional
layers to handle the noise vector $\epsilon$. $\epsilon$ has a fixed
length of 128. The symbols in Tables~\ref{tab:Architectures-MNIST-CelebA},
\ref{tab:Architectures-MNIST-CelebA-Blocks}, \ref{tab:4-Architectures-CIFAR10},
\ref{tab:4-Architectures-CIFAR10-Blocks} have the following meanings:
$c_{i}$ is input channel, $c_{o}$ is output channel, $k$ is kernel
size, $s$ is stride, $p$ is padding, $p_{o}$ is output padding,
$m$ is momentum, and $b$ is bias.

The architectures of $\Filter$ in VIF are adapted from those in AIF
by changing the middle layer between the encoder and decoder to produce
the latent mean $\mu_{z}$ and standard deviation $\sigma_{z}$ that
characterize $q_{\Filter}(z|x)$.

\paragraph{Training settings}

If not otherwise specified, we train the generator $\Generator$,
the filter $\Filter$, and the parameterized triggers $(m_{i},p_{i})$
using an Adam optimizer \cite{Kingma_Ba_14Adam} (learning rate $=1e^{-3}$,
$\beta_{1}=0.5$, $\beta_{2}=0.9$) for 600 epochs with batch size
equal to 128. For trigger synthesis (Section~\ref{sec:Finding-Triggers}),
in Eqs.~\ref{eq:NeuralCleanse_loss}, the norm is L2, $\delta=0$,
$\lambda_{0}$ varies from $1e^{-3}$ to $1$ with the multiplicative
step size $\approx$ 0.3. For VIF (Section~\ref{subsec:Variational-Input-Filtering}),
in Eq.~\ref{eq:VIF_loss}, the norm is L2, $\lambda_{1}=1.0$ and
$\lambda_{2}=0.003$. An analysis of different values of $\lambda_{2}$
is provided in Section~\ref{subsec:Different-coefficients-of-DKL}.
For AIF (Section~\ref{subsec:Adversarial-Input-Filtering}), $\Filter$
and $\Generator$ are optimized alternately with the learning rate
for $\Generator$ is $3e^{-4}$. In Eq.~\ref{eq:AIF_gen_loss}, $\delta=0.05$,
$\lambda_{0}=0.01$, $\lambda_{3}=0.3$. In Eq.~\ref{eq:AIF_loss},
$\lambda_{1}=0.1$, $\lambda_{4}=0.3$, $\lambda_{5}=0.01$. To ensure
that $\Generator$ and $\Filter$ are in good states before adversarial
learning is conducted, we pretrain $\Generator$ and $\Filter$ for
100 epochs each. The pretraining losses for $\Generator$ and $\Filter$
are the terms $\Loss_{\text{gen}}$ in Eq.~\ref{eq:AIF_gen_loss}
and $\Loss_{\text{IF}}$ in Eq.~\ref{eq:AIF_loss}, respectively.
The training data in $\Data_{\text{val}}$ are augmented with random
flipping, random crop (padding size = 5), and random rotation (degree
= 10).

\section{Additional Results of Benchmark Attacks\label{sec:Additional-Results-of-Benchmark-Attacks}}

\begin{figure}[t]
\begin{centering}
\includegraphics[width=0.85\textwidth]{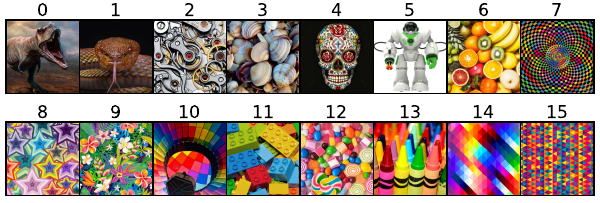}
\par\end{centering}
\caption{A list of Trojan-triggering images that we tried. The images are sorted
by their training Trojan accuracy in ascending order. The last 5 images
(11-15) were selected to be triggers for image-BI+ in our experiments.\label{fig:BI-trigger-images}}
\end{figure}

\begin{figure}[t]
\begin{centering}
\includegraphics[width=0.7\textwidth]{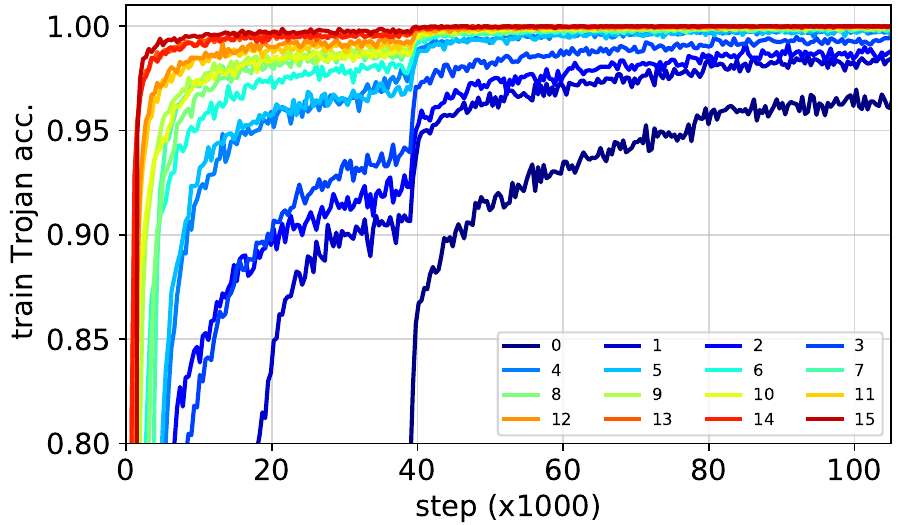}
\par\end{centering}
\caption{Training Trojan accuracy curves of BI on CIFAR10 w.r.t. different
triggers in Fig.~\ref{fig:BI-trigger-images}\label{fig:BI-curves}}
\end{figure}

\subsection{BadNet+ and Noise/Image-BI+\label{subsec:BadNet-and-Noise-Image-BI}}

\paragraph{BadNet+}

\emph{BadNet+} is a variant of BadNet \cite{gu2017badnets} that uses
$M$ different image patches $p_{0},...,p_{M-1}$ ($p_{m}\in\mathbb{I}^{c\times h_{p}\times w_{p}}$,
$0\leq m<M$) as Trojan triggers. Each patch $p_{m}$ is associated
with a 2-tuple $l_{m}=(i,j)$ specifying the location of this patch
in an input image, where $0\leq i<h-h_{p}$ and $0\leq j<w-w_{p}$.
The pixel values and locations of the patches are generated randomly
during construction. If not otherwise specified, we set $M=20$ and
set the patch size $h_{p}\times w_{p}$ to be $5\times5$ for MNIST,
CIFAR10, GTSRB, and $8\times8$ for CelebA.

\paragraph{Blended Injection+}

\emph{Blended Injection+ (BI+)} is similar to BadNet+ except that
it uses \emph{full-size images} $\rho_{0},...,\rho_{M-1}$ ($\rho_{m}\in\mathbb{I}^{c\times h\times w}$,
$0\leq m<M$) as triggers instead of patches. Given a clean image
$x$ and a Trojan-triggering image $\rho_{m}$, the corresponding
Trojan image $\tilde{x}$ is computed as follows:
\[
\tilde{x}=(1-\alpha)\cdot x+\alpha\cdot\rho_{m}
\]
where $\alpha$ is the blending ratio set to 0.1 by default.

$\rho_{0},...,\rho_{M-1}$ can be either random noises or real images,
resulting in two sub-versions of BI+, namely \emph{noise-BI+} and
\emph{image-BI+}. Choosing good Trojan-triggering images for image-BI+
is non-trivial. We tried various real images (Fig.~\ref{fig:BI-trigger-images})
and found that they lead to very different attack success rates (aka
\emph{Trojan accuracies}) (Fig.~\ref{fig:BI-curves}). The best ones
often contain colorful, repetitive patterns (e.g., \emph{``candies''}
or \emph{``crayons''} images). Besides, we also observed that training
image-BI+ with $M\geq10$ is difficult since the classifier usually
needs a lot of time to remember real images. Therefore, we selected
only 5 images with the highest training Trojan accuracies (images
11-15 in Fig.~\ref{fig:BI-trigger-images}) to be used as triggers
for image-BI+ in our experiments.

noise-BI+, by contrast, achieves almost perfect Trojan accuracies
even when $M$ is big ($M\approx100$). We think the main reason behind
this phenomenon is that a random noise image usually have much more
distinct patterns than a real image. Although blending with clean
input images may destroy some patterns in the noise image, many other
patterns are still unaffected and can successfully cause the classifier
to output the target class.

\begin{table}
\begin{centering}
\resizebox{\textwidth}{!}{%
\par\end{centering}
\begin{centering}
\begin{tabular}{cccccccccccccccccc}
\hline 
\multirow{2}{*}{Dataset} &  & Benign &  & \multicolumn{2}{c}{BadNet+} &  & \multicolumn{2}{c}{noise-BI+} &  & \multicolumn{2}{c}{image-BI+} &  & \multicolumn{2}{c}{InpAwAtk} &  & \multicolumn{2}{c}{WaNet}\tabularnewline
 &  & Clean &  & Clean & Trojan &  & Clean & Trojan &  & Clean & Trojan &  & Clean & Trojan &  & Clean & Trojan\tabularnewline
\cline{1-1} \cline{3-3} \cline{5-6} \cline{6-6} \cline{8-9} \cline{9-9} \cline{11-12} \cline{12-12} \cline{14-15} \cline{15-15} \cline{17-18} \cline{18-18} 
MNIST &  & 99.57 &  & 99.47 & 99.97 &  & 99.35 & 100.0 &  & 99.37 & 100.0 &  & 99.33 & 99.30 &  & 99.50 & 99.07\tabularnewline
CIFAR10 &  & 94.71 &  & 94.83 & 100.0 &  & 94.57 & 100.0 &  & 95.13 & 99.90 &  & 94.57 & 99.40 &  & 94.27 & 99.67\tabularnewline
GTSRB &  & 99.63 &  & 99.42 & 100.0 &  & 99.45 & 100.0 &  & 99.34 & 100.0 &  & 98.97 & 99.66 &  & 99.16 & 99.42\tabularnewline
CelebA &  & 78.82 &  & 79.57 & 100.0 &  & 78.32 & 100.0 &  & 78.82 & 100.0 &  & 78.17 & 100.0 &  & 78.18 & 99.97\tabularnewline
\hline 
\end{tabular}}
\par\end{centering}
\caption{Clean and Trojan accuracies of \emph{single-target} attacks on the
defense test set ($\protect\Data_{\text{test}}$) of different datasets.\label{tab:single-target-atk-results-on-Dtest}}
\end{table}

\begin{table}
\begin{centering}
\resizebox{.82\textwidth}{!}{%
\par\end{centering}
\begin{centering}
\begin{tabular}{ccccccccccccc}
\hline 
\multirow{2}{*}{Dataset} &  & \multicolumn{2}{c}{BadNet+} &  & \multicolumn{2}{c}{noise-BI+} &  & \multicolumn{2}{c}{image-BI+} &  & \multicolumn{2}{c}{InpAwAtk}\tabularnewline
 &  & Clean & Trojan &  & Clean & Trojan &  & Clean & Trojan &  & Clean & Trojan\tabularnewline
\cline{1-1} \cline{3-4} \cline{4-4} \cline{6-7} \cline{7-7} \cline{9-10} \cline{10-10} \cline{12-13} \cline{13-13} 
MNIST &  & 99.37 & 98.54 &  & 99.57 & 99.38 &  & 99.53 & 99.25 &  & 99.23 & 97.64\tabularnewline
CIFAR10 &  & 94.63 & 94.30 &  & 94.32 & 93.77 &  & 94.70 & 94.06 &  & 94.53 & 94.10\tabularnewline
GTSRB &  & 99.63 & 99.08 &  & 99.58 & 99.06 &  & 99.37 & 99.08 &  & 99.16 & 99.29\tabularnewline
\hline 
\end{tabular}}
\par\end{centering}
\caption{Clean and Trojan accuracies of different \emph{all-target} attacks
on the defense test set ($\protect\Data_{\text{test}}$) of different
datasets.\label{tab:all-target-atk-results-on-Dtest}}
\end{table}

\subsection{Results of Benchmark Attacks on $\protect\Data_{\text{test}}$}

For completeness, we provide results of the benchmark attacks on $\Data_{\text{test}}$
in Table~\ref{tab:single-target-atk-results-on-Dtest} (for single-target
mode) and in Table~\ref{tab:all-target-atk-results-on-Dtest} (for
all-target mode). The results in Table~\ref{tab:single-target-atk-results-on-Dtest}
are quite similar to the results on $\Data_{\text{val}}\cup\Data_{\text{test}}$
in Table~\ref{tab:main-attack-results}. Since we were not successful
in training the all-target version of WaNet, we exclude this attack
from the results in Table~\ref{tab:all-target-atk-results-on-Dtest}.

\section{Additional Results of Baseline Defenses\label{subsec:Additional-Results-of-Baseline-Defenses}}

\subsection{Network Pruning}

\begin{table}[t]
\begin{centering}
\resizebox{\textwidth}{!}{%
\par\end{centering}
\begin{centering}
\begin{tabular}{ccdabcdabcdabcdabcdab}
\hline 
\multirow{2}{*}{Dataset} &  & \multicolumn{3}{c}{BadNet+} &  & \multicolumn{3}{c}{noise-BI+} &  & \multicolumn{3}{c}{image-BI+} &  & \multicolumn{3}{c}{InpAwAtk} &  & \multicolumn{3}{c}{WaNet}\tabularnewline
 &  & \multicolumn{1}{c}{1\%} & \multicolumn{1}{c}{5\%} & \multicolumn{1}{c}{10\%} &  & \multicolumn{1}{c}{1\%} & \multicolumn{1}{c}{5\%} & \multicolumn{1}{c}{10\%} &  & \multicolumn{1}{c}{1\%} & \multicolumn{1}{c}{5\%} & \multicolumn{1}{c}{10\%} &  & \multicolumn{1}{c}{1\%} & \multicolumn{1}{c}{5\%} & \multicolumn{1}{c}{10\%} &  & \multicolumn{1}{c}{1\%} & \multicolumn{1}{c}{5\%} & \multicolumn{1}{c}{10\%}\tabularnewline
\hhline{-~---~---~---~---~---}MNIST &  & 37.38 & 22.73 & 14.98 &  & 14.21 & 11.00 & 6.75 &  & 14.69 & 6.31 & 4.72 &  & 3.32 & 3.32 & 3.32 &  & 1.18 & 1.18 & 1.18\tabularnewline
CIFAR10 &  & 100.0 & 100.0 & 100.0 &  & 99.33 & 87.26 & 87.26 &  & 99.74 & 99.74 & 99.74 &  & 56.52 & 56.52 & 56.52 &  & 96.85 & 96.59 & 96.59\tabularnewline
GTSRB &  & 100.0 & 100.0 & 100.0 &  & 100.0 & 100.0 & 100.0 &  & 99.87 & 99.87 & 99.87 &  & 18.38 & 18.38 & 18.38 &  & 98.86 & 98.86 & 98.86\tabularnewline
CelebA &  & 84.08 & 63.35 & 50.64 &  & 100.0 & 99.97 & 99.97 &  & 99.87 & 99.87 & 99.80 &  & 100.0 & 100.0 & 99.97 &  & 98.22 & 88.48 & 53.10\tabularnewline
\hline 
\end{tabular}}
\par\end{centering}
\caption{Trojan accuracies of $\protect\Classifier$ pruned by Network Pruning
at 1\%, 5\%, and 10\% decrease in clean accuracy. \emph{Smaller values
are better}. Results are computed on $\protect\Data'_{\text{test}}$.\label{tab:Network-Pruning-results}}
\end{table}

\begin{figure}[t]
\begin{centering}
\subfloat[MNIST]{\begin{centering}
\includegraphics[width=0.232\textwidth]{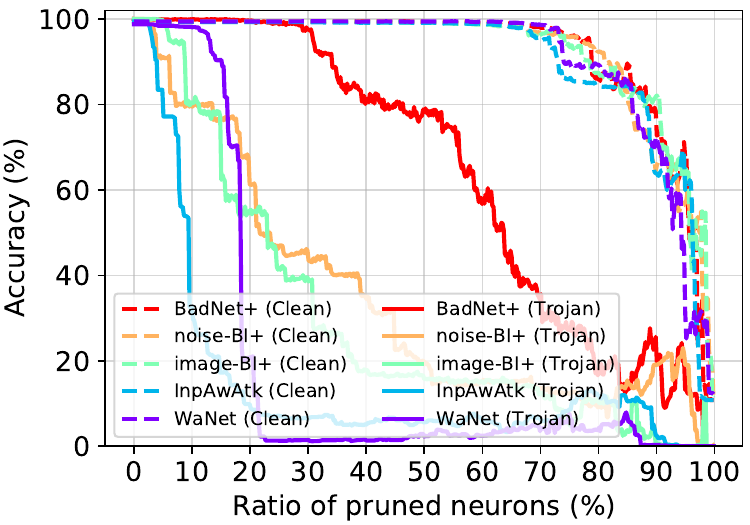}
\par\end{centering}
}\subfloat[CIFAR10]{\begin{centering}
\includegraphics[width=0.232\textwidth]{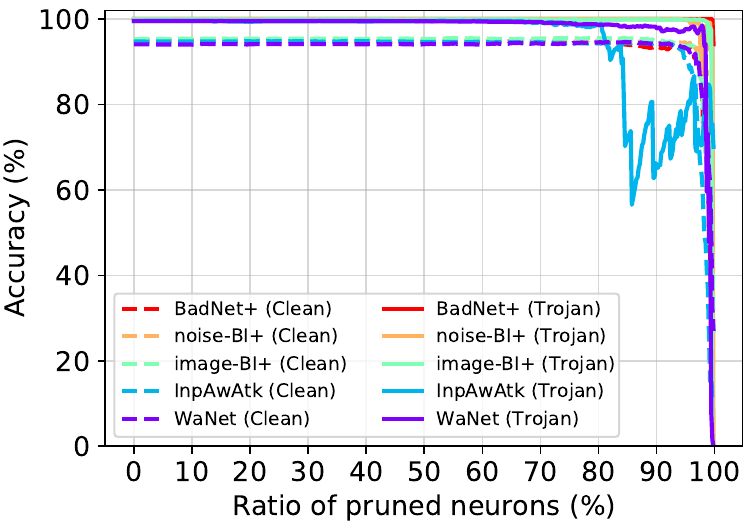}
\par\end{centering}
}\subfloat[GTSRB]{\begin{centering}
\includegraphics[width=0.232\textwidth]{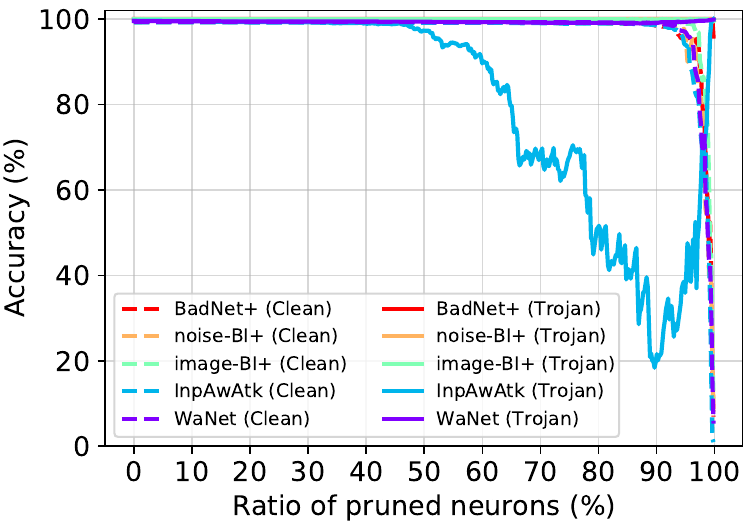}
\par\end{centering}
}\subfloat[CelebA]{\begin{centering}
\includegraphics[width=0.232\textwidth]{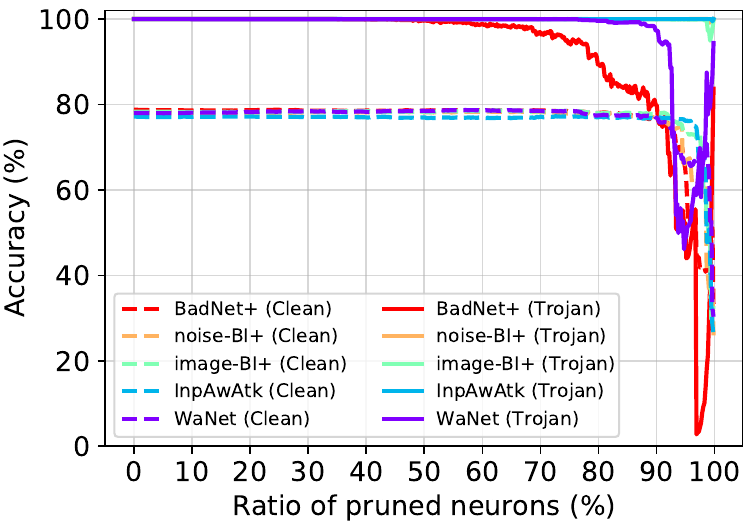}
\par\end{centering}
}
\par\end{centering}
\caption{Clean accuracy (dashed) and Trojan accuracy (solid) curves of $\protect\Classifier$
pruned by Network Pruning for different attacks and datasets which
correspond to the results in Table~\ref{tab:Network-Pruning-results}.\label{fig:Network-Pruning-curves}}
\end{figure}

We provide the Trojan accuracies of Network Pruning (NP) \cite{liu2018fine}
at 1\%, 5\%, and 10\% decrease in clean accuracy in Table~\ref{tab:Network-Pruning-results}
and the corresponding pruning curves in Fig.~\ref{fig:Network-Pruning-curves}.
It is clear that NP is a very ineffective Trojan mitigation method
since the classifier pruned by NP still achieves nearly 100\% Trojan
accuracies on CIFAR10, GTSRB, and CelebA even when experiencing about
10\% decrease in clean accuracy.

\subsection{STRIP}

\begin{figure}[t]
\begin{centering}
\includegraphics[width=0.7\textwidth]{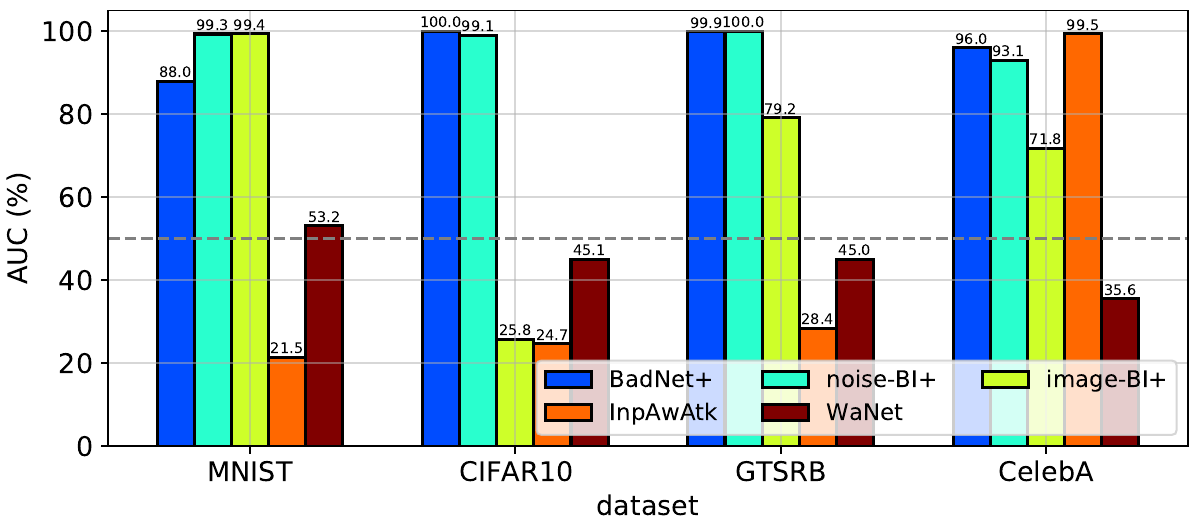}
\par\end{centering}
\caption{AUCs of STRIP against different attacks on different datasets. \emph{Larger
values are better}.\label{fig:AUCs-of-STRIP}}
\end{figure}

\begin{table}[t]
\begin{centering}
\resizebox{\textwidth}{!}{%
\par\end{centering}
\begin{centering}
\begin{tabular}{ccdabcdabcdabcdabcdab}
\hline 
\multirow{2}{*}{Dataset} &  & \multicolumn{3}{c}{BadNet+} &  & \multicolumn{3}{c}{noise-BI+} &  & \multicolumn{3}{c}{image-BI+} &  & \multicolumn{3}{c}{InpAwAtk} &  & \multicolumn{3}{c}{WaNet}\tabularnewline
 &  & \multicolumn{1}{c}{1\%} & \multicolumn{1}{c}{5\%} & \multicolumn{1}{c}{10\%} &  & \multicolumn{1}{c}{1\%} & \multicolumn{1}{c}{5\%} & \multicolumn{1}{c}{10\%} &  & \multicolumn{1}{c}{1\%} & \multicolumn{1}{c}{5\%} & \multicolumn{1}{c}{10\%} &  & \multicolumn{1}{c}{1\%} & \multicolumn{1}{c}{5\%} & \multicolumn{1}{c}{10\%} &  & \multicolumn{1}{c}{1\%} & \multicolumn{1}{c}{5\%} & \multicolumn{1}{c}{10\%}\tabularnewline
\hhline{-~---~---~---~---~---}MNIST &  & 88.45 & 64.40 & 36.25 &  & 13.95 & 0.00 & 0.00 &  & 16.40 & 0.05 & 0.00 &  & 99.95 & 99.80 & 99.15 &  & 99.85 & 97.55 & 92.40\tabularnewline
CIFAR10 &  & 0.00 & 0.00 & 0.00 &  & 10.35 & 2.35 & 0.95 &  & 99.45 & 97.30 & 95.75 &  & 99.55 & 97.90 & 96.20 &  & 100.0 & 99.25 & 96.90\tabularnewline
GTSRB &  & 1.05 & 0.20 & 0.00 &  & 0.00 & 0.00 & 0.00 &  & 79.20 & 58.30 & 44.75 &  & 99.85 & 98.80 & 97.15 &  & 99.50 & 96.00 & 91.65\tabularnewline
CelebA &  & 17.50 & 11.20 & 7.70 &  & 33.80 & 19.80 & 14.70 &  & 75.75 & 62.35 & 54.45 &  & 1.90 & 1.20 & 1.00 &  & 99.80 & 98.55 & 96.20\tabularnewline
\hline 
\end{tabular}}
\par\end{centering}
\caption{False negative rates (FNRs) of STRIP at 1\%, 5\%, and 10\% false positive
rate (FPR) for different attacks and datasets. \emph{Smaller values
are better}.\label{tab:FNRs-of-STRIP}}
\end{table}

\begin{figure}[t]
\begin{centering}
\resizebox{\textwidth}{!}{%
\par\end{centering}
\begin{centering}
\begin{tabular}{>{\centering}m{0.08\textwidth}ccccc}
 & BadNet+ & noise-BI+ & image-BI+ & InpAwAtk & WaNet\tabularnewline
MNIST & \includegraphics[width=0.3\textwidth,valign=m]{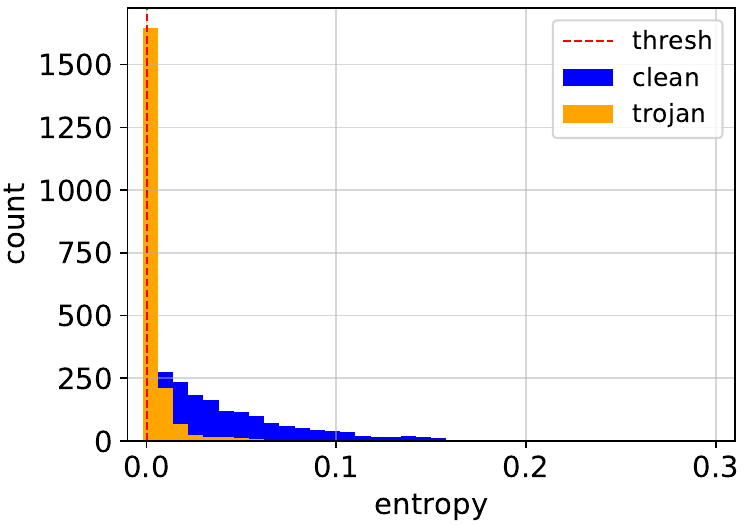} & \includegraphics[width=0.3\textwidth,valign=m]{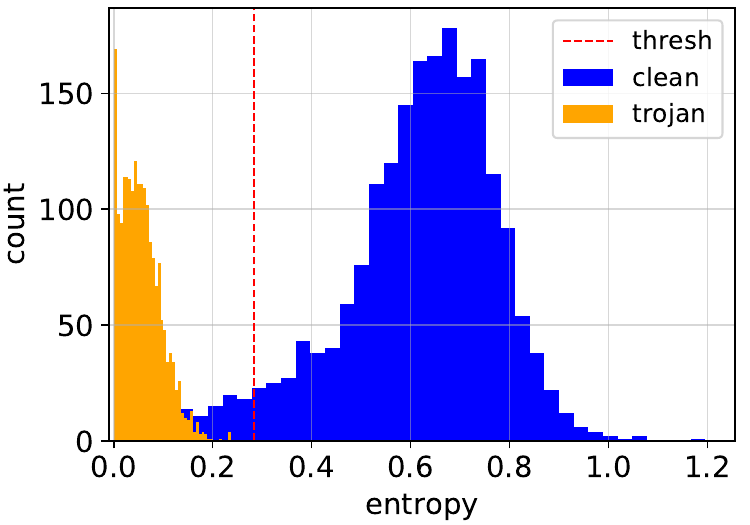} & \includegraphics[width=0.3\textwidth,valign=m]{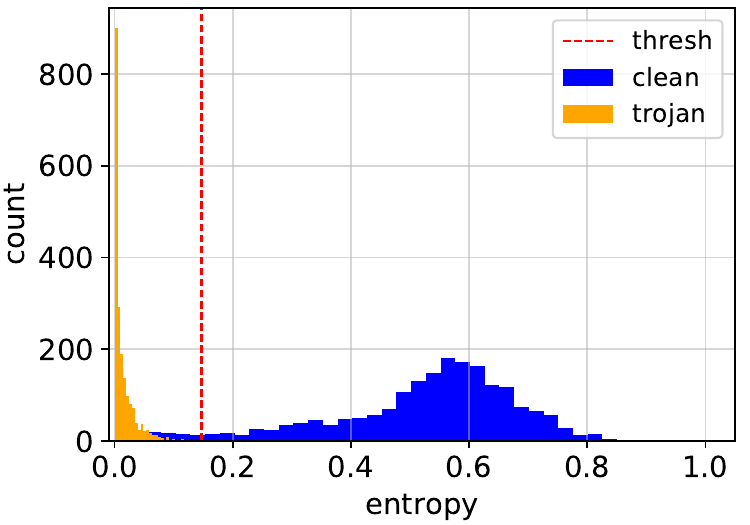} & \includegraphics[width=0.3\textwidth,valign=m]{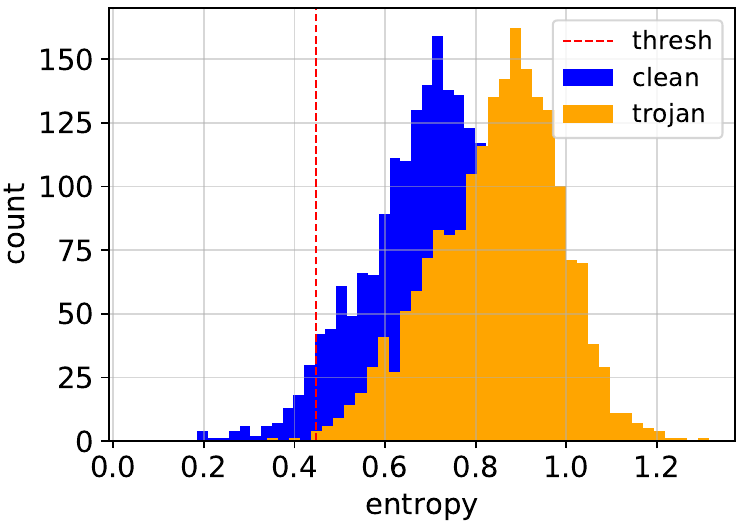} & \includegraphics[width=0.3\textwidth,valign=m]{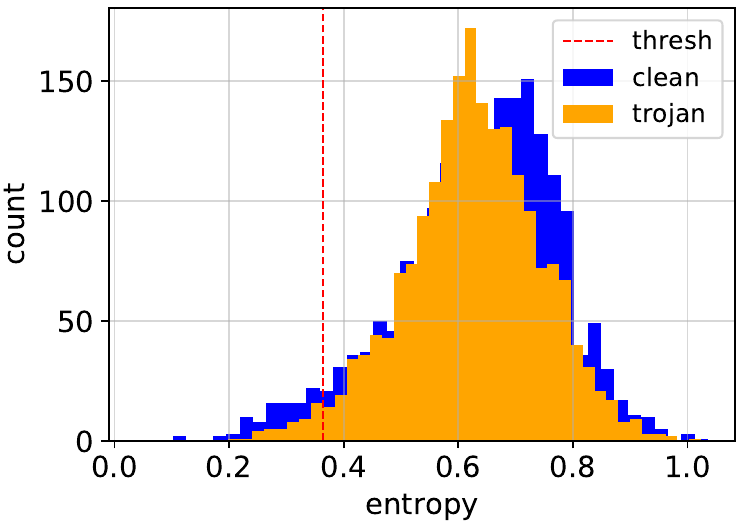}\tabularnewline
CIFAR10 & \includegraphics[width=0.3\textwidth,valign=m]{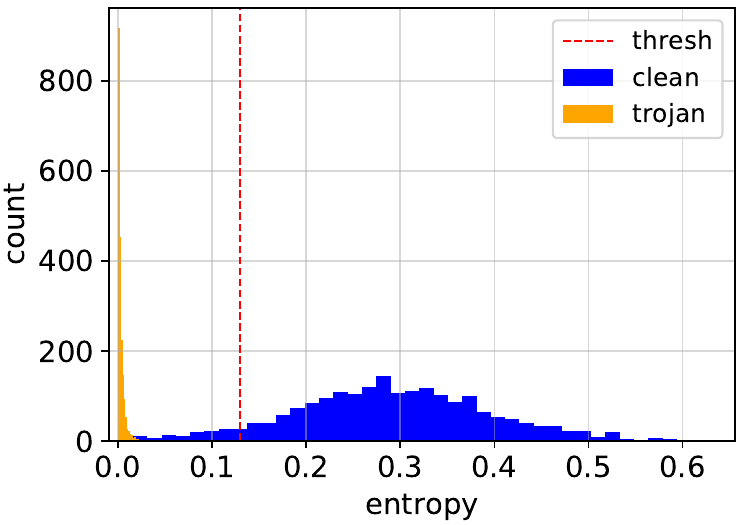} & \includegraphics[width=0.3\textwidth,valign=m]{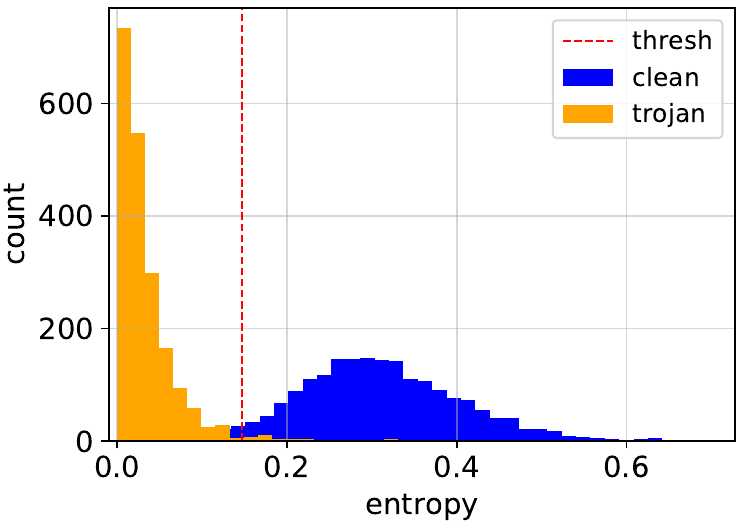} & \includegraphics[width=0.3\textwidth,valign=m]{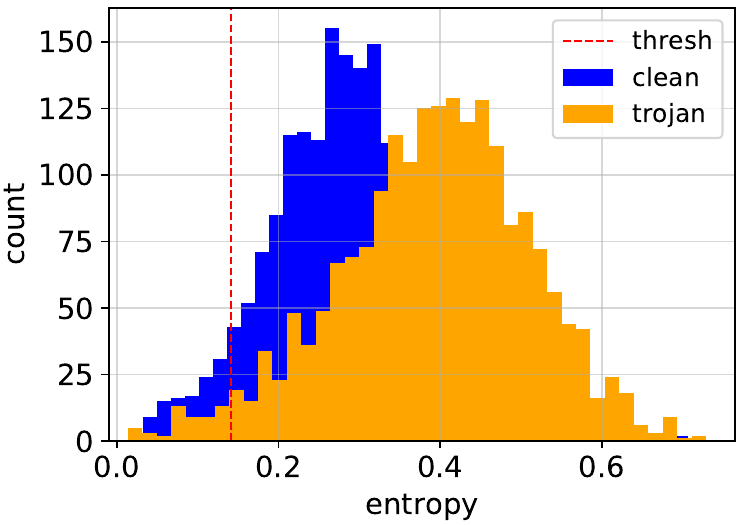} & \includegraphics[width=0.3\textwidth,valign=m]{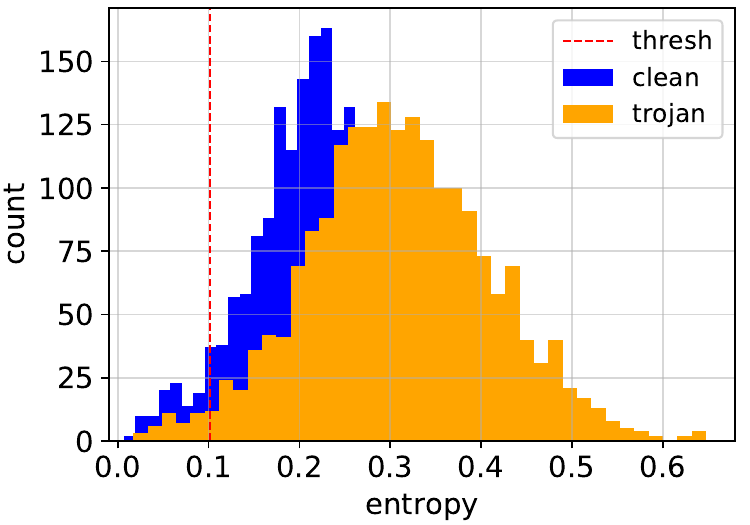} & \includegraphics[width=0.3\textwidth,valign=m]{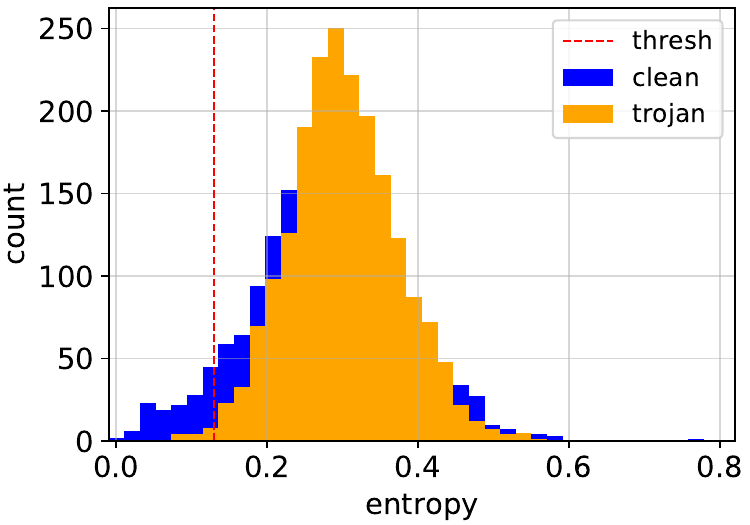}\tabularnewline
GTSRB & \includegraphics[width=0.3\textwidth,valign=m]{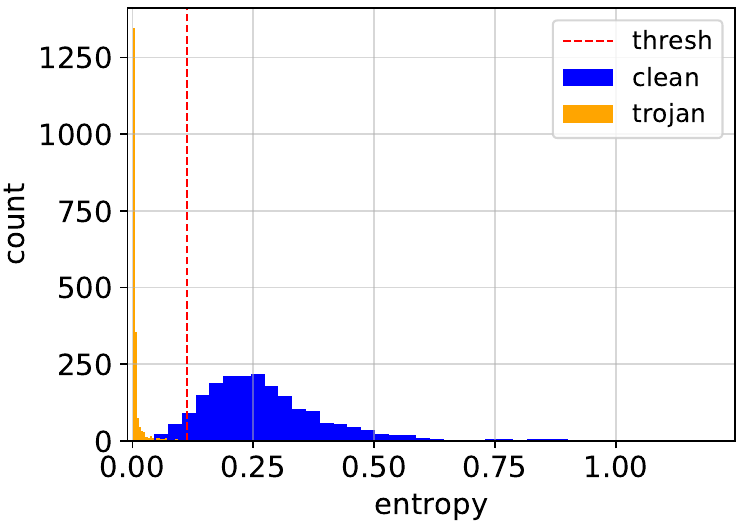} & \includegraphics[width=0.3\textwidth,valign=m]{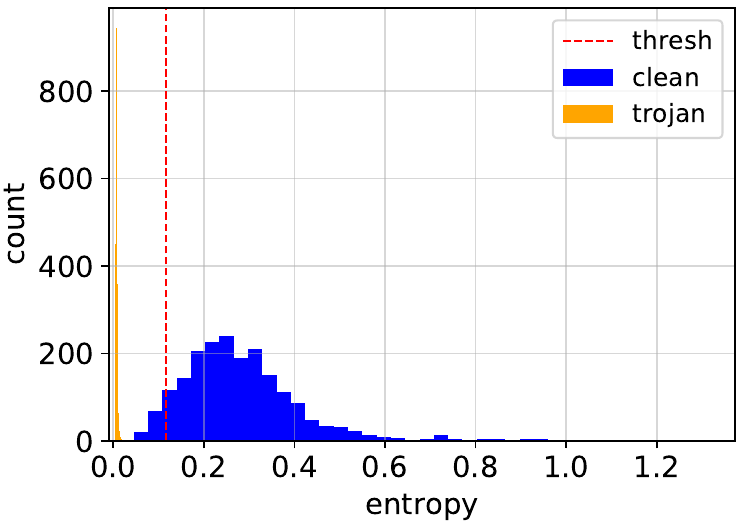} & \includegraphics[width=0.3\textwidth,valign=m]{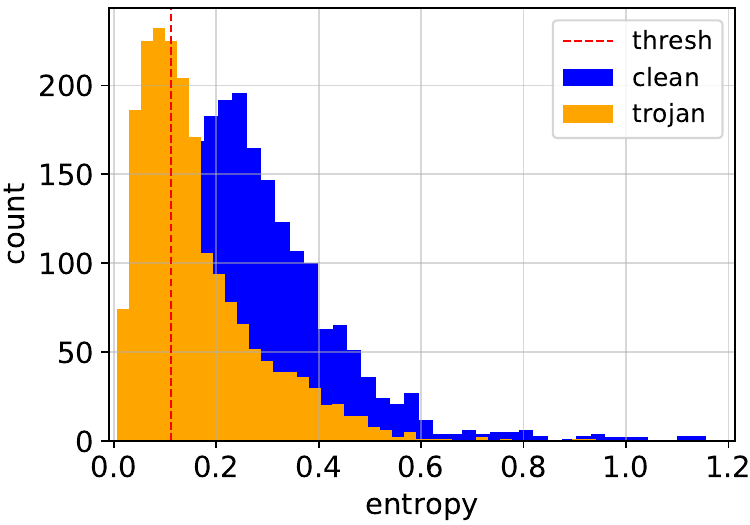} & \includegraphics[width=0.3\textwidth,valign=m]{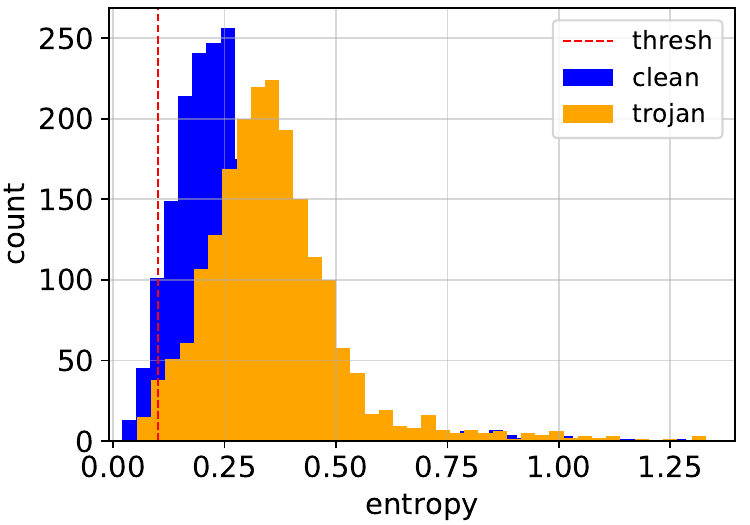} & \includegraphics[width=0.3\textwidth,valign=m]{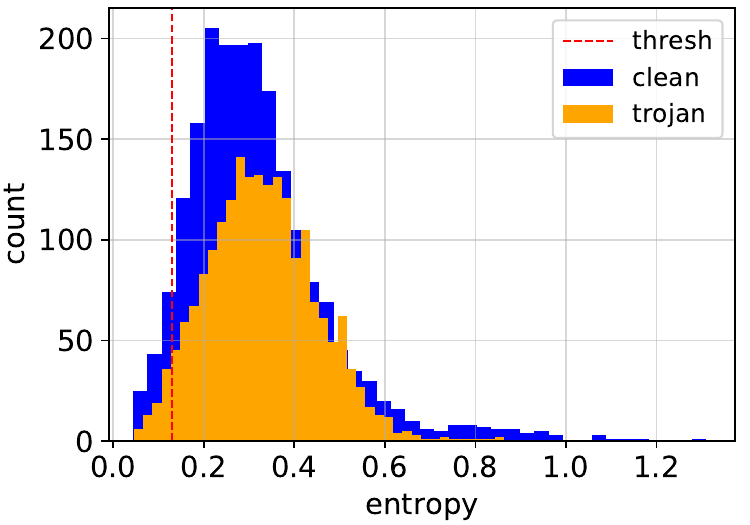}\tabularnewline
CelebA & \includegraphics[width=0.3\textwidth,valign=m]{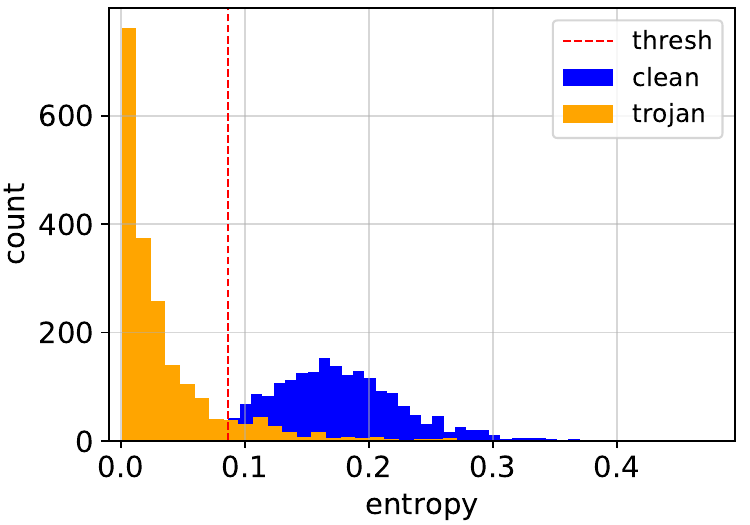} & \includegraphics[width=0.3\textwidth,valign=m]{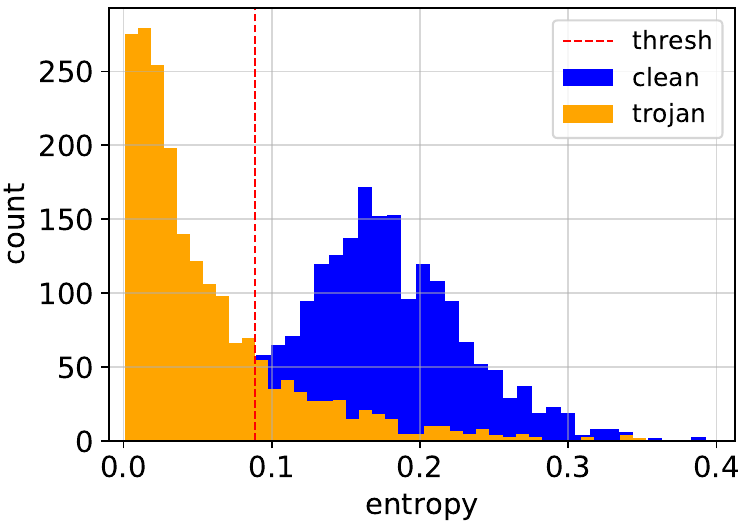} & \includegraphics[width=0.3\textwidth,valign=m]{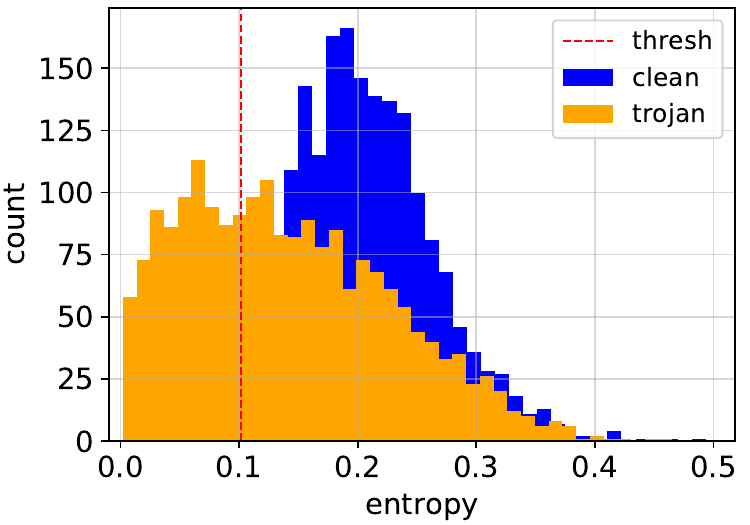} & \includegraphics[width=0.3\textwidth,valign=m]{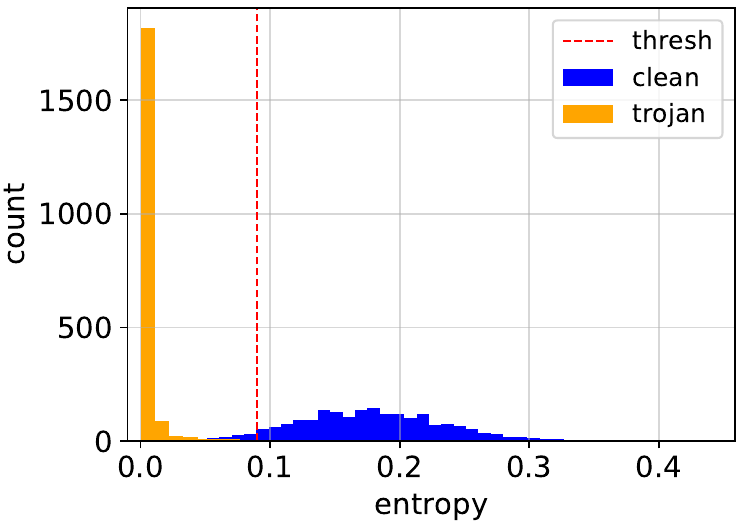} & \includegraphics[width=0.3\textwidth,valign=m]{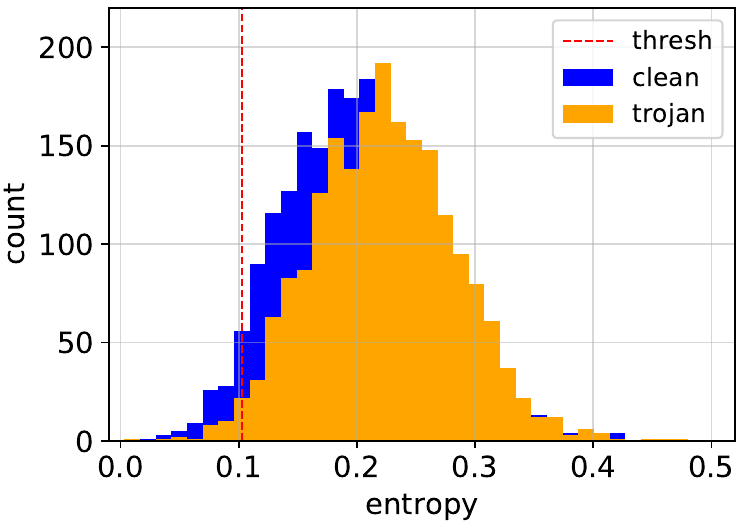}\tabularnewline
\end{tabular}}
\par\end{centering}
\caption{Histograms of entropies computed by STRIP for different attacks and
datasets. The vertical red dashed line in each plot indicates the
threshold at 5\% false positive rate.\label{fig:Hist-of-STRIP}}
\end{figure}

In Table~\ref{tab:FNRs-of-STRIP}, we report the false negative rates
(FNRs) of STRIP at 1\%, 5\%, and 10\% false positive rate (FPR). We
also provide the AUCs and the entropy histograms of STRIP in Figs.~\ref{fig:AUCs-of-STRIP}
and \ref{fig:Hist-of-STRIP}, respectively. Note that AUCs are only
suitable for experimental purpose not practical use since in real-world
scenarios, we still have to compute thresholds based on FPRs on clean
data. STRIP achieves very high FNRs on MNIST, CIFAR10, and GTSRB when
defending against InpAwAtk and WaNet (Table~\ref{tab:FNRs-of-STRIP})
which corresponds to low AUCs (Fig.~\ref{fig:AUCs-of-STRIP}).

\subsection{Neural Cleanse\label{subsec:Neural-Cleanse-supp}}

\begin{table}[t]
\begin{centering}
\resizebox{\textwidth}{!}{%
\par\end{centering}
\begin{centering}
\begin{tabular}{ccdabcdabcdabcdabcdab}
\hline 
\multirow{2}{*}{Dataset} &  & \multicolumn{3}{c}{BadNet+} &  & \multicolumn{3}{c}{noise-BI+} &  & \multicolumn{3}{c}{image-BI+} &  & \multicolumn{3}{c}{InpAwAtk} &  & \multicolumn{3}{c}{WaNet}\tabularnewline
 &  & \multicolumn{1}{c}{1\%} & \multicolumn{1}{c}{5\%} & \multicolumn{1}{c}{10\%} &  & \multicolumn{1}{c}{1\%} & \multicolumn{1}{c}{5\%} & \multicolumn{1}{c}{10\%} &  & \multicolumn{1}{c}{1\%} & \multicolumn{1}{c}{5\%} & \multicolumn{1}{c}{10\%} &  & \multicolumn{1}{c}{1\%} & \multicolumn{1}{c}{5\%} & \multicolumn{1}{c}{10\%} &  & \multicolumn{1}{c}{1\%} & \multicolumn{1}{c}{5\%} & \multicolumn{1}{c}{10\%}\tabularnewline
\hhline{-~---~---~---~---~---}MNIST &  & 0.00 & 0.00 & 0.00 &  & 0.00 & 0.00 & 0.00 &  & 0.00 & 0.00 & 0.00 &  & 0.00 & 0.00 & 0.00 &  & - & - & -\tabularnewline
CIFAR10 &  & 0.00 & 0.00 & 0.00 &  & 7.37 & 1.11 & 0.70 &  & 0.00 & 0.00 & 0.00 &  & - & - & - &  & - & - & -\tabularnewline
GTSRB &  & 61.94 & 48.23 & 48.23 &  & 78.71 & 1.03 & 1.03 &  & 51.74 & 51.74 & 51.74 &  & 49.02 & 38.33 & 38.33 &  & 2.75 & 1.58 & 1.58\tabularnewline
CelebA &  & 75.97 & 26.73 & 7.63 &  & 99.95 & 98.75 & 95.30 &  & 99.75 & 96.21 & 83.62 &  & 99.87 & 98.42 & 93.31 &  & - & - & -\tabularnewline
\hline 
\end{tabular}}
\par\end{centering}
\caption{Trojan accuracies of the Trojan classifier $\protect\Classifier$
pruned by Neural Cleanse Pruning at 1\%, 5\%, and 10\% decrease in
clean accuracy for different attacks and datasets. \emph{Smaller values
are better}. Some results for InpAwAtk and WaNet are not available
because Neural Cleanse fails to classify $\protect\Classifier$ as
Trojan-infected in these cases (Fig.~\ref{fig:Neural-Cleanse-result-bars}).\label{tab:NeuralCleansePruning-results}}
\end{table}

\begin{figure}[t]
\begin{centering}
\subfloat[MNIST]{\begin{centering}
\includegraphics[width=0.232\textwidth]{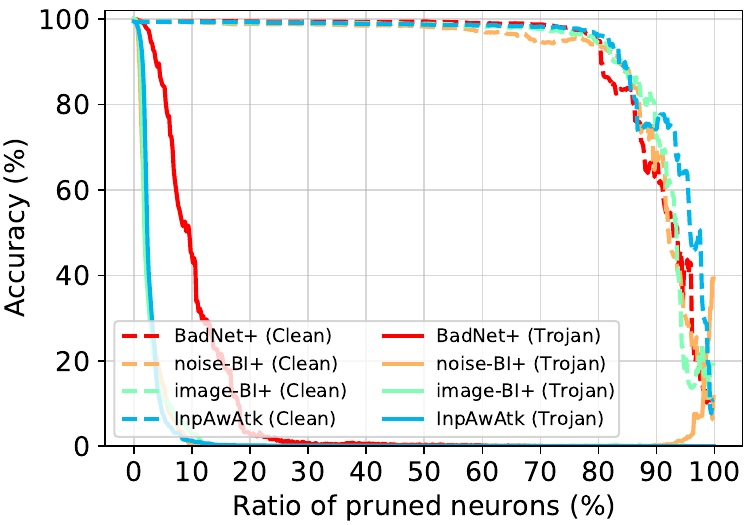}
\par\end{centering}
}\subfloat[CIFAR10]{\begin{centering}
\includegraphics[width=0.232\textwidth]{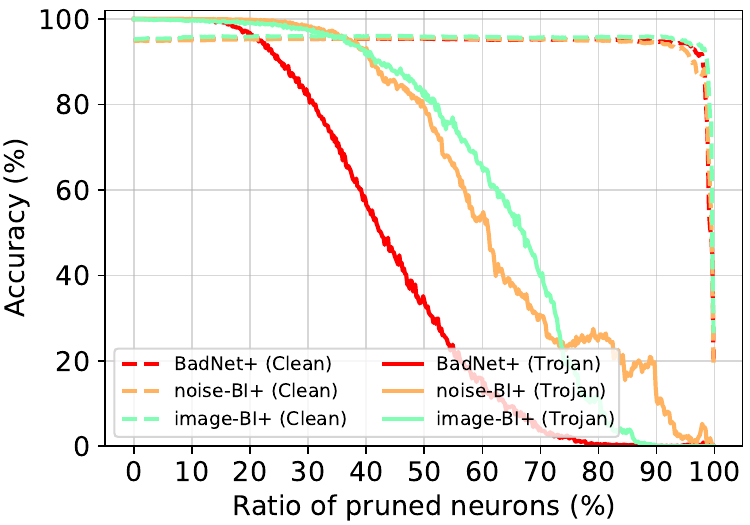}
\par\end{centering}
}\subfloat[GTSRB]{\begin{centering}
\includegraphics[width=0.232\textwidth]{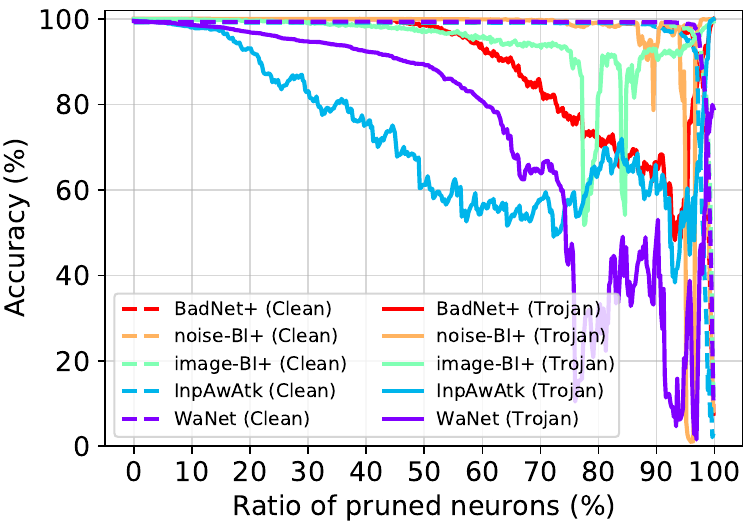}
\par\end{centering}
}\subfloat[CelebA]{\begin{centering}
\includegraphics[width=0.232\textwidth]{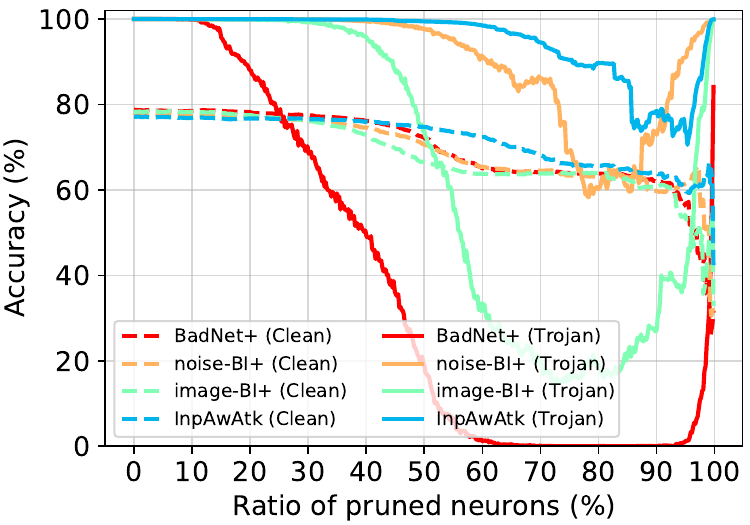}
\par\end{centering}
}
\par\end{centering}
\caption{Clean accuracy (dashed) and Trojan accuracy (solid) curves of the
Trojan classifier $\protect\Classifier$ pruned by Neural Cleanse
Pruning for different attacks and datasets which correspond to the
results in Table~\ref{tab:NeuralCleansePruning-results}.\label{fig:NeuralCleansePruning-curves}}
\end{figure}

\begin{table}[t]
\begin{centering}
\resizebox{\textwidth}{!}{%
\par\end{centering}
\begin{centering}
\begin{tabular}{ccdabcdabcdabcdabcdab}
\hline 
\multirow{2}{*}{Dataset} &  & \multicolumn{3}{c}{BadNet+} &  & \multicolumn{3}{c}{noise-BI+} &  & \multicolumn{3}{c}{image-BI+} &  & \multicolumn{3}{c}{InpAwAtk} &  & \multicolumn{3}{c}{WaNet}\tabularnewline
 &  & \multicolumn{1}{c}{1\%} & \multicolumn{1}{c}{5\%} & \multicolumn{1}{c}{10\%} &  & \multicolumn{1}{c}{1\%} & \multicolumn{1}{c}{5\%} & \multicolumn{1}{c}{10\%} &  & \multicolumn{1}{c}{1\%} & \multicolumn{1}{c}{5\%} & \multicolumn{1}{c}{10\%} &  & \multicolumn{1}{c}{1\%} & \multicolumn{1}{c}{5\%} & \multicolumn{1}{c}{10\%} &  & \multicolumn{1}{c}{1\%} & \multicolumn{1}{c}{5\%} & \multicolumn{1}{c}{10\%}\tabularnewline
\hhline{-~---~---~---~---~---}MNIST &  & 9.75 & 4.75 & 3.25 &  & 0.00 & 0.00 & 0.00 &  & 0.00 & 0.00 & 0.00 &  & 0.80 & 0.30 & 0.15 &  & - & - & -\tabularnewline
CIFAR10 &  & 99.50 & 78.90 & 0.65 &  & 90.55 & 39.30 & 0.05 &  & 95.50 & 56.90 & 0.40 &  & - & - & - &  & - & - & -\tabularnewline
GTSRB &  & 100.0 & 0.40 & 0.00 &  & 0.00 & 0.00 & 0.00 &  & 0.03 & 0.00 & 0.00 &  & 99.48 & 92.17 & 86.65 &  & 3.02 & 0.40 & 0.32\tabularnewline
CelebA &  & 2.25 & 0.00 & 0.00 &  & 17.70 & 0.65 & 0.00 &  & 50.20 & 11.60 & 2.35 &  & 88.25 & 42.15 & 13.60 &  & - & - & -\tabularnewline
\hline 
\end{tabular}}
\par\end{centering}
\caption{False negative rates (FNRs) computed by Neural Cleanse Input Checking
at 1\%, 5\%, and 10\% false positive rate (FPR) for different attacks
and datasets. \emph{Smaller values are better}.\label{tab:NeuralCleanseFiltering-results}}
\end{table}

After classifying $\Classifier$ as Trojan-infected, Neural Cleanse
(NC) mitigates Trojans via pruning $\Classifier$ or via input checking.
We refer to these two methods as \emph{Neural Cleanse Pruning (NCP)}
and \emph{Neural Cleanse Input Checking (NCIC)}. Both methods build
a set of synthetic Trojan images by blending all clean images in $\Data_{\text{val}}$
with the synthetic trigger corresponding to the detected target class.
NCP ranks neurons in the second last layer of $\Classifier$ according
to their average activation gaps computed on the synthetic Trojan
images and the corresponding clean images in $\Data_{\text{val}}$
in descending order. It gradually prunes the neurons with the highest
ranks first until certain decrease in clean accuracy is met. NCIC,
on the other hand, picks the top 1\% of the neurons in the second
last layer of $\Classifier$ with largest average activations on the
synthetic

Trojan images to form a characteristic group of Trojan neurons. Given
an input image $x$, NCIC considers the mean activations of the neurons
in the group w.r.t. $x$ as a score for detecting whether $x$ contains
Trojan triggers or not. If the score is greater than a threshold,
$x$ is considered as a Trojan image, otherwise, a clean image. The
threshold is chosen based on the scores of all clean images in $\Data_{\text{val}}$.
We provide the results of NCP in Table~\ref{tab:NeuralCleansePruning-results},
Fig.~\ref{fig:NeuralCleansePruning-curves} and the results of NCIC
in Table~\ref{tab:NeuralCleanseFiltering-results}. At 5\% decrease
in clean accuracy, NCP reduces the Trojan accuracies of all the attacks
except WaNet to almost 0\% on MNIST and CIFAR10. However, NCP is ineffective
against these attacks especially image-BI+ and InpAwAtk on GTSRB and
CelebA. At 10\% FPR, NCIC achieves nearly perfect FNRs against BadNet+,
noise-BI+, and image-BI+ on all datasets but is also ineffective against
InpAwAtk on GTSRB and CelebA. Note that both NCP and NCIC have almost
no effect against WaNet on MNIST, CIFAR10, and CelebA since NC misclassifies
the Trojan classifiers w.r.t. this attack as benign.

\subsection{Februus\label{subsec:Additional-Results-Februus}}

\begin{figure}[t]
\begin{centering}
\includegraphics[width=0.7\textwidth]{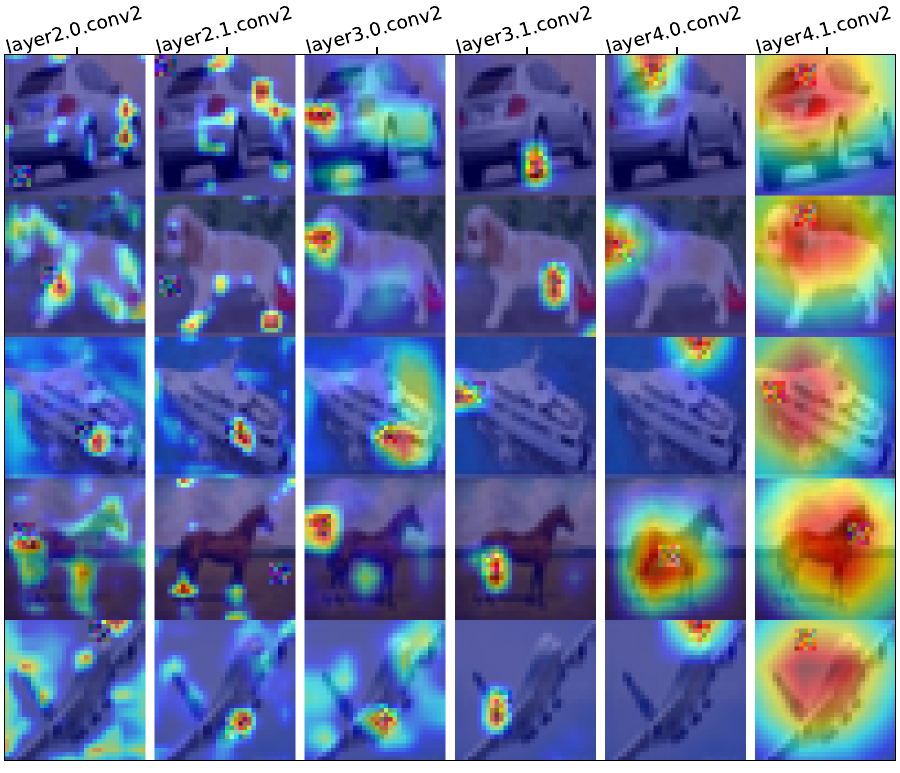}
\par\end{centering}
\caption{Examples of heatmaps computed by Februus's GradCAM at different layers
of $\protect\Classifier$ ordered from bottom (layer2.0.conv2) to
top (layer4.1.conv2). The dataset is CIFAR10 and the classifier is
a PreactResNet18 \cite{he2016identity}. The Trojan attack is BadNet+.
It is clear that only some layers give reasonably good results.\label{fig:Februus-heatmap}}
\end{figure}

\begin{figure}[t]
\begin{centering}
\subfloat[Decrease in clean accuracy ($\downarrow$C)]{\begin{centering}
\includegraphics[width=0.96\textwidth]{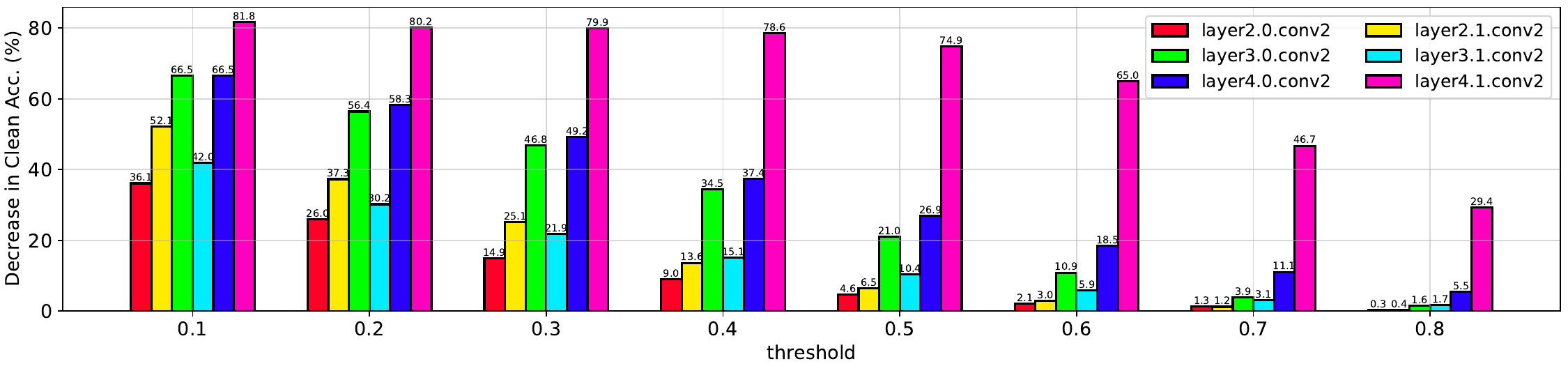}
\par\end{centering}
}
\par\end{centering}
\begin{centering}
\subfloat[Trojan accuracy (T)]{\begin{centering}
\includegraphics[width=0.96\textwidth]{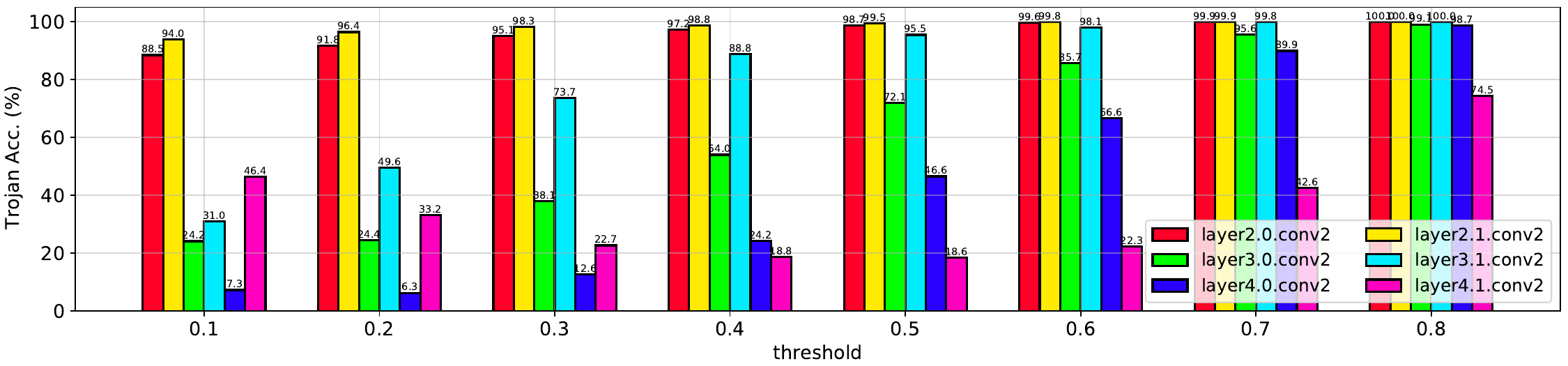}
\par\end{centering}
}
\par\end{centering}
\begin{centering}
\subfloat[Decrease in recovery accuracy ($\downarrow$R)]{\begin{centering}
\includegraphics[width=0.96\textwidth]{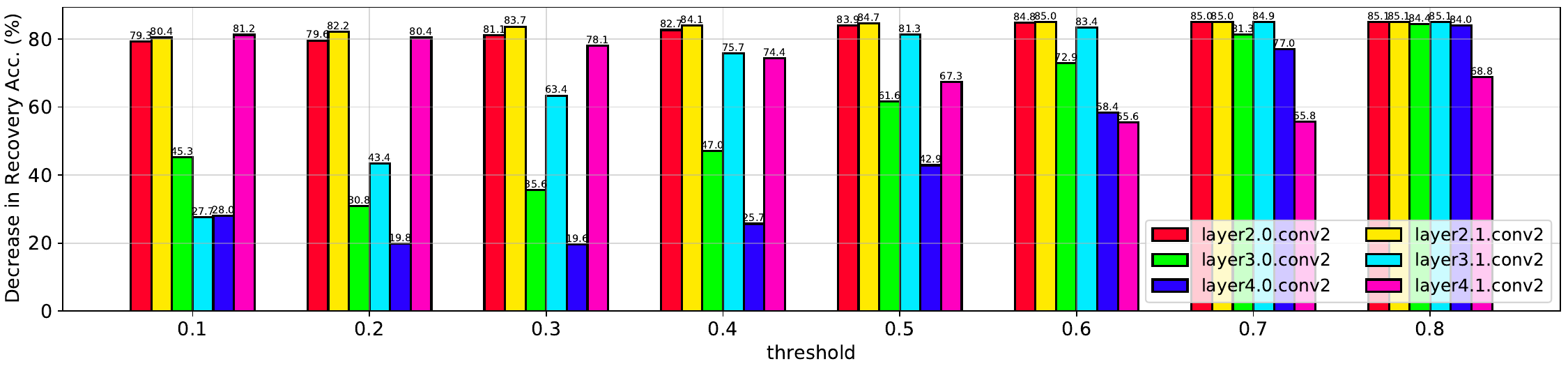}
\par\end{centering}
}
\par\end{centering}
\caption{Decreases in clean accuracy (a), Trojan accuracies (b), and decreases
in recovery accuracy (c) of Februus when mitigating BadNet+'s Trojans
on CIFAR10 w.r.t. different heatmap layers and binary thresholds.\label{fig:Results-Feb-layers-threshs}}
\end{figure}

We reimplement Februus based on the official code provided by the
authors\footnote{Februus: \url{https://github.com/AdelaideAuto-IDLab/Februus}}.
Since there is no script for training the inpainting GAN in the authors'
code, we use the ``inpaint'' function from OpenCV instead. Februus
has 2 main hyperparameters that need to be tuned which are: i) the
convolutional layer of $\Classifier$ at which GradCAM computes heatmaps
(\emph{``heatmap layer''} for short) and ii) the threshold for converting
GradCAM heatmaps into binary masks (\emph{``binary threshold''}
for short). As shown in Fig.~\ref{fig:Results-Feb-layers-threshs},
the performance of Februus greatly depends on these hyperparameters.
Increasing the binary threshold means smaller areas are masked and
inpainted, which usually leads to smaller decreases in clean accuracy
(smaller $\downarrow$Cs) yet higher Trojan accuracies (higher Ts).
Meanwhile, choosing top layers of $\Classifier$ to compute heatmap
(e.g., layer4) usually causes bigger $\downarrow$Cs yet lower Ts
since the selected regions are often broader (Fig.~\ref{fig:Februus-heatmap}).
For simplicity, we choose the (layer, threshold) setting that gives
the smallest decrease in recovery accuracy ($\downarrow$R) of Februus
when defending against BadNet+ and apply this setting to all other
attacks.

\subsection{Neural Attention Distillation}

We reimplement Neural Attention Distillation (NAD) based on the official
code provided by the authors\footnote{NAD: \url{https://github.com/bboylyg/NAD}}.
We finetune the original Trojan classifier $\Classifier$ to obtain
the teacher $\mathtt{T}$ in 10 epochs. After that, we distill knowledge
from $\mathtt{T}$ to $\Classifier$ in 20 more epochs. In both cases,
the batch size is 64 and the optimizer is Adam with an learning rate
of 0.0001 divided by 10 after 10 epochs. We note that in the original
paper, the authors reported that they used an initial learning rate
of 0.1 and divided it by 10 after every 2 epochs during knowledge
distillation. However, in our experiment, we found that such an initial
learning rate is too large for distillation and can cause a significant
drop in the clean accuracy of $\Classifier$ which is hard to be recovered
even if the learning rate is decayed later.

\section{Additional Results and Ablation Studies of Our Defenses\label{sec:Additional-Results-of-Our-Defenses}}

\begin{table}[t]
\begin{centering}
\resizebox{.96\textwidth}{!}{%
\par\end{centering}
\begin{centering}
\begin{tabular}{|c|c|dab|dab|dab|dab|}
\hline 
\multirow{2}{*}{Dataset} & \multirow{2}{*}{Defense} & \multicolumn{3}{c|}{BadNet+} & \multicolumn{3}{c|}{noise-BI+} & \multicolumn{3}{c|}{image-BI+} & \multicolumn{3}{c|}{InpAwAtk}\tabularnewline
\cline{3-14} \cline{4-14} \cline{5-14} \cline{6-14} \cline{7-14} \cline{8-14} \cline{9-14} \cline{10-14} \cline{11-14} \cline{12-14} \cline{13-14} \cline{14-14} 
 &  & \multicolumn{1}{c}{$\downarrow$C} & \multicolumn{1}{c}{T} & \multicolumn{1}{c|}{$\downarrow$R} & \multicolumn{1}{c}{$\downarrow$C} & \multicolumn{1}{c}{T} & \multicolumn{1}{c|}{$\downarrow$R} & \multicolumn{1}{c}{$\downarrow$C} & \multicolumn{1}{c}{T} & \multicolumn{1}{c|}{$\downarrow$R} & \multicolumn{1}{c}{$\downarrow$C} & \multicolumn{1}{c}{T} & \multicolumn{1}{c|}{$\downarrow$R}\tabularnewline
\hline 
\hline 
 & IF & 0.00 & 75.90 & 76.30 & -0.03 & 99.30 & 99.47 & 0.13 & 6.96 & 6.96 & 0.00 & 95.74 & 96.17\tabularnewline
MNIST & VIF & \textbf{-0.07} & 49.67 & 50.17 & \textbf{-0.07} & 4.06 & 4.23 & 0.07 & 0.20 & 0.23 & \textbf{-0.23} & 63.58 & 64.21\tabularnewline
 & AIF & 0.47 & 19.31 & \textbf{20.27} & 0.13 & \textbf{0.03} & \textbf{0.10} & \textbf{0.03} & \textbf{0.03} & \textbf{0.13} & -0.13 & \textbf{23.90} & \textbf{24.53}\tabularnewline
\hline 
\hline 
 & IF & \textbf{3.93} & 1.57 & \textbf{8.03} & \textbf{2.63} & 1.17 & \textbf{3.30} & \textbf{3.60} & 18.90 & 22.20 & \textbf{3.83} & 9.97 & 14.63\tabularnewline
CIFAR10 & VIF & 8.13 & 1.87 & 12.73 & 6.27 & 1.33 & 6.72 & 6.83 & 5.67 & 11.90 & 7.96 & \textbf{5.95} & 16.87\tabularnewline
 & AIF & 5.67 & \textbf{1.23} & 9.13 & 4.40 & \textbf{0.97} & 5.47 & 5.13 & \textbf{5.10} & \textbf{9.63} & 5.43 & 6.73 & \textbf{14.33}\tabularnewline
\hline 
\hline 
 & IF & 0.29 & \textbf{0.11} & \textbf{1.58} & 0.13 & \textbf{0.21} & \textbf{1.18} & \textbf{-0.26} & 61.57 & 62.25 & 0.11 & 3.97 & 4.50\tabularnewline
GTSRB & VIF & 0.47 & 0.32 & 3.36 & 0.32 & 0.37 & 1.26 & 0.16 & 6.28 & 8.23 & 0.11 & \textbf{0.60} & \textbf{1.58}\tabularnewline
 & AIF & \textbf{0.11} & 0.29 & 2.08 & \textbf{-0.05} & 0.29 & 1.74 & 0.11 & \textbf{1.21} & \textbf{3.23} & \textbf{-0.16} & 2.02 & 2.39\tabularnewline
\hline 
\end{tabular}}
\par\end{centering}
\caption{Decreases in clean accuracy ($\downarrow$Cs), Trojan accuracies (Ts),
and recovery accuracies ($\downarrow$Rs) of our filtering defenses
(IF, VIF, AIF) against different \emph{all-target} Trojan attacks
on different datasets. \emph{Smaller values are better}. For a particular
dataset, attack, and metric, the best defense is highlighted in bold.\label{tab:VIF-AIF-all-target}}
\end{table}

\begin{table}[t]
\begin{centering}
\begin{tabular}{|c|c|da|da|da|da|}
\hline 
\multirow{2}{*}{Dataset} & \multirow{2}{*}{Defense} & \multicolumn{2}{c|}{BadNet+} & \multicolumn{2}{c|}{noise-BI+} & \multicolumn{2}{c|}{image-BI+} & \multicolumn{2}{c|}{InpAwAtk}\tabularnewline
\cline{3-10} \cline{4-10} \cline{5-10} \cline{6-10} \cline{7-10} \cline{8-10} \cline{9-10} \cline{10-10} 
 &  & \multicolumn{1}{c}{FPR} & \multicolumn{1}{c|}{FNR} & \multicolumn{1}{c}{FPR} & \multicolumn{1}{c|}{FNR} & \multicolumn{1}{c}{FPR} & \multicolumn{1}{c|}{FNR} & \multicolumn{1}{c}{FPR} & \multicolumn{1}{c|}{FNR}\tabularnewline
\hline 
\hline 
\multirow{3}{*}{MNIST} & IFtC & \textbf{0.20} & 77.83 & 0.10 & 99.63 & 0.23 & 7.22 & \textbf{0.20} & 97.44\tabularnewline
 & VIFtC & \textbf{0.20} & 49.97 & \textbf{0.07} & 4.19 & \textbf{0.07} & 0.27 & 0.33 & 64.85\tabularnewline
 & AIFtC & 0.70 & \textbf{19.24} & 0.27 & \textbf{0.07} & 0.30 & \textbf{0.07} & 0.40 & \textbf{24.83}\tabularnewline
\hline 
\hline 
\multirow{3}{*}{CIFAR10} & IFtC & \textbf{7.13} & 1.00 & \textbf{6.10} & \textbf{0.70} & \textbf{6.70} & 18.57 & \textbf{6.30} & 9.97\tabularnewline
 & VIFtC & 12.10 & 1.30 & 10.14 & 1.12 & 10.50 & \textbf{5.40} & 11.13 & \textbf{5.83}\tabularnewline
 & AIFtC & 9.07 & \textbf{0.87} & 8.03 & 1.10 & 8.47 & \textbf{5.40} & 8.90 & 6.70\tabularnewline
\hline 
\hline 
\multirow{3}{*}{GTSRB} & IFtC & 0.50 & \textbf{0.08} & 0.29 & 0.29 & \textbf{0.42} & 62.57 & \textbf{0.34} & 3.97\tabularnewline
 & VIFtC & 0.68 & 0.32 & 0.58 & 0.47 & 0.84 & 6.15 & 0.58 & \textbf{0.55}\tabularnewline
 & AIFtC & \textbf{0.34} & 0.42 & \textbf{0.26} & \textbf{0.26} & 0.79 & \textbf{0.92} & 0.45 & 2.05\tabularnewline
\hline 
\end{tabular}
\par\end{centering}
\caption{FPRs and FNRs of our FtC defenses against different \emph{all-target}
attacks. \emph{Smaller values are better}. For a particular dataset,
attack, and metric, the best defense is highlighted in bold.\label{tab:VIFtC-AIFtC-all-target-full}}
\end{table}

\subsection{Results of Our Defenses against All-target Attacks\label{subsec:Results-of-Our-Defenses-All-target}}

In Tables~\ref{tab:VIF-AIF-all-target} and \ref{tab:VIFtC-AIFtC-all-target-full},
we show the results our filtering and FtC defenses against different
\emph{all-target} attacks. On CIFAR10 and GTSRB, VIF/VIFtC and AIF/AIFtC
are comparable. However, on MNIST, AIF/AIFtC is clearly better than
VIF/VIFtC.

\begin{table}[t]
\begin{centering}
\resizebox{\textwidth}{!}{%
\par\end{centering}
\begin{centering}
\begin{tabular}{|c|c|dab|dab|dab|dab|dab|}
\hline 
\multirow{2}{*}{Dataset} & \multirow{2}{*}{Def.} & \multicolumn{3}{c|}{BadNet+} & \multicolumn{3}{c|}{noise-BI+} & \multicolumn{3}{c|}{image-BI+} & \multicolumn{3}{c|}{InpAwAtk} & \multicolumn{3}{c|}{WaNet}\tabularnewline
\cline{3-17} \cline{4-17} \cline{5-17} \cline{6-17} \cline{7-17} \cline{8-17} \cline{9-17} \cline{10-17} \cline{11-17} \cline{12-17} \cline{13-17} \cline{14-17} \cline{15-17} \cline{16-17} \cline{17-17} 
 &  & \multicolumn{1}{c}{$\downarrow$C} & \multicolumn{1}{c}{T} & \multicolumn{1}{c|}{$\downarrow$R} & \multicolumn{1}{c}{$\downarrow$C} & \multicolumn{1}{c}{T} & \multicolumn{1}{c|}{$\downarrow$R} & \multicolumn{1}{c}{$\downarrow$C} & \multicolumn{1}{c}{T} & \multicolumn{1}{c|}{$\downarrow$R} & \multicolumn{1}{c}{$\downarrow$C} & \multicolumn{1}{c}{T} & \multicolumn{1}{c|}{$\downarrow$R} & \multicolumn{1}{c}{$\downarrow$C} & \multicolumn{1}{c}{T} & \multicolumn{1}{c|}{$\downarrow$R}\tabularnewline
\hline 
\hline 
\multirow{2}{*}{MNIST} & IP & 0.37 & \textbf{1.29} & \textbf{3.60} & \textbf{0.03} & 3.69 & \textbf{10.92} & \textbf{0.13} & 1.81 & \textbf{11.72} & 0.30 & 1.14 & 2.10 & \textbf{0.20} & 1.66 & 1.70\tabularnewline
 & IF & \textbf{\emph{0.27}} & \emph{2.47} & \emph{4.99} & \emph{0.10} & \textbf{\emph{0.16}} & \emph{13.52} & \textbf{\emph{0.13}} & \textbf{\emph{1.29}} & \emph{12.02} & \textbf{\emph{0.21}} & \textbf{\emph{0.96}} & \textbf{\emph{2.08}} & \emph{0.23} & \textbf{\emph{0.34}} & \textbf{\emph{0.61}}\tabularnewline
\hline 
\hline 
\multirow{2}{*}{CIFAR10} & IP & 4.23 & 2.74 & 8.60 & 3.60 & \textbf{0.78} & 4.97 & \textbf{4.27} & \textbf{35.85} & \textbf{33.60} & 5.20 & 20.48 & 22.80 & 5.37 & 34.85 & 30.37\tabularnewline
 & IF & \textbf{\emph{4.15}} & \textbf{\emph{2.30}} & \textbf{\emph{7.79}} & \textbf{\emph{3.32}} & \emph{1.01} & \textbf{\emph{4.43}} & \emph{4.76} & \emph{37.48} & \emph{34.30} & \textbf{\emph{4.47}} & \textbf{\emph{16.35}} & \textbf{\emph{18.96}} & \textbf{\emph{3.21}} & \textbf{\emph{4.82}} & \textbf{\emph{6.80}}\tabularnewline
\hline 
\hline 
\multirow{2}{*}{GTSRB} & IP & 0.18 & \textbf{0.00} & 2.79 & 0.24 & \textbf{0.00} & 1.76 & \textbf{0.37} & 52.61 & 52.37 & 0.42 & 2.06 & 3.76 & 10.73 & 62.33 & 61.33\tabularnewline
 & IF & \textbf{\emph{0.13}} & \textbf{\emph{0.00}} & \textbf{\emph{2.55}} & \textbf{\emph{0.13}} & \emph{0.03} & \textbf{\emph{1.52}} & \textbf{\emph{0.37}} & \textbf{\emph{52.27}} & \textbf{\emph{51.95}} & \textbf{\emph{0.03}} & \textbf{\emph{0.66}} & \textbf{\emph{3.60}} & \textbf{\emph{0.08}} & \textbf{\emph{9.83}} & \textbf{\emph{9.62}}\tabularnewline
\hline 
\hline 
\multirow{2}{*}{CelebA} & IP & \textbf{2.83} & 9.98 & 4.23 & 2.73 & 25.57 & 14.72 & 2.43 & 73.63 & 35.82 & \textbf{2.31} & 15.09 & 7.81 & \textbf{2.28} & 76.49 & 36.76\tabularnewline
 & IF & \emph{4.21} & \textbf{\emph{8.62}} & \textbf{\emph{4.75}} & \textbf{\emph{2.57}} & \textbf{\emph{13.83}} & \textbf{\emph{6.00}} & \textbf{\emph{2.25}} & \textbf{\emph{59.39}} & \textbf{\emph{27.94}} & \emph{2.86} & \textbf{\emph{11.95}} & \textbf{\emph{6.07}} & \emph{2.43} & \textbf{\emph{15.21}} & \textbf{\emph{4.75}}\tabularnewline
\hline 
\end{tabular}}
\par\end{centering}
\caption{Trojan filtering results (in \%) of Input Processing \cite{liu2017neural}
and our Input Filtering. For a particular dataset, attack, and metric,
the best among the 2 defenses are highlighted in bold. Results taken
from Table~\ref{tab:VIF-AIF-full-results} are shown in italic.\label{tab:IF-IP-results}}
\end{table}

\subsection{Comparison between Input Filtering and Input Processing \cite{liu2017neural}\label{subsec:Comparing-IF-with-IP}}

In Fig.~\ref{tab:IF-IP-results}, we compare the performances of
our Input Filtering (IF) and Input Processing (IP) against all the
benchmark attacks on all the datasets. It is clear that IF outperforms
IP in terms of Trojan accuracy in most cases, especially under InputAwAtk
and WaNet. For example, IP achieves very high (poor) Trojan accuracies
of 34.85\%, 62.33\%, and 76.49\% against WaNet on CIFAR10, GTSRB,
and CelebA, respectively while our IF achieves only 4.82\%, 9.83\%,
and 15.21\%. These results empirically verify the importance of the
term $-\log p_{\Classifier}(y|x^{\circ})$ in the loss of IF. This
term ensures that the filtered output $x^{\circ}$computed by $\Filter$
cannot cause harm to $\Classifier$ even when it look very similar
to the original input $x$.

\begin{figure}[t]
\begin{centering}
\subfloat[Reconstruction losses\label{fig:Rec-losses-4-archs}]{\begin{centering}
\includegraphics[width=0.46\textwidth]{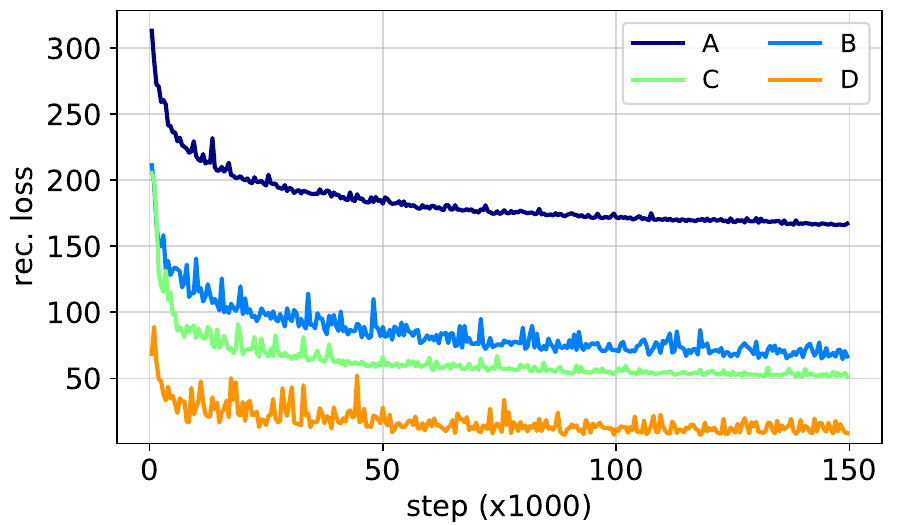}
\par\end{centering}
}\subfloat[Reconstructed images\label{fig:Rec-imgs-4-archs}]{\begin{centering}
\includegraphics[width=0.46\textwidth]{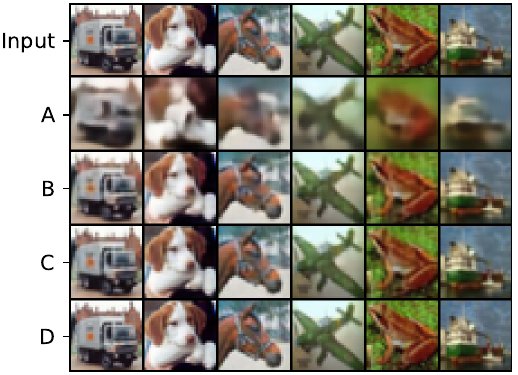}
\par\end{centering}
}
\par\end{centering}
\caption{Reconstruction losses (a) and reconstructed images (b) on CIFAR10
corresponding to the 4 autoencoders described in Table~\ref{tab:4-Architectures-CIFAR10}.}
\end{figure}

\begin{table}[t]
\begin{centering}
\resizebox{\textwidth}{!}{%
\par\end{centering}
\begin{centering}
\begin{tabular}{|c|dab|dab|dab|dab|dab|}
\hline 
\multirow{2}{*}{Arch.} & \multicolumn{3}{c|}{BadNet+} & \multicolumn{3}{c|}{noise-BI+} & \multicolumn{3}{c|}{image-BI+} & \multicolumn{3}{c|}{InpAwAtk} & \multicolumn{3}{c|}{WaNet}\tabularnewline
\cline{2-16} \cline{3-16} \cline{4-16} \cline{5-16} \cline{6-16} \cline{7-16} \cline{8-16} \cline{9-16} \cline{10-16} \cline{11-16} \cline{12-16} \cline{13-16} \cline{14-16} \cline{15-16} \cline{16-16} 
 & \multicolumn{1}{c}{$\downarrow$C} & \multicolumn{1}{c}{T} & \multicolumn{1}{c|}{$\downarrow$R} & \multicolumn{1}{c}{$\downarrow$C} & \multicolumn{1}{c}{T} & \multicolumn{1}{c|}{$\downarrow$R} & \multicolumn{1}{c}{$\downarrow$C} & \multicolumn{1}{c}{T} & \multicolumn{1}{c|}{$\downarrow$R} & \multicolumn{1}{c}{$\downarrow$C} & \multicolumn{1}{c}{T} & \multicolumn{1}{c|}{$\downarrow$R} & \multicolumn{1}{c}{$\downarrow$C} & \multicolumn{1}{c}{T} & \multicolumn{1}{c|}{$\downarrow$R}\tabularnewline
\hline 
\hline 
A & 40.02 & 5.96 & 41.10 & 40.90 & 14.52 & 42.43 & 35.73 & \textbf{3.56} & 37.92 & 38.13 & 3.52 & 38.97 & 32.65 & 6.81 & 32.94\tabularnewline
B & 9.23 & 2.74 & 13.37 & 8.73 & 1.96 & 10.32 & 10.48 & 10.59 & 18.87 & 9.75 & \textbf{1.96} & 13.91 & 9.94 & 4.22 & 11.73\tabularnewline
C & 7.70 & \textbf{2.52} & \textbf{11.27} & 6.43 & \textbf{1.22} & \textbf{7.10} & 7.53 & 10.52 & \textbf{16.50} & 7.67 & 3.07 & \textbf{12.38} & 7.97 & \textbf{3.96} & \textbf{10.67}\tabularnewline
D & \textbf{0.27} & 100.0 & 84.83 & \textbf{-0.02} & 100.0 & 84.57 & \textbf{0.21} & 99.81 & 84.96 & \textbf{-0.10} & 98.85 & 83.33 & \textbf{-0.07} & 98.96 & 83.35\tabularnewline
\hline 
\end{tabular}
\par\end{centering}
\begin{centering}
}
\par\end{centering}
\caption{Trojan filtering results (in \%) of VIF against different attacks
on CIFAR10 w.r.t. different architectures of $\protect\Filter$. For
a particular dataset, attack, and metric, the best defense is highlighted
in bold.\label{tab:Main-results-4-archs}}
\end{table}

\begin{figure}[t]
\begin{centering}
\subfloat[Crossentropy loss\label{fig:XentY-by-arch}]{\begin{centering}
\includegraphics[width=0.32\textwidth]{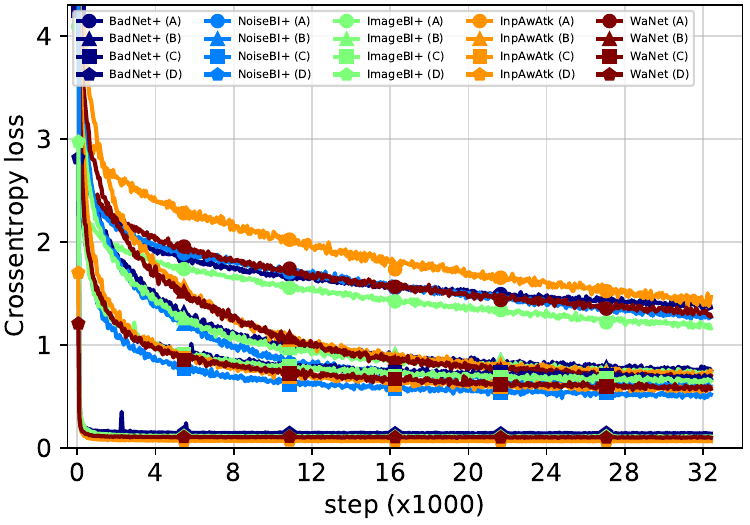}
\par\end{centering}
}\subfloat[Reconstruction loss\label{fig:RecX-by-arch}]{\begin{centering}
\includegraphics[width=0.32\textwidth]{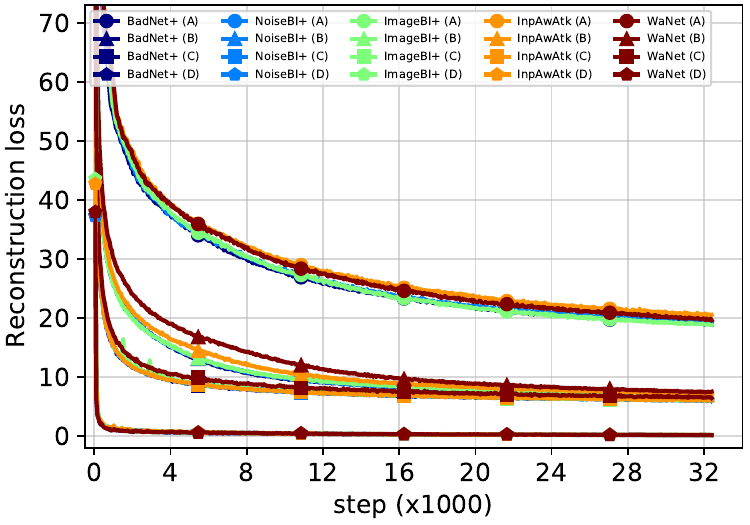}
\par\end{centering}
}\subfloat[KL divergence\label{fig:kldZ-by-arch}]{\begin{centering}
\includegraphics[width=0.32\textwidth]{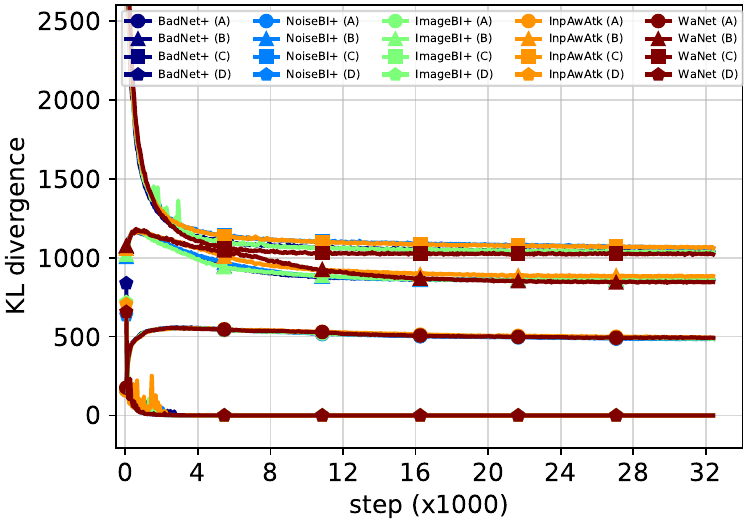}
\par\end{centering}
}
\par\end{centering}
\caption{Training curves of VIF against InpAwAtk on CIFAR10 ($\protect\Data_{\text{val}}$)
w.r.t. the 4 architectures of $\protect\Filter$ in Table~\ref{tab:4-Architectures-CIFAR10}.\label{fig:train-curves-by-arch}}
\end{figure}

\begin{figure}[t]
\centering{}\subfloat[Filtered Clean Acc.]{\begin{centering}
\includegraphics[width=0.32\textwidth]{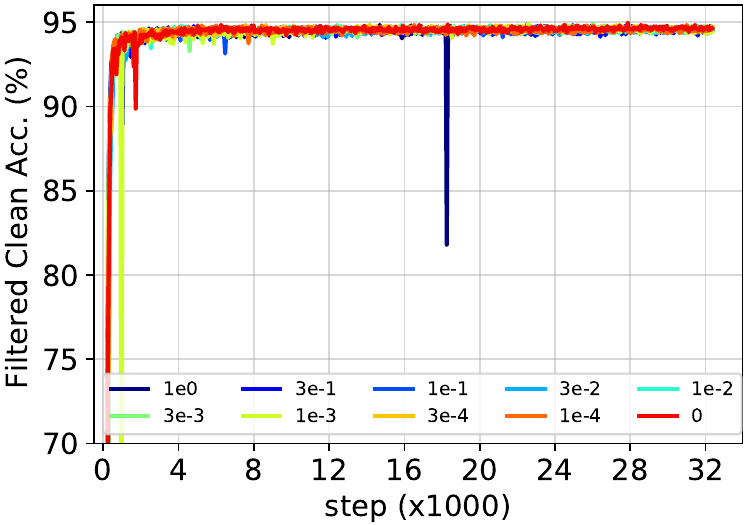}
\par\end{centering}
}\subfloat[Filtered Trojan Acc.]{\begin{centering}
\includegraphics[width=0.32\textwidth]{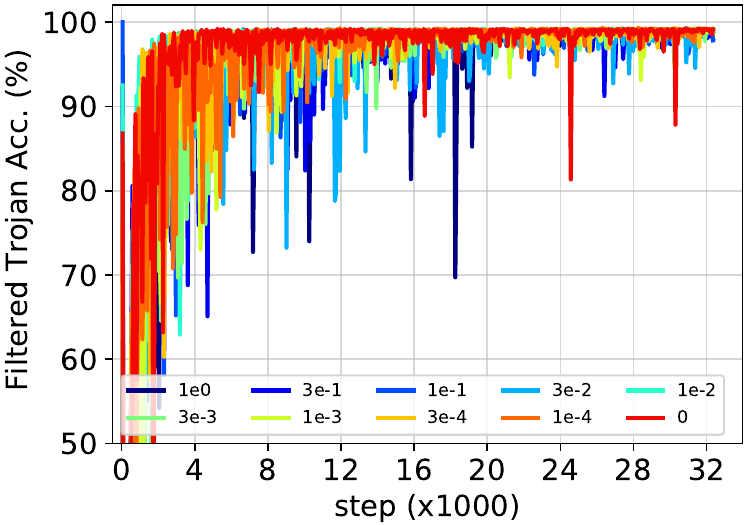}
\par\end{centering}
}\subfloat[Filtered Recovery Acc.]{\begin{centering}
\includegraphics[width=0.32\textwidth]{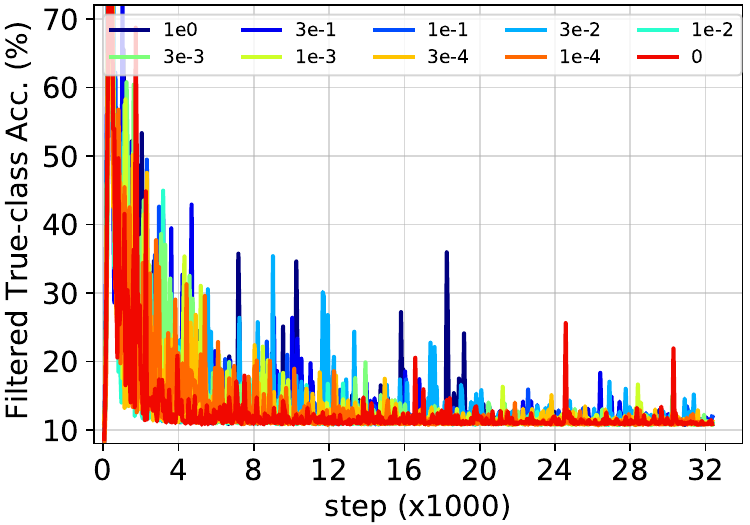}
\par\end{centering}
}\caption{Test results of VIF using the architecture D (Table~\ref{tab:4-Architectures-CIFAR10})
for $\protect\Filter$ against InpAwAtk on CIFAR10 ($\protect\Data_{\text{test}}$)
w.r.t. different coefficient values for $D_{\text{KL}}$ ($\lambda_{2}$
in Eq.~\ref{eq:VIF_loss}).\label{fig:Results-of-VIF-w-arch-D}}
\end{figure}

\subsection{Different architectures of the filter $\protect\Filter$\label{subsec:Different-architectures-of-F}}

A major factor that affects the performance of $\Filter$ is its architecture.
We consider an architecture of $\Filter$ to be more complex than
others if $\Filter$ achieves smaller reconstruction loss on $\Data_{\text{val}}$
with this architecture. In Section~\ref{sec:Our-Proposed-Defenses},
we argued that an optimal filter should be neither too simple nor
too complex. Here, we empirically verify this intuition by examining
4 different architectures of $\Filter$ for CIFAR10 marked as A, B,
C, D (Table~\ref{tab:4-Architectures-CIFAR10}). Their complexities
are greater in alphabetical order as shown in Figs.~\ref{fig:Rec-losses-4-archs},
\ref{fig:Rec-imgs-4-archs}. The architecture D has skip-connections
between its encoder and decoder while A, B, C do not. 

Denote $\Filter$ with the architectures A, B, C, D as $\Filter_{\text{A}}$,
$\Filter_{\text{B}}$, $\Filter_{\text{C}}$, $\Filter_{\text{D}}$
respectively. As shown in Table~\ref{tab:Main-results-4-archs},
$\Filter_{\text{C}}$ is the best in terms of recovery accuracy although
C is neither the simplest (A) nor the most complex architecture (D).
$\Filter_{\text{A}}$ achieves reasonably low Trojan accuracies comparable
to those of $\Filter_{\text{B}}$ and $\Filter_{\text{C}}$ but the
worst clean accuracies. $\Filter_{\text{D}}$, by contrast, experiences
almost no decrease in clean accuracy but achieves very high Trojan
accuracies. The reason is that $\Filter_{\text{D}}$ simply copies
all information from an input to the output via its skip-connections
rather than learning compressed latent representations of input images.
One can verify this by observing that the $D_{\text{KL}}$ (Eq.~\ref{eq:VIF_loss})
of $\Filter_{\text{D}}$ is almost 0 (Fig.~\ref{fig:kldZ-by-arch}).
In this case, changing $\lambda_{2}$ has no effect on the Trojan
accuracy of $\Filter_{\text{D}}$ as shown in Fig.~\ref{fig:Results-of-VIF-w-arch-D}.
However, it is still possible to lower the Trojan accuracy of $\Filter_{\text{D}}$
without removing the skip-connections in D. For example, we can treat
the latent representations corresponding to the skip-connections as
random variables and apply the $D_{\text{KL}}$ to these variables
like what we do with the middle representation. Or we can compress
the whole $\Filter_{\text{D}}$ by using advanced network compression
techniques \cite{molchanov2017variational,louizos2017bayesian,van2020bayesian}.
These ideas are out of scope of this paper so we leave them for future
work.

\subsection{Explicit Normalization of Triggers in AIF\label{subsec:Explicit-Normalization-of-Triggers-in-AIF}}

\subsubsection{Theoretical Derivation}

During training AIF, we observed that under the influence of $\Loss_{\text{VIF-gen}}$
(Eq.~\ref{eq:AIF_gen_loss}), the norm $\|m\|$ of a synthetic trigger
mask $m$ usually increases overtime despite the fact that $\Loss_{\text{VIF-gen}}$
also contains a norm regularization term. The reason is that $\Generator$
is encouraged to output bigger Trojan triggers to fool $\Filter$.
However, too big trigger causes the generated Trojan image $\tilde{x}$
to be very different from the input image $x$, which affects the
learning of $\Filter$. One way to deal with this problem is \emph{explicitly
normalizing} $m$ so that its norm is always upper-bounded by $\delta$.
Denoted by $\bar{m}$ the $\delta$-normalized version of $m$. $\bar{m}$
can be computed as follows:
\begin{equation}
\bar{m}=\begin{cases}
m & \text{if}\ \|m\|\leq\delta\\
\delta\frac{m}{\|m\|} & \text{if}\ \|m\|>\delta
\end{cases}\label{eq:normalized_mask}
\end{equation}
or more compactly, 
\begin{align}
\bar{m} & =m\times\left(1-\frac{\max(\|m\|-\delta,0)}{\|m\|}\right)\nonumber \\
 & =m\times\left(1-\frac{\text{ReLU}(\|m\|-\delta)}{\|m\|}\right)\label{eq:ReLU_normalized_mask}
\end{align}
In case the norm is L2, the second expression in Eq.~\ref{eq:normalized_mask}
can be seen as the projection of $m$ onto the surface of a sphere
of radius $\delta$. By replacing $\text{ReLU}(\cdot)$ in Eq.~\ref{eq:ReLU_normalized_mask}
with $\text{Softplus}(\cdot)$, we obtain a soft version of $\bar{m}$:
\begin{align}
\bar{m}_{s} & =m\times\left(1-\frac{\text{Softplus}(\|m\|-\delta,\tau)}{\|m\|}\right)\nonumber \\
 & =m\times\left(1-\frac{\tau\log(1+\exp((\|m\|-\delta)/\tau))}{\|m\|}\right)\label{eq:Softplus_normalized_mask}
\end{align}
where $\tau>0$ is a temperature. $\bar{m}_{s}$ with smaller $\tau$
approximates $\bar{m}$ better. Generally, using $\bar{m}_{s}$ gives
better gradient update than using $\bar{m}$. However, since Softplus
is an upper-bound of ReLU, $\text{Softplus}(\|m\|-\delta,\tau)$ can
be greater than $\|m\|$, which causes $\bar{m}_{s}$ to be negative.
To avoid that, we slightly modify Eq.~\ref{eq:Softplus_normalized_mask}
into the equation below:
\begin{equation}
\bar{m}_{s}=m\times\left(1-\frac{\tau\log(1+\exp((\|m\|-\delta)/\tau))}{\|m\|+\Omega}\right)\label{eq:Adjusted_Softplus_normalized_mask}
\end{equation}
where $\Omega>0$ is an added term to ensure that $\bar{m}_{s}\geq0$.
$\Omega$ can be computed from $\tau$ and $\delta$ by solving the
following inequality:
\begin{align*}
 & \tau\log(1+\exp((\|m\|-\delta)/\tau))\leq\|m\|+\Omega\\
\Leftrightarrow & 1+\exp\left(\frac{\|m\|-\delta}{\tau}\right)\leq\exp\left(\frac{\|m\|+\Omega}{\tau}\right)\\
\Leftrightarrow & 1\leq\exp\left(\frac{\|m\|}{\tau}\right)\left(\exp\frac{\Omega}{\tau}-\exp\left(\frac{-\delta}{\tau}\right)\right)\\
\Leftrightarrow & \exp\left(\frac{-\|m\|}{\tau}\right)\leq\exp\frac{\Omega}{\tau}-\exp\left(\frac{-\delta}{\tau}\right)
\end{align*}
Since $\exp\left(\frac{-\|m\|}{\tau}\right)$ is always smaller than
1, we can choose $\Omega$ so that: 
\begin{align*}
 & \exp\frac{\Omega}{\tau}-\exp\left(\frac{-\delta}{\tau}\right)=1\\
\Leftrightarrow & \Omega=\tau\log\left(1+\exp\left(\frac{-\delta}{\tau}\right)\right)
\end{align*}
In our experiments, we set $\delta=0.05$ and $\tau=0.01$. We also
observed that the hard normalization (Eq.~\ref{eq:ReLU_normalized_mask})
and the soft normalization (Eq.~\ref{eq:Adjusted_Softplus_normalized_mask})
of $m$ give roughly the same performance.

\begin{table}[t]
\begin{centering}
\resizebox{\textwidth}{!}{%
\par\end{centering}
\begin{centering}
\begin{tabular}{|c|c|dab|dab|dab|dab|dab|}
\hline 
\multirow{2}{*}{Dataset} & \multirow{2}{*}{Norm.} & \multicolumn{3}{c|}{BadNet+} & \multicolumn{3}{c|}{noise-BI+} & \multicolumn{3}{c|}{image-BI+} & \multicolumn{3}{c|}{InpAwAtk} & \multicolumn{3}{c|}{WaNet}\tabularnewline
\cline{3-17} \cline{4-17} \cline{5-17} \cline{6-17} \cline{7-17} \cline{8-17} \cline{9-17} \cline{10-17} \cline{11-17} \cline{12-17} \cline{13-17} \cline{14-17} \cline{15-17} \cline{16-17} \cline{17-17} 
 &  & \multicolumn{1}{c}{$\downarrow$C} & \multicolumn{1}{c}{T} & \multicolumn{1}{c|}{$\downarrow$R} & \multicolumn{1}{c}{$\downarrow$C} & \multicolumn{1}{c}{T} & \multicolumn{1}{c|}{$\downarrow$R} & \multicolumn{1}{c}{$\downarrow$C} & \multicolumn{1}{c}{T} & \multicolumn{1}{c|}{$\downarrow$R} & \multicolumn{1}{c}{$\downarrow$C} & \multicolumn{1}{c}{T} & \multicolumn{1}{c|}{$\downarrow$R} & \multicolumn{1}{c}{$\downarrow$C} & \multicolumn{1}{c}{T} & \multicolumn{1}{c|}{$\downarrow$R}\tabularnewline
\hline 
\hline 
\multirow{2}{*}{CIFAR10} & w/ & \textbf{\emph{5.60}} & \emph{2.37} & \textbf{\emph{9.03}} & \textbf{\emph{4.87}} & \emph{1.14} & \textbf{\emph{6.02}} & \textbf{\emph{5.23}} & \emph{1.96} & \textbf{\emph{7.10}} & \textbf{\emph{5.28}} & \emph{5.30} & \emph{11.87} & \textbf{\emph{4.30}} & \textbf{\emph{1.22}} & \textbf{\emph{5.67}}\tabularnewline
 & wo/ & 5.90 & \textbf{2.15} & 10.00 & 5.60 & \textbf{0.78} & 7.50 & 6.60 & \textbf{1.56} & 7.53 & 6.20 & \textbf{2.56} & \textbf{9.37} & 5.80 & 2.22 & 8.10\tabularnewline
\hline 
\hline 
\multirow{2}{*}{GTSRB} & w/ & \textbf{\emph{-0.16}} & \textbf{\emph{0.00}} & \textbf{\emph{1.87}} & \textbf{\emph{0.05}} & \textbf{\emph{0.00}} & \textbf{\emph{0.81}} & \textbf{\emph{0.13}} & \emph{7.47} & \emph{9.54} & \textbf{\emph{-0.03}} & \emph{0.05} & \textbf{\emph{1.37}} & \textbf{\emph{-0.05}} & \emph{0.50} & \textbf{\emph{0.42}}\tabularnewline
 & wo/ & 0.13 & 0.08 & 3.18 & 0.24 & 0.00 & 1.26 & 0.45 & \textbf{0.92} & \textbf{5.44} & 0.45 & \textbf{0.00} & 2.60 & 0.21 & \textbf{0.11} & 0.45\tabularnewline
\hline 
\end{tabular}}
\par\end{centering}
\caption{Trojan filtering results (in \%) of AIF with and without explicit
trigger normalization against different attacks on CIFAR10 and GTSRB.
For a particular attack, dataset and metric, the best result is highlighted
in bold. Results taken from Table~\ref{tab:VIF-AIF-full-results}
are shown in italic.\label{tab:Explicit-Trigger-Norm-results}}
\end{table}

\begin{figure}[t]
\subfloat[Clean Accuracy]{\begin{centering}
\includegraphics[width=0.237\textwidth]{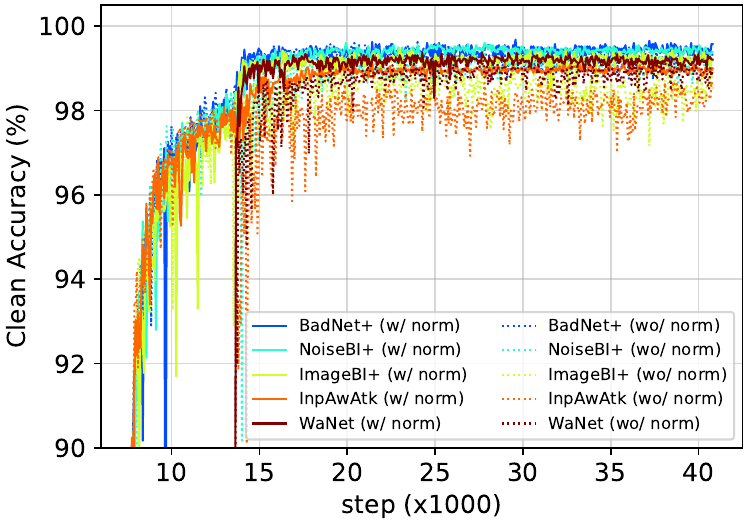}
\par\end{centering}
}\subfloat[Trojan Accuracy]{\begin{centering}
\includegraphics[width=0.237\textwidth]{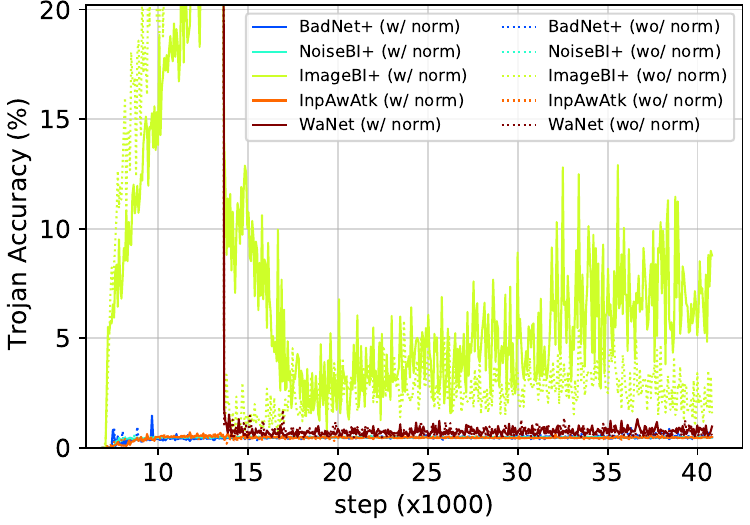}
\par\end{centering}
}\subfloat[Recovery Accuracy]{\begin{centering}
\includegraphics[width=0.237\textwidth]{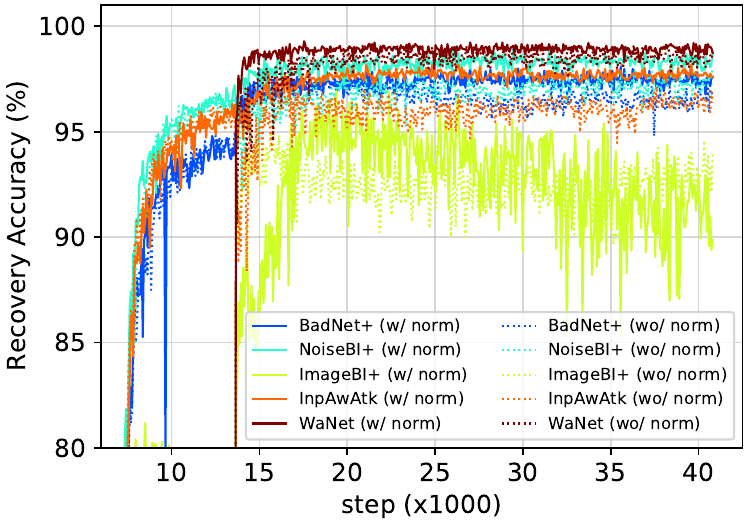}
\par\end{centering}
}\subfloat[Trojan Acc. on Sync. Data]{\begin{centering}
\includegraphics[width=0.237\textwidth]{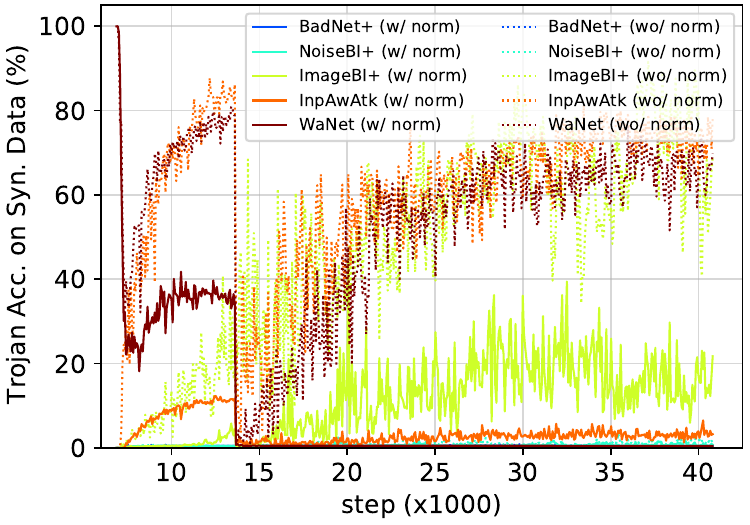}
\par\end{centering}
}

\subfloat[$\log p(y|x^{\circ})$ ($\protect\Filter$)]{\begin{centering}
\includegraphics[width=0.237\textwidth]{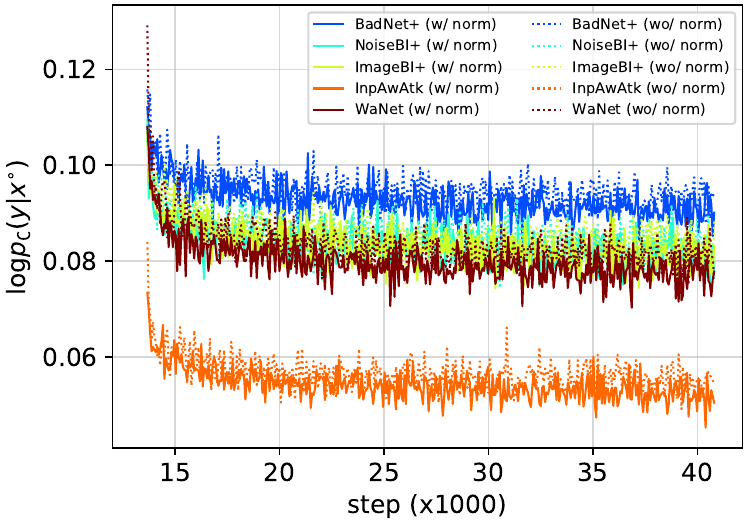}
\par\end{centering}
}\subfloat[$\|x^{\circ}-x\|$ ($\protect\Filter$)]{\begin{centering}
\includegraphics[width=0.237\textwidth]{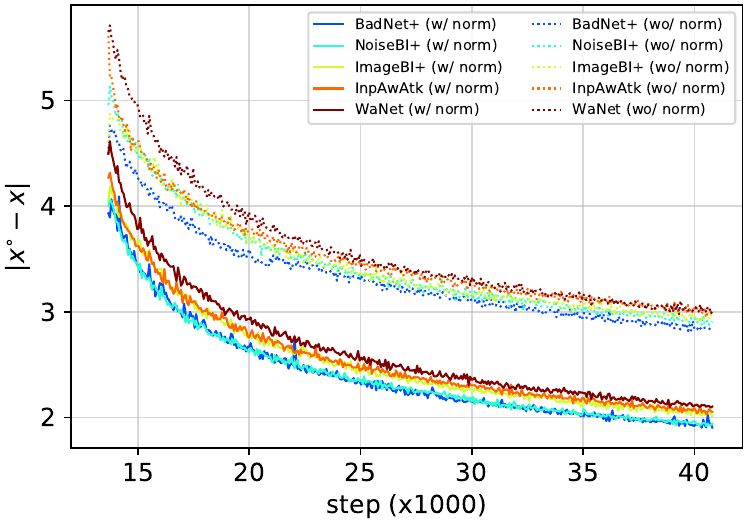}
\par\end{centering}
}\subfloat[$\log p(y|\tilde{x}^{\circ})$ ($\protect\Filter$)]{\begin{centering}
\includegraphics[width=0.237\textwidth]{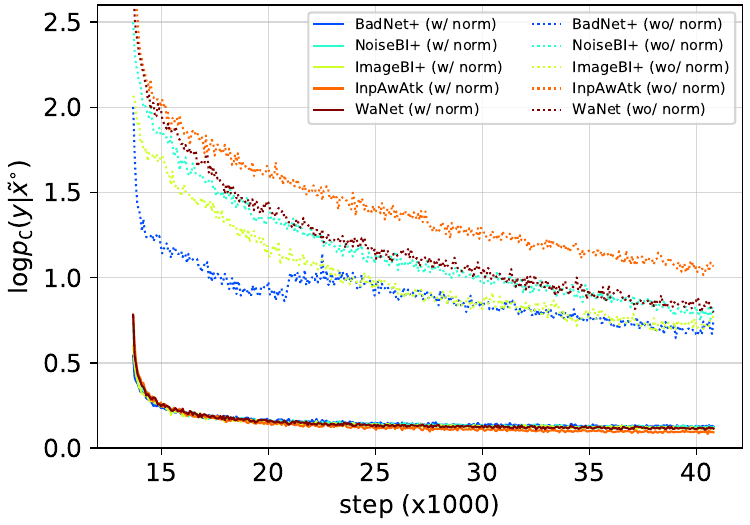}
\par\end{centering}
}\subfloat[$\|\tilde{x}^{\circ}-x\|$ ($\protect\Filter$)]{\begin{centering}
\includegraphics[width=0.237\textwidth]{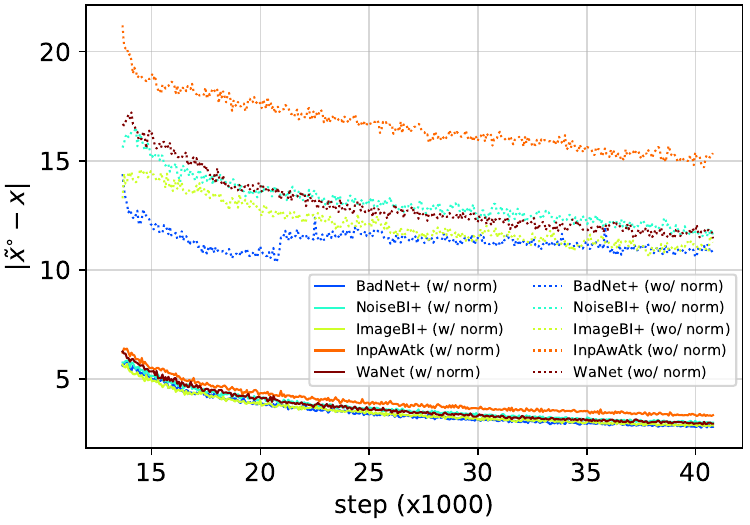}
\par\end{centering}
}

\subfloat[$\log p(k|\tilde{x}^{\circ})$ ($\protect\Generator$)]{\begin{centering}
\includegraphics[width=0.237\textwidth]{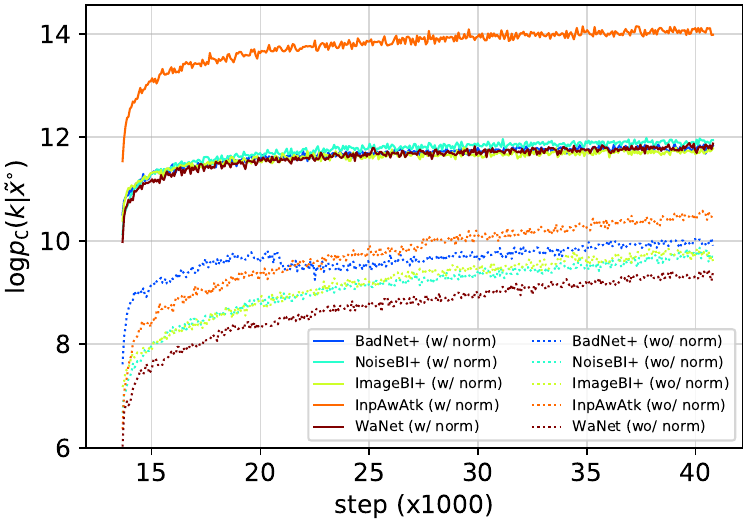}
\par\end{centering}
}\subfloat[$\log p(k|\tilde{x})$ ($\protect\Generator$)]{\begin{centering}
\includegraphics[width=0.237\textwidth]{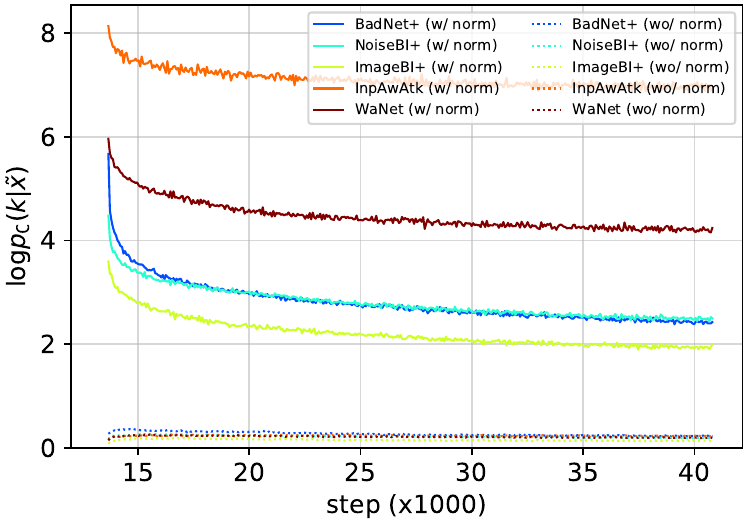}
\par\end{centering}
}\subfloat[$\|m\|$ ($\protect\Generator$)]{\begin{centering}
\includegraphics[width=0.237\textwidth]{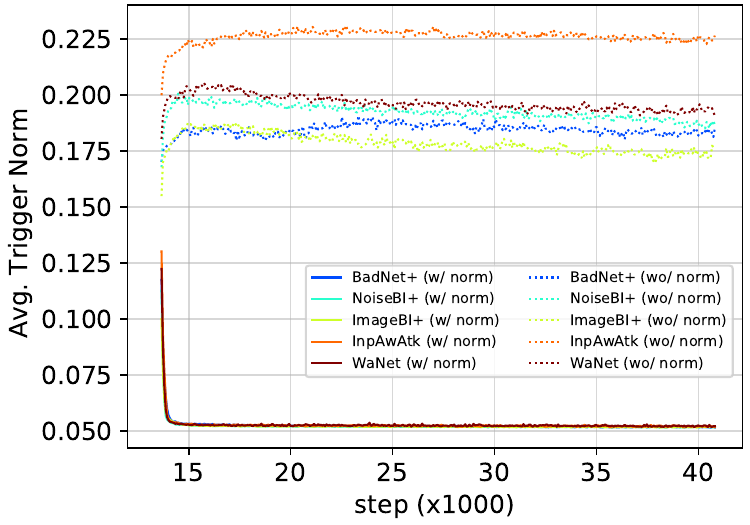}
\par\end{centering}
}

\caption{Test result curves and training loss curves of $\protect\Filter$,
$\protect\Generator$ of VIF with and without explicit trigger normalization
against different attacks on GTSRB. These plots correspond to the
results in the bottom 2 rows in Table~\ref{tab:Explicit-Trigger-Norm-results}.\label{fig:Explicit-trigger-norm-AIF-plots}}
\end{figure}

\subsubsection{Empirical Results}

From Table~\ref{tab:Explicit-Trigger-Norm-results}, we see that
AIF with explicit trigger normalization (denoted as AIF-w) always
achieve smaller $\downarrow$Cs and sometimes achieve larger Ts than
the counterpart \emph{without} explicit trigger normalization (denoted
as AIF-wo). In general, AIF-w usually achieves lower $\downarrow$Rs
than AIF-wo and is considered to be better so we set it as default.
Fig.~\ref{fig:Explicit-trigger-norm-AIF-plots} provides a deeper
insight into the results in Table~\ref{tab:Explicit-Trigger-Norm-results}.
Without explicit trigger normalization, the generator $\Generator$
can easily fool $\Filter$ (low crossentropy losses in Fig.~\ref{fig:Explicit-trigger-norm-AIF-plots}i)
by just increasing the norm of the synthetic triggers (Fig.~\ref{fig:Explicit-trigger-norm-AIF-plots}k).
This makes $\tilde{x}$ more different from $x$ and causes more difficulty
for $\Filter$ to force $\tilde{x}^{\circ}$ close to $x$ (Fig.~\ref{fig:Explicit-trigger-norm-AIF-plots}h)
as well as correcting the label of $\tilde{x}$ (Fig.~\ref{fig:Explicit-trigger-norm-AIF-plots}g).
The large difference between $\tilde{x}$ and $x$ also negatively
affects the performance of $\Filter$ on reconstructing clean images
(Fig.~\ref{fig:Explicit-trigger-norm-AIF-plots}f). As a result,
AIF-wo achieves much poorer Ts on synthetic Trojan images (Fig.~\ref{fig:Explicit-trigger-norm-AIF-plots}d)
than AIF-w. On ground-truth Trojan images, AIF-wo performs as well
as AIF-w in terms of T (Fig.~\ref{fig:Explicit-trigger-norm-AIF-plots}b)
and worse than AIF-w in terms of $\downarrow$C (Fig.~\ref{fig:Explicit-trigger-norm-AIF-plots}a)
and $\downarrow$R (Fig.~\ref{fig:Explicit-trigger-norm-AIF-plots}d).

\section{Qualitative Results of Our Defenses\label{sec:Qualitative-Results-of-Our-Defenses}}

\subsection{Against Benchmark Attacks\label{subsec:Qualitative-Results-Against-Benchmark-Attacks}}

\begin{figure}[t]
\begin{centering}
\subfloat[Filtered images\label{fig:Filtered-Images-MNIST}]{\begin{centering}
\includegraphics[width=0.99\textwidth]{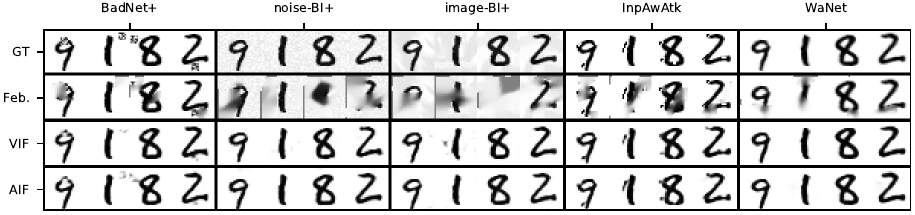}
\par\end{centering}
}
\par\end{centering}
\begin{centering}
\subfloat[Counter-triggers\label{fig:Countertriggers-MNIST}]{\begin{centering}
\includegraphics[width=0.99\textwidth]{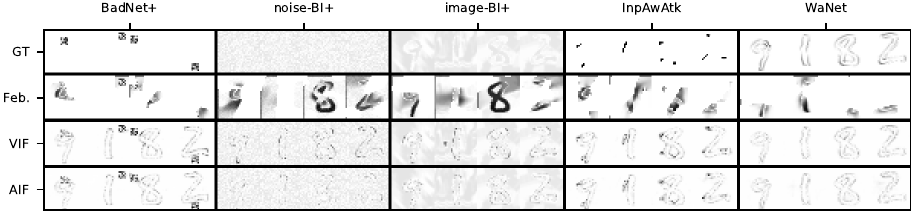}
\par\end{centering}
}
\par\end{centering}
\caption{(a): Ground-truth (GT) Trojan images of different attacks and the
corresponding filtered images computed by Februus (Feb.), VIF, and
AIF on MNIST. (b): GT triggers and counter-triggers w.r.t. the filtered
images in (a).}
\end{figure}

\begin{figure}[t]
\begin{centering}
\subfloat[Filtered images\label{fig:Filtered-Images-CIFAR10}]{\begin{centering}
\includegraphics[width=0.99\textwidth]{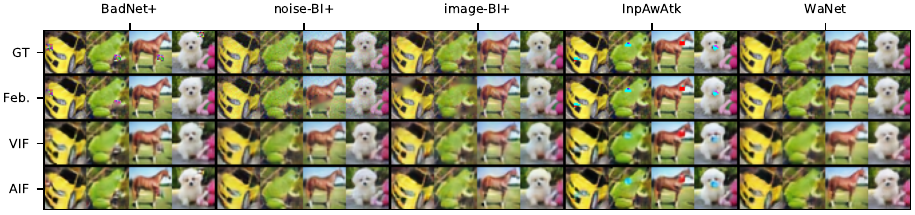}
\par\end{centering}
}
\par\end{centering}
\begin{centering}
\subfloat[Counter-triggers\label{fig:Countertriggers-CIFAR10}]{\begin{centering}
\includegraphics[width=0.99\textwidth]{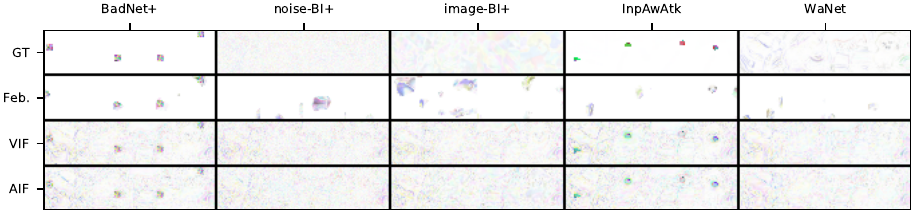}
\par\end{centering}
}
\par\end{centering}
\caption{(a): Ground-truth (GT) Trojan images of different attacks and the
corresponding filtered images computed by Februus (Feb.), VIF, and
AIF on CIFAR10. (b): GT triggers and counter-triggers w.r.t. the filtered
images in (a).}
\end{figure}

\begin{figure}[t]
\begin{centering}
\subfloat[Filtered images\label{fig:Filtered-Images-GTSRB}]{\begin{centering}
\includegraphics[width=0.99\textwidth]{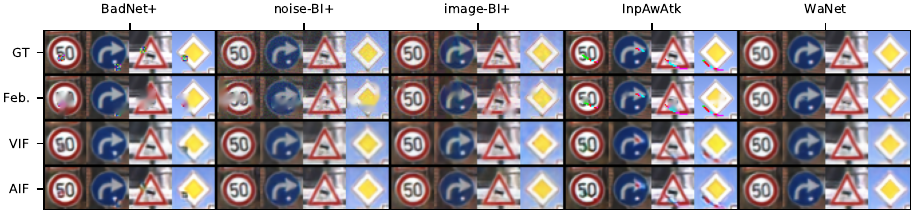}
\par\end{centering}
}
\par\end{centering}
\begin{centering}
\subfloat[Counter-triggers\label{fig:Countertriggers-GTSRB}]{\begin{centering}
\includegraphics[width=0.99\textwidth]{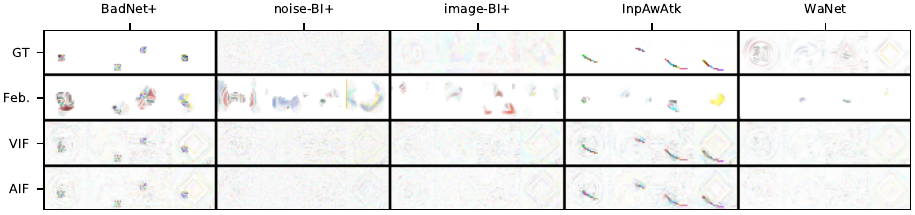}
\par\end{centering}
}
\par\end{centering}
\caption{(a): Ground-truth (GT) Trojan images of different attacks and the
corresponding filtered images computed by Februus (Feb.), VIF, and
AIF on GTSRB. (b): GT triggers and counter-triggers w.r.t. the filtered
images in (a).}
\end{figure}

\begin{figure}[t]
\begin{centering}
\subfloat[Filtered images\label{fig:Filtered-Images-CelebA}]{\begin{centering}
\includegraphics[width=0.99\textwidth]{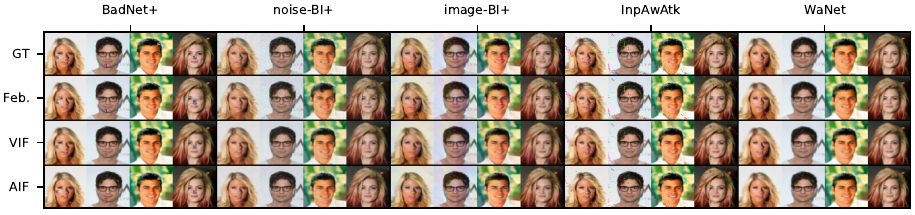}
\par\end{centering}
}
\par\end{centering}
\begin{centering}
\subfloat[Counter-triggers\label{fig:Countertriggers-CelebA}]{\begin{centering}
\includegraphics[width=0.99\textwidth]{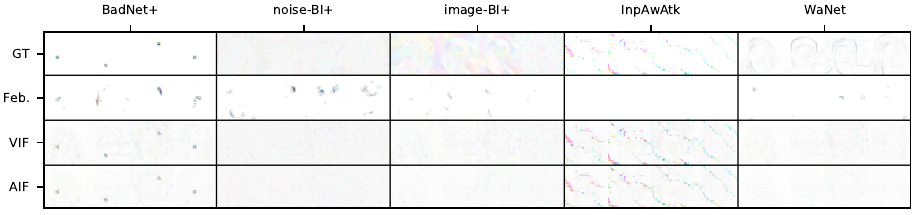}
\par\end{centering}
}
\par\end{centering}
\caption{(a): Ground-truth (GT) Trojan images of different attacks and the
corresponding filtered images computed by Februus (Feb.), VIF, and
AIF on CelebA. (b): GT triggers and counter-triggers w.r.t. the filtered
images in (a).}
\end{figure}

In Figs.~\ref{fig:Filtered-Images-MNIST}, \ref{fig:Filtered-Images-CIFAR10},
\ref{fig:Filtered-Images-GTSRB}, \ref{fig:Filtered-Images-CelebA},
we show some ground-truth (GT) Trojan images $\tilde{x}$ and their
filtered counterparts $\tilde{x}^{\circ}$ computed by Februus (Feb.)
and our filtering defenses (VIF, AIF) for different Trojan attacks
and datasets. We also show the corresponding ``counter-triggers''
of $\tilde{x}^{\circ}$ defined as $|\tilde{x}^{\circ}-\tilde{x}|$
in Figs.~\ref{fig:Countertriggers-MNIST}, \ref{fig:Countertriggers-CIFAR10},
\ref{fig:Countertriggers-GTSRB}, \ref{fig:Countertriggers-CelebA}
in comparison with the ground-truth Trojan triggers $|\tilde{x}-x|$.
It is apparent that VIF and AIF correctly filter the true triggers
of all the attacks without knowing them while Februus fails to filter
the true triggers of noise-BI+, image-BI+, InpAwAtk and WaNet. The
failure of Februus comes from the fact that GradCAM is unable to find
regions containing full-sized, distributed triggers of noise/image-BI+
or isomorphic, input-specific triggers of InpAwAtk/WaNet. For BadNet+
and InpAwAtk, our filtering defenses mainly blur the triggers of these
attacks instead of completely removing the triggers but this is enough
to deactivate the triggers.

\subsection{Against Attacks with Large-Norm Triggers\label{subsec:Qualitative-Results-Against-Attacks-with-Large-Norm-Triggers}}

\begin{figure}[t]
\begin{centering}
\subfloat[Filtered images (BadNet+)]{\begin{centering}
\includegraphics[width=0.45\textwidth]{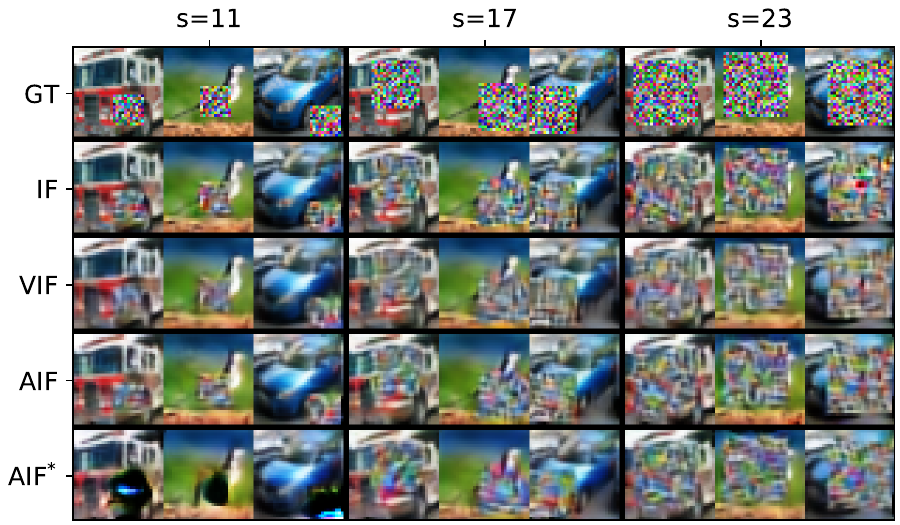}
\par\end{centering}
}\hspace{0.02\textwidth}\subfloat[Filtered images (noise-BI+)]{\begin{centering}
\includegraphics[width=0.45\textwidth]{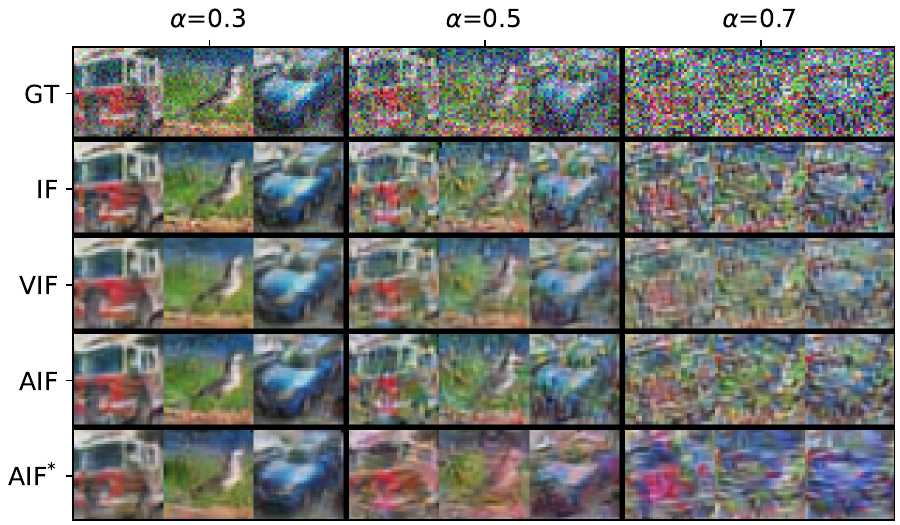}
\par\end{centering}
}
\par\end{centering}
\begin{centering}
\subfloat[Counter-triggers (BadNet+)]{\begin{centering}
\includegraphics[width=0.45\textwidth]{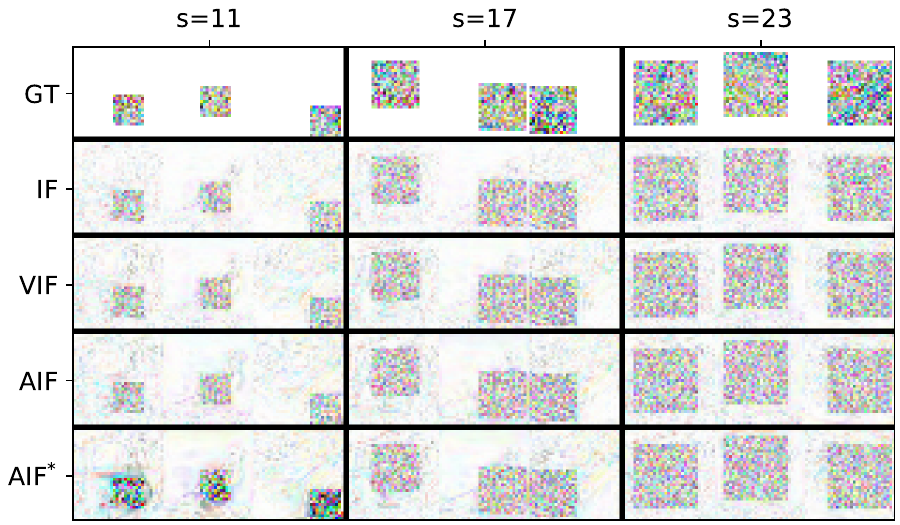}
\par\end{centering}
}\hspace{0.02\textwidth}\subfloat[Counter-triggers (noise-BI+)]{\begin{centering}
\includegraphics[width=0.45\textwidth]{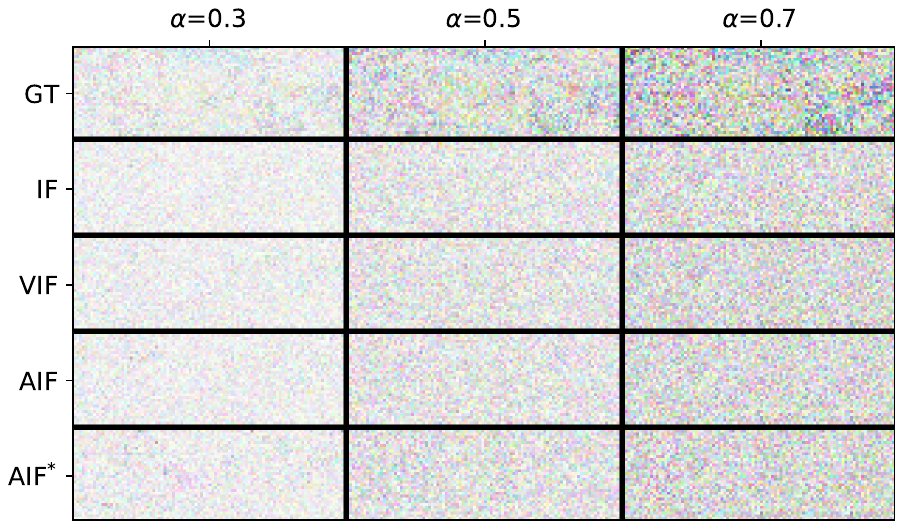}
\par\end{centering}
}
\par\end{centering}
\caption{(a)/(b): Ground-truth (GT) Trojan images of BadNet+/noise-BI+ and
the corresponding filtered images computed by IF, VIF, AIF, and AIF
without explicit trigger normalization (AIF$^{*}$). (c)/(d): GT triggers
and counter-triggers w.r.t. the filtered images in (a)/(b).\label{fig:Large-norm-triggers-visualization}}
\end{figure}

In Fig.~\ref{fig:Large-norm-triggers-visualization}, we visualize
the filtered images and their corresponding counter-triggers computed
by our methods for Trojan images of BadNet+ with different trigger
sizes and of noise-BI+ with different blending ratios. In general,
our filtering defenses can effectively deactivate triggers embedded
in the Trojan images via modifying the trigger pixels but cannot fully
reconstruct the original clean images. These qualitative results correspond
to the quantitative results in Tables~\ref{tab:Large-norm-results-BadNet},
\ref{fig:Large-norm-results-noiseBI}.

\end{document}